\documentclass[a4paper, cits, 11pt]{article}

\usepackage{jinstpubAthena}
\usepackage{graphicx}
\usepackage{amsmath}
\usepackage{wrapfig}
\usepackage{color}
\usepackage{float}
\usepackage[usenames,dvipsnames]{xcolor}
\usepackage{makeidx}
\usepackage[utf8]{inputenc}
\usepackage{multicol}
\usepackage{multirow}
\usepackage{longtable}
\usepackage{subfloat}

\newcommand{\cm}{\operatorname{cm}}
\newcommand{\mm}{\operatorname{mm}}
\newcommand{\nm}{\operatorname{nm}}
\newcommand{\um}{\operatorname{{\mu}m}}
\newcommand{\ns}{\operatorname{ns}}
\newcommand{\ep}{\textit{e}+\textit{p}}

\newcommand{\eA}{\mbox{\textit{e}+A}}
\newcommand{\epA}{\mbox{\textit{e}+\textit{p}\,/A}}

\newcommand{\pA}{\textit{p}+A}

\newcommand{\eAu}{\mbox{\textit{e}+Au}}

\newcommand{\pp}{\mbox{\textit{p}+\textit{p}}}

\title{ATHENA Detector Proposal
\\
\vspace{0.5cm}
{\Large
A Totally Hermetic Electron Nucleus Apparatus \\ \vspace{-2mm}
proposed for IP6 at the Electron-Ion Collider}
}

\author{The ATHENA Collaboration$^*$\\}
\newcommand{\ACU}{1}
\newcommand{\AGH}{2}
\newcommand{\AANSL}{3}
\newcommand{\ANL}{4}
\newcommand{\BHU}{5}
\newcommand{\BARI}{6}
\newcommand{\BARICNR}{7}
\newcommand{\BARIP}{8}
\newcommand{\BARIU}{9}
\newcommand{\BERHAM}{10}
\newcommand{\BIRMINGHAM}{11}
\newcommand{\BOLOGNA}{12}
\newcommand{\BOMBAY}{13}
\newcommand{\BNL}{14}
\newcommand{\BRUNEL}{15}
\newcommand{\CATANIA}{16}
\newcommand{\CATANIAU}{17}
\newcommand{\CUA}{18}
\newcommand{\CCNU}{19}
\newcommand{\CFNS}{20}
\newcommand{\CIAE}{21}
\newcommand{\CIDL}{22}
\newcommand{\COSENZA}{23}
\newcommand{\COSENZAU}{24}
\newcommand{\CUGW}{25}
\newcommand{\CTU}{26}
\newcommand{\DL}{27}
\newcommand{\DAV}{28}
\newcommand{\DELHI}{29}
\newcommand{\DUKE}{30}
\newcommand{\FERRARA}{31}
\newcommand{\FIT}{32}
\newcommand{\FIU}{33}
\newcommand{\FSU}{34}
\newcommand{\BARIF}{35}
\newcommand{\FRAS}{36}
\newcommand{\FUDAN}{37}
\newcommand{\GENOVA}{38}
\newcommand{\GSU}{39}
\newcommand{\GLASGOW}{40}
\newcommand{\GOA}{41}
\newcommand{\GSI}{42}
\newcommand{\CUH}{43}
\newcommand{\IFJ}{44}
\newcommand{\IMP}{45}
\newcommand{\INDORE}{46}
\newcommand{\IOP}{47}
\newcommand{\SACLAY}{48}
\newcommand{\JAMMU}{49}
\newcommand{\JLAB}{50}
\newcommand{\KANSAS}{51}
\newcommand{\CUK}{52}
\newcommand{\KENT}{53}
\newcommand{\UKY}{54}
\newcommand{\LLR}{55}
\newcommand{\LNS}{56}
\newcommand{\LANCASTER}{57}
\newcommand{\LBNL}{58}
\newcommand{\LLNL}{59}
\newcommand{\LEHIGH}{60}
\newcommand{\LIVERPOOL}{61}
\newcommand{\LANL}{62}
\newcommand{\MADRAS}{63}
\newcommand{\JAIPUR}{64}
\newcommand{\MANITOBA}{65}
\newcommand{\MEPHI}{66}
\newcommand{\UMICH}{67}
\newcommand{\MSU}{68}
\newcommand{\MAU}{69}
\newcommand{\NCKU}{70}
\newcommand{\NISER}{71}
\newcommand{\NPI}{72}
\newcommand{\ODU}{73}
\newcommand{\PADOVA}{74}
\newcommand{\PANJAB}{75}
\newcommand{\IJCLAB}{76}
\newcommand{\PATNA}{77}
\newcommand{\PAVIA}{78}
\newcommand{\PAVIAU}{79}
\newcommand{\KOLKATA}{80}
\newcommand{\RICE}{81}
\newcommand{\RIKEN}{82}
\newcommand{\ROMAONE}{83}
\newcommand{\ROMATWO}{84}
\newcommand{\ROMATWOU}{85}
\newcommand{\RAL}{86}
\newcommand{\SCNU}{87}
\newcommand{\SDU}{88}
\newcommand{\SMU}{89}
\newcommand{\SBU}{90}
\newcommand{\TEMPLE}{91}
\newcommand{\TIRUP}{92}
\newcommand{\TORINO}{93}
\newcommand{\TORINOU}{94}
\newcommand{\TRIESTE}{95}
\newcommand{\TRIESTEU}{96}
\newcommand{\TSINGHUA}{97}
\newcommand{\UCB}{98}
\newcommand{\UCD}{99}
\newcommand{\UCLA}{100}
\newcommand{\UCR}{101}
\newcommand{\UIC}{102}
\newcommand{\UMASS}{103}
\newcommand{\TORINOPO}{104}
\newcommand{\USTC}{105}
\newcommand{\CUTN}{106}
\newcommand{\UTA}{107}
\newcommand{\UVA}{108}
\newcommand{\WSU}{109}
\newcommand{\WPI}{110}
\newcommand{\YALE}{111}
\newcommand{\YORK}{112}

\author[\BNL]{J.~Adam}
\author[\AGH]{L.~Adamczyk}
\author[\BOLOGNA]{N.~Agrawal}
\author[\UMICH]{C.~Aidala}
\author[\JLAB]{W.~Akers}
\author[\TORINO]{M.~Alekseev}
\author[\WPI]{M.M.~Allen}
\author[\ROMAONE]{F.~Ameli}
\author[\LLNL]{A.~Angerami}
% ...
\affiliation[*]{\normalfont{The name of a country or region in the affiliation does not indicate political recognition.}}

\emailAdd{Silvia.DallaTorre@ts.infn.it}

\abstract{ATHENA has been designed as a general purpose detector capable of delivering the full scientific scope of the Electron-Ion Collider. Careful technology choices provide fine tracking and momentum resolution, high performance electromagnetic and hadronic calorimetry, hadron identification over a wide kinematic range, and near-complete hermeticity.

This article describes the detector design and its expected performance in the most relevant physics channels. It includes an evaluation of detector technology choices, the technical challenges to realizing the detector and the R\&D required to meet those challenges.
}

\usepackage{subcaption}
\usepackage{placeins}
\usepackage{soul}
\usepackage{rotating}
\usepackage{colortbl}
\usepackage{enumitem}
\usepackage{hvfloat}
\usepackage{lineno}

\usepackage{caption}
\setlist{noitemsep}

\usepackage[toc, nogroupskip, nonumberlist, nopostdot]{glossaries}
\usepackage{glossary-mcols}

\makenoidxglossaries
\newacronym{ip6}{IP6}{Interaction Point 6}
\newacronym{eic}{EIC}{Electron-Ion Collider}
\newacronym{pid}{PID}{Particle Identification}
\newacronym{maps}{MAPS}{Monolithic Active Pixel Sensor}
\newacronym{mpgd}{MPGD}{Micro-Pattern Gaseous Detector}
\newacronym{micromegas}{Micromegas}{Micro-mesh Gas Detector}
\newacronym{gem}{GEM}{Gas Electron Multiplier}
\newacronym{lgad}{LGAD}{Low Gain Avalanche Device}
\newacronym{aclgad}{AC-LGAD}{AC-coupled LGAD}
\newacronym{dirc}{DIRC}{Detection of Internally Reflected Cherenkov light}
\newacronym{hpdirc}{hpDIRC}{high-performance DIRC}
\newacronym{rich}{RICH}{Ring Imaging Cherenkov}
\newacronym{drich}{dRICH}{dual-radiator RICH}
\newacronym{pfrich}{pfRICH}{proximity-focusing RICH}
\newacronym{btof}{bToF}{barrel Time-of-Flight}
\newacronym{becal}{bECal}{barrel Electromagnetic Calorimeter}
\newacronym{necal}{nECal}{electron-endcap Electromagnetic Calorimeter}
\newacronym{pecal}{pECal}{hadron-endcap Electromagnetic Calorimeter}
\newacronym{bhcal}{bHCal}{barrel Hadron Calorimeter}
\newacronym{nhcal}{nHCal}{electron-endcap Hadron Calorimeter}
\newacronym{phcal}{pHCal}{hadron-endcap Hadron Calorimeter}
\newacronym{scifi}{SciFi}{Scintillating Fibers}
\newacronym{sciglass}{SciGlass}{Scintillating Glass}
\newacronym{dis}{DIS}{Deep Inelastic Scattering}
\newacronym{sidis}{SIDIS}{Semi-Inclusive DIS}
\newacronym{dvcs}{DVCS}{Deeply Virtual Compton Scattering}
\newacronym{pdf}{PDF}{Parton Distribution Function}
\newacronym{npdfs}{nPDFs}{nuclear Parton Distribution Functions}
\newacronym{gpds}{GPDs}{Generalized Parton Distributions}
\newacronym{tmds}{TMDs}{Transverse Momentum Distributions}
\newacronym{murwell}{$\mu$RWell}{micro-Resistive Well}
\newacronym{acts}{ACTS}{A Common Tracking Software}
\newacronym{htc}{HTC}{High-Throughput Computing}
\newacronym{osg}{OSG}{Open Science Grid}
\newacronym{rcs}{RCS}{Rapid Cycling Synchrotron}
\newacronym{sipm}{SiPM}{Silicon Photomultiplier}
\newacronym{omds}{OMDs}{Off-Momentum Detectors}
\newacronym{rps}{RPs}{Roman Pots}
\newacronym{zdc}{ZDC}{Zero-Degree Calorimeter}
\newacronym{sr}{SR}{Synchrotron Radiation}
\newacronym{felix}{FELIX}{Front End Link eXchange}
\newacronym{feb}{FEB}{Front-End Board}
\newacronym{fee}{FEE}{Front-End Electronics}
\newacronym{fep}{FEP}{Front-End Processor}
\newacronym{hpc}{HPC}{High-Performance Computing}
\newacronym{lappd}{LAPPD}{Large Area Picosecond Photodetector}
\newacronym{hrppd}{HRPPD}{High Resolution Picosecond Photodetector}
\newacronym{nc}{NC}{Neutral Current}
\newacronym{cc}{CC}{Charged Current}
\newacronym{jb}{JB}{Jacquet-Blondel}
\newacronym{met}{MET}{Missing Transverse Energy}
\newacronym{ff}{FF}{Fragmentation Function}
\newacronym{emt}{EMT}{Energy-Momentum Tensor}
\newacronym{tcs}{TCS}{Timelike Compton Scattering}
\newacronym{bh}{BH}{Bethe-Heitler}
\newacronym{bsa}{BSA}{Beam-Spin Asymmetry}
\newacronym{dvmp}{DVMP}{Deeply Virtual Meson Production}
\newacronym{vm}{VM}{Vector-Meson}
\newacronym{cnm}{CNM}{Cold Nuclear Matter}
\newacronym{tof}{ToF}{Time-of-Flight}
\newacronym{pmt}{PMT}{Photomultiplier Tube}
\newacronym{daq}{DAQ}{Data Acquisition}
\newacronym{asic}{ASIC}{Application-Specific Integrated Circuit}
\newacronym{fpga}{FPGA}{Field-Programmable Gate Array}
\newacronym{ai}{AI}{Artificial Intelligence}
\newacronym{mcppmt}{MCP-PMT}{Micro-Channel Plate PMT}
\newacronym{mip}{MIP}{Minimum Ionizing Particle}
\newacronym{ir}{IR}{Interaction Region}
\newacronym{da}{DA}{Double Angle}
\newacronym{NLO}{NLO}{Next-to-Leading Order}
\newacronym{gwp}{GWP}{Global Warming Power}
\newacronym{trd}{TRD}{Transition Radiation Detector}
\newacronym{cdr}{CDR}{Conceptual Design Report}
\newacronym{tdr}{TDR}{Technical Design Report}
\newacronym{minitpc}{miniTPC}{mini Time Projection Chamber}
\glsaddall

\begin{document}

\maketitle
\flushbottom
\emergencystretch 3em
\renewcommand*{\arraystretch}{1.4}

\section{Introduction}
The Electron-Ion Collider (EIC) will be the world’s first collider of polarized electrons with polarized protons and light nuclei.  It will also be the world’s first collider of polarized electrons with heavy nuclei.  Its purpose will be to explore the quark and gluon structure of protons and nuclei, elucidating the origins of nuclear spin and nuclear mass, and shedding light on emergent phenomena involving dense systems of gluons.  

A large international scientific community has grown around the EIC since its inception. The EIC Users Group~\cite{eicug} was formed in 2016 to coordinate efforts toward developing the science case and detector concepts required to realize the facility. It currently represents more than 1300 scientists worldwide.

The EIC is to be built at the Brookhaven National Laboratory and will be hosted jointly by Brookhaven and the Thomas Jefferson National Accelerator Facility, bringing together two world class laboratories with long standing expertise in building hadron and electron beam accelerators.  It will incorporate the existing Relativistic Heavy Ion Collider and will instrument two interactions points, designated IP6 and IP8, providing locations for two detectors. 

On March 6, 2021, the two host laboratories issued a \emph{Call for Collaboration Proposals for Detectors at the EIC}.  The first detector is to be located at IP6 and falls within the scope of the Department of Energy (DoE) funded project. The call stipulated that the project detector should be based on the reference design developed by the EIC Users Group, which is described in a Yellow Report~\cite{AbdulKhalek:2021gbh} and was included in the EIC Conceptual Design Report (CDR)~\cite{EIC-CDR}.  The project detector must satisfy all the science requirements of the DoE ``mission need'' statement that was informed by the EIC community White Paper~\cite{Accardi:2012qut} and the National Academies of Science (NAS) assessment of EIC science~\cite{NASRep:2018}.  Proposals for a second detector, to be located at IP8, could be complementary in technology choices; optimized for particular areas of EIC science or address science beyond that described in the White Paper and NAS report.

In response to this call, a kick-off meeting was held on March 12-13, 2021, with the aim of forming a collaboration to design a novel, powerful, general-purpose detector that meets all the science requirements within the given cost envelope. Originally named EIC@IP6, this effort was joined by many EIC enthusiasts who had been instrumental in realizing the Yellow Report and the CDR, and who had been active participants in the preceding Generic Detector R\&D Program~\cite{EIC-RD}. In the months that followed, this effort led to the formation of the ATHENA collaboration comprising 94 institutions from 13 countries, with 36\% of participants from North America, 34\% from Europe, and 30\% from Asia. Design activities were organized around ten Working Groups (WGs).  Six detector WGs were established focusing on different aspects of the design, and four physics WGs evaluated the detector performance against the science requirements. A separate software WG supported the development of realistic simulations of the detector design in concert with the detector WGs.  A proposal committee formed of three subgroups: integration and global design, costing and editing; was charged to distil this work into a coherent detector proposal.  This article presents the outcome of this combined effort, describing the design and performance of the proposed ATHENA detector.

\subsection{EIC Physics Scope}


The EIC will be a world-wide unique facility to address fundamental questions regarding visible matter in the universe.  The EIC Yellow Report~\cite{AbdulKhalek:2021gbh} poses the   overarching questions as follows:
\begin{itemize}
    \item{How do the nucleonic properties such as mass and spin emerge from partons and their underlying interactions?}
    \item{How are partons inside the nucleon distributed in both momentum and position space?}
    \item{How do color-charged quarks and gluons, and jets, interact with a nuclear medium? How do the confined hadronic states emerge from these quarks and gluons? How do the quark-gluon interactions create nuclear binding?}
    \item{How does a dense nuclear environment affect the dynamics of quarks and gluons, their correlations, and their interactions? What happens to the gluon density in nuclei? Does it saturate at high energy, giving rise to gluonic matter or a gluonic phase with universal properties in all nuclei and even in nucleons?}
\end{itemize}

\begin{table}[htb]
\begin{tabular}{p{9.5cm}p{5cm}}
\hline
 & \\
{\bf Neutral-current Inclusive DIS:} \epA\,$\longrightarrow e'+X;$ for this process, it is essential to detect the 
scattered electron, $e'$, with high precision. All other final state particles ($X$) are ignored. The scattered electron is critical for all processes
to determine the event kinematics. The key kinematic variable in this process are $x$ and $Q^2$ where
$x$ is the momentum fraction of the quark (w.r.t.~the nucleon) on which the photon scatters. $Q^2$ is the squared momentum transfer to the
electron $Q^2 = -q^2$, equal to the virtuality of the exchanged photon. Large values of $Q^2$ provide a hard scale to the process, which allows one to resolve quarks and gluons in the proton. &
\vspace{-15pt}
\includegraphics[width=5cm]{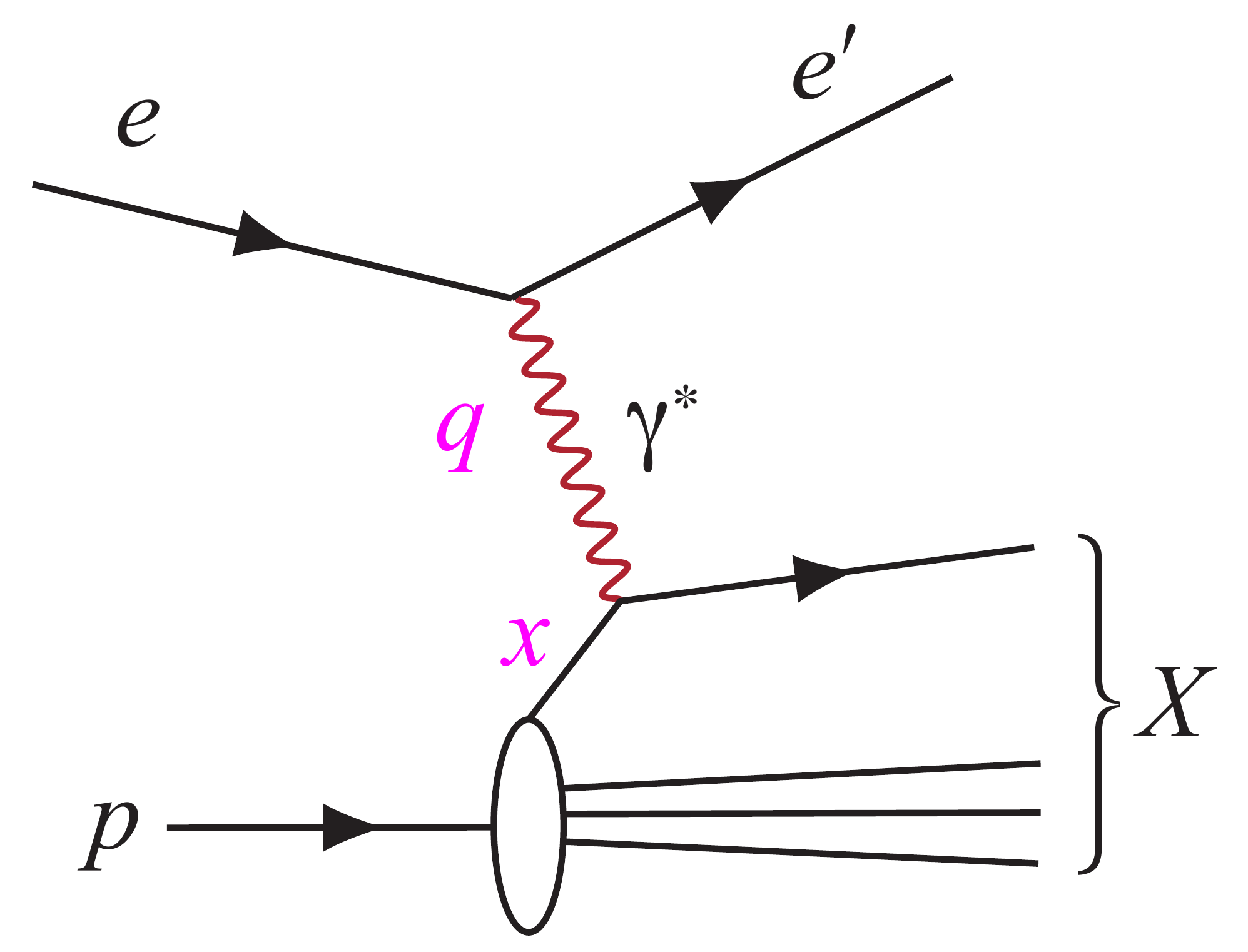}\\
& \\
{\bf Charged-current Inclusive DIS:} \epA\,$\longrightarrow \nu + X;$ at high enough momentum transfer 
$Q^2$, the electron-quark interaction is mediated by the exchange of a $W^{\pm}$ gauge boson 
instead of 
the virtual photon. In this case the event kinematic cannot be reconstructed from the 
scattered electron, but needs to be reconstructed from the final state particles. & 
\vspace{-10pt}
\includegraphics[width=5cm]{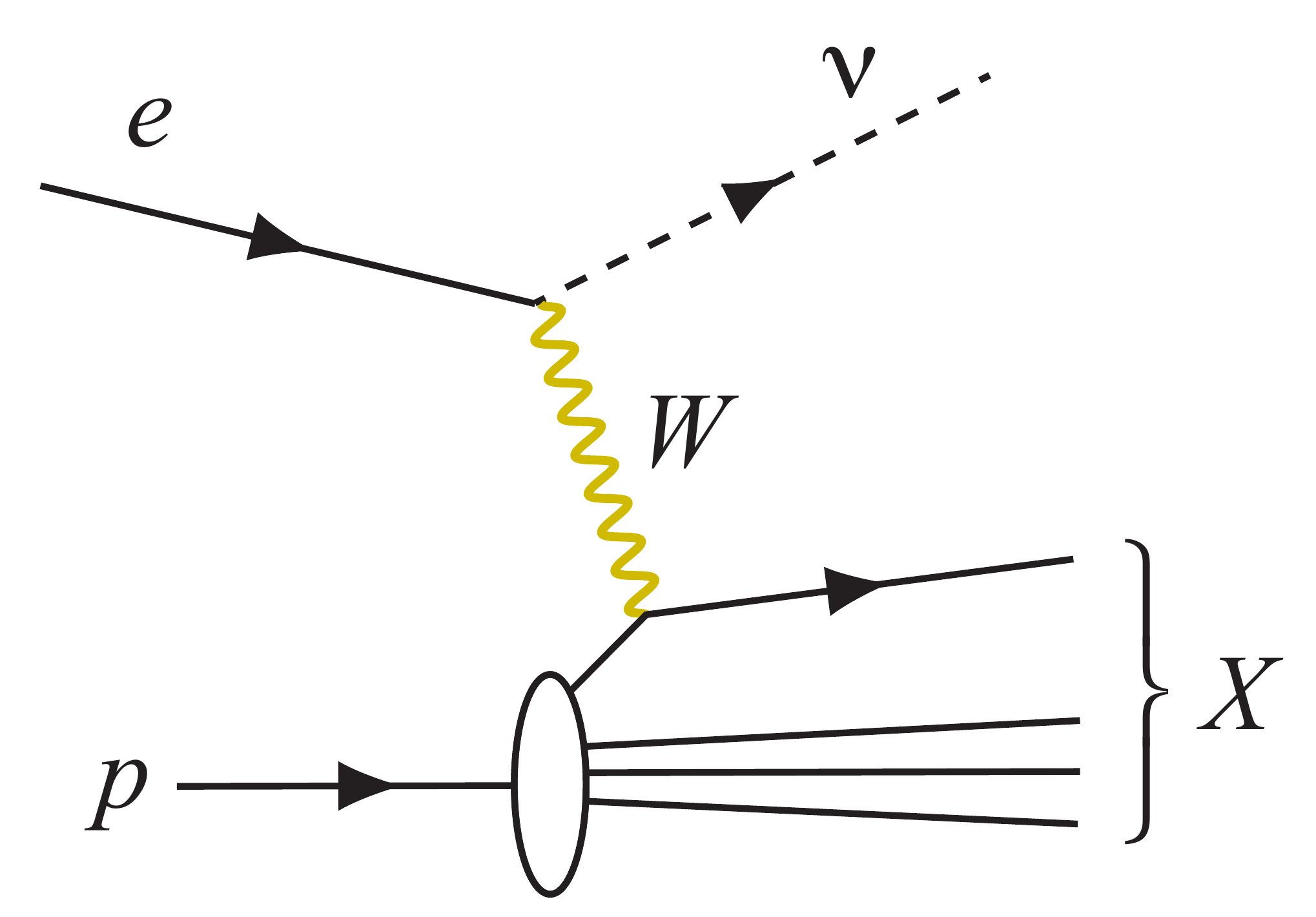} \\ 
& \\
{\bf Semi-inclusive DIS:} \epA\,$\longrightarrow e'+h^{\pm,0}+X$, which requires measurement of  
{\it at least one} identified hadron in coincidence with the scattered electron. &
\vspace{-10pt}
\includegraphics[width=5cm]{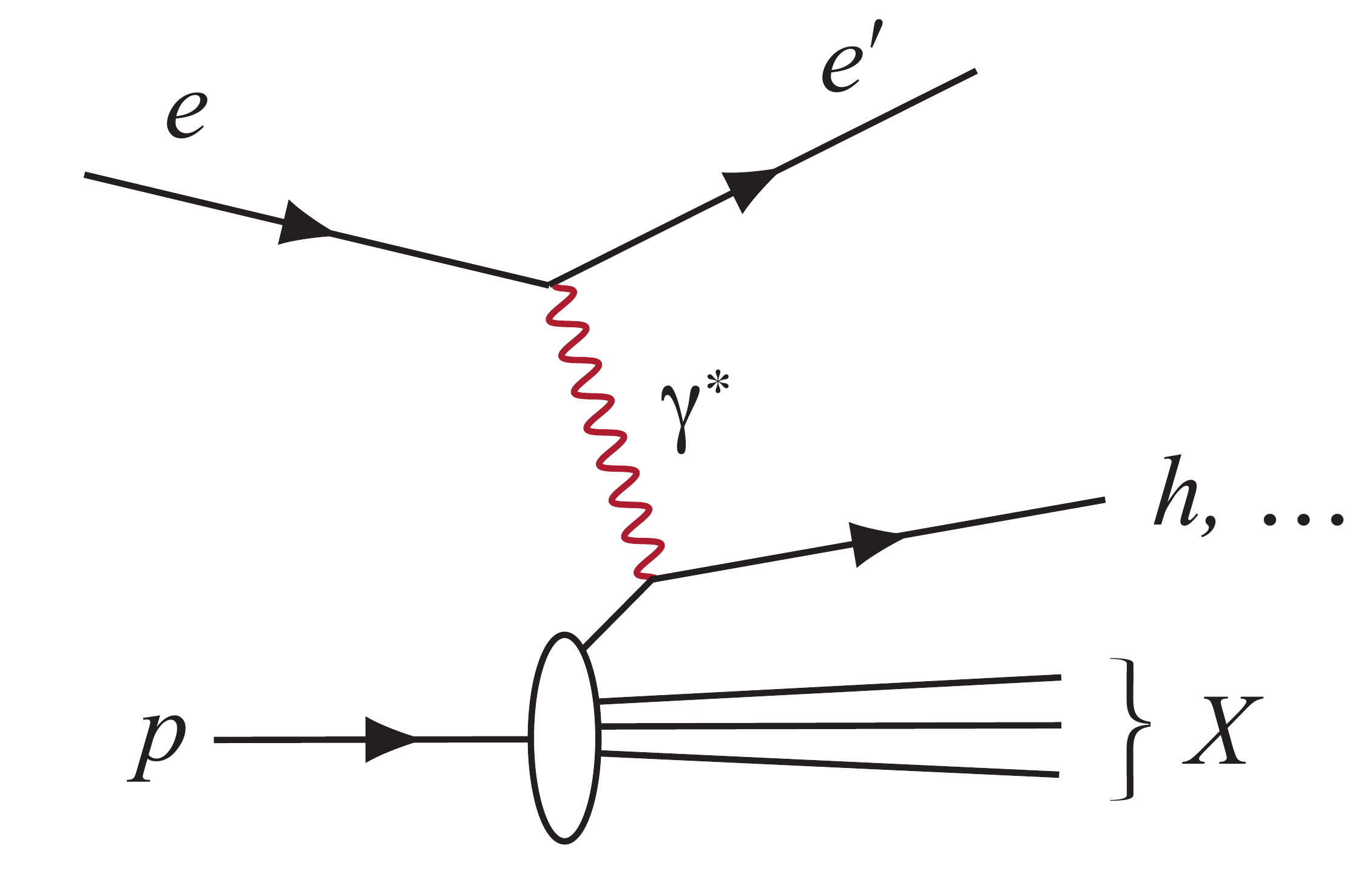} \\
& \\

{\bf Exclusive DIS:} \epA\,$\longrightarrow e'+p\mathrm{/A'}+\gamma/h^{\pm,0}/VM$, which require the 
measurement of {\it all} particles in the event with high precision. A key kinematic variable for this process is $t = (p^\prime - p)^2$, the invariant square of the momentum transfer of the scattered proton or ion, which is crucial for all parton imaging studies.&
\vspace{-15pt}
\includegraphics[width=5cm]{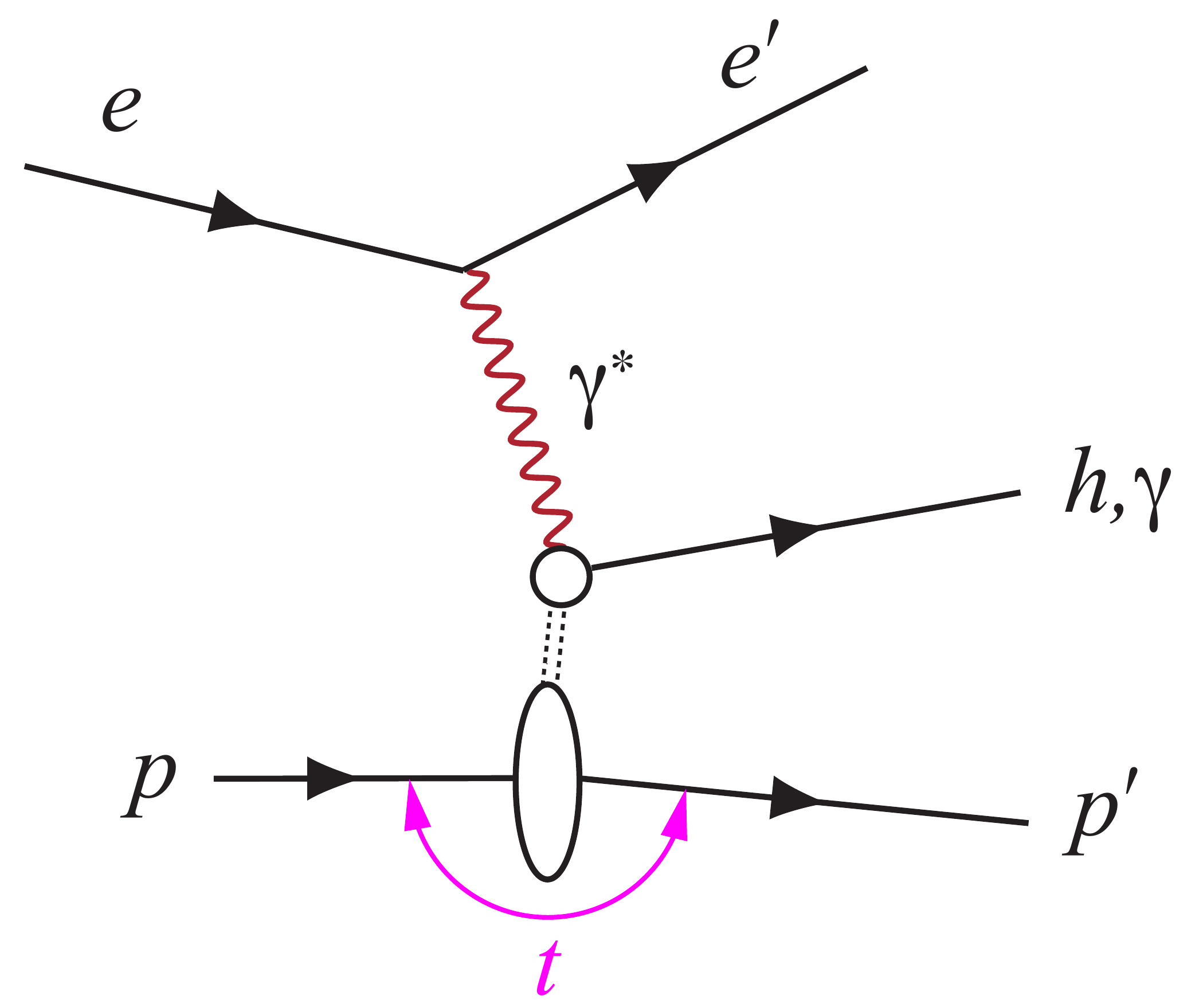} \\
\hline
& \\
\end{tabular}
\caption{\label{DIS.processes} Different categories of processes measured at an EIC (Initial state: Colliding electron ($e$), proton ($p$), and nuclei (A). Final state: Scattered electron ($e'$), neutrino ($\nu$), photon ($\gamma$), hadron ($h$), and hadronic final state ($X$)). Key kinematic variables are indicated in magenta.}
\end{table}

\FloatBarrier

The EIC facility is designed~\cite{EIC-CDR} to collide electrons with a variety of ions from protons up to the heaviest stable nuclei at center-of-mass energies ranging from 20 to 140 GeV.
The colliding beam electrons, protons, and light ions can be spin-polarized at the level of 70\%.  The luminosity is expected to reach $10^{34}\,\mathrm{cm}^{-2}\mathrm{s}^{-1}$ for electron-proton collisions.
The interaction region will have an integrated detector capable of nearly 100\% kinematic coverage and the possibility of a second interaction region is foreseen.

Core physics processes to address the above science questions are shown in Table~\ref{DIS.processes}, which also introduces their main kinematic variables.
The main classes of observables are differential production cross sections, correlations, and spin asymmetries, which in turn give insight into the distributions of quarks and gluons and their dynamics.
In the case of nucleon spin, for example, it is well known that only a small fraction is accounted for by quark spins, leaving large gaps in the knowledge and understanding of the roles of gluons and orbital angular momenta.
The EIC will enable a vast ``tomography'' program of inclusive, semi-inclusive, and exclusive DIS measurements that will provide precision data sensitive to both, the spatial and momentum distributions of quarks and gluons in nucleons and nuclei, including their spin dependencies.
The capability to measure collisions of electrons with a variety of ions over a wide range of center-of-mass energies will offer unprecedented possibilities to study the quark and gluon structure of nuclear matter, hadron formation in and transport through this environment, as well as tantalizing prospects to probe the highly occupied gluon states of heavy nuclei at low-$x$, where gluon self-interactions are predicted to give rise to new degrees of freedom and phenomena.

We would like to refer the reader to Refs.~\cite{Boer:2011fh,Proceedings:2020eah}, reporting on programs at the Institute for Nuclear Theory dedicated to EIC physics, as well as the EIC White Paper~\cite{Accardi:2012qut} and Yellow Report~\cite{AbdulKhalek:2021gbh} for comprehensive overviews of these and other EIC physics opportunities.

\subsection{Detector Overview}

As an entirely new detector, ATHENA has been designed to accommodate all necessary subsystems without compromising on performance, while leaving room for future upgrades. 
Central to the proposal is a new, large-bore magnet with a maximum field strength of 3\,T. 
Particle tracking and vertex reconstruction will be performed by a combination of next-generation silicon pixel sensors and state-of-the-art 
micro-pattern gas detectors. The combination of magnetic field strength and high resolution, low mass tracking technologies optimizes momentum resolution and vertex reconstruction. The large bore of the magnet allows for layered, complementary, state-of-the-art particle identification technologies. A novel hybrid imaging/sampling electromagnetic calorimeter is proposed for the barrel region of the detector, along with a high resolution crystal calorimeter in the electron-going direction. 
The hadron endcap will have calorimetry, tracking and \gls{pid} detectors that are optimized for high-momentum hadron identification and high-energy jet reconstruction. We have striven for hermeticity by closely integrating the far-forward and far-backward detectors with the central detector to achieve maximal kinematic coverage and to optimize the detection of particles at small scattering angles. Careful choice between cutting-edge and mature detector technologies achieves the necessary detector performance while minimizing risk and providing a cost-effective solution that is achievable on the required timescale. Scalable modern technology choices assure optimum performance for multi-year operation from day one.

The integrated ATHENA detector is shown in Fig.~\ref{fig:ATHENA-Layout-Main}. 
To achieve the required level of performance, the inner subsystems must comprise very low-mass detectors and support structures. This is accomplished by an inner tracking system (vertex layers, barrel layers and disks) 
based on low-power consumption silicon \gls{maps} technology complemented by cylindrical \gls{micromegas} layers at larger radii in the barrel and \gls{gem} rings in the forward/backward direction.  The required level of particle identification is achieved with a wide range of complementary technologies comprising a \gls{hpdirc} detector in the barrel augmented by an \gls{aclgad} time-of-flight layer at smaller radii; a \gls{drich} in the forward region and by a single-volume \gls{pfrich} in the electron endcap. In the forward region, a \gls{murwell} tracker is positioned behind the \gls{drich} to improve tracking and pointing accuracy.  In the barrel, tracking information behind the \gls{hpdirc} is provided by silicon pixel sensors in the imaging part of the \gls{becal}.  
The \gls{becal} is a hybrid of imaging calorimetry employing silicon pixel sensors and sampling calorimetry based on \gls{scifi} embedded in lead.
The complete barrel tracking and barrel electromagnetic calorimeter subsystems are contained within the superconducting solenoid. 

The forward (proton-going direction) calorimeters consist of a W/\gls{scifi} \gls{pecal}) augmented by an iron-scintillator sampling \gls{phcal}. The backward (electron-going direction) electromagnetic calorimeter (nECal) has the most stringent resolution requirements since it must measure scattered electrons with high precision. This is achieved by using lead tungstate (PbWO$_4$) crystals in the inner part and \gls{sciglass} in the outer part. This is completed by an iron-scintillator sampling \gls{nhcal}. The barrel and endcap hadronic calorimeters sit outside the solenoid and serve as flux returns for the magnet. 

Table~\ref{tab:subsystemTable} 
lists the ATHENA subsystems in the central detector and serves as a reference for Chapter~\ref{chapter:detector} where the technology choices and performances are discussed.
Crucial to addressing many of the science questions is the detection of both electrons and hadrons scattered at small angles close to the beam.  The arrangement and technology choices of far-forward and far-backward detectors are also discussed in Chapter~\ref{chapter:detector}.

\begin{sidewaysfigure}
    \centering
    \makeatletter
    \if@twoside
        \rotatebox[origin=c]{180}{%
        \begin{minipage}{\textwidth}
        \input{Introduction/detector_figure}
        \end{minipage}
        }
    \else
        \input{Introduction/detector_figure}
    \fi
    \makeatother
\end{sidewaysfigure}

\begin{sidewaystable}
\begin{minipage}{0.99\textwidth}
%
%
\centering
\scriptsize
\renewcommand{\arraystretch}{1.4}
\caption{Complete list of ATHENA subsystems in the central detector ordered from small to large radii (barrel) and increasing distance from the interaction point (forward and backward regions). The PID range in momentum is quoted for $3\sigma$ separation.}
\label{tab:subsystemTable}
\begin{tabular}{|p{3mm}|p{3.4cm}|p{4cm}|p{6.4cm}|p{3.2cm}|p{3cm}|}
%
%
\hline
\multicolumn{1}{|c}{} &
\multicolumn{1}{|c}{\textbf{Detector}} &
\multicolumn{1}{|c}{\textbf{Purpose}} &
\multicolumn{1}{|c}{\textbf{Technology}} &
\multicolumn{1}{|c|}{\textbf{Acceptance}} &
\multicolumn{1}{|c|}{\textbf{PID Range (GeV/$\mathbf{c}$)} } \\
\hline \hline
%
%
\parbox{3mm}{\multirow{8}{*}{\rotatebox{90}{\textbf{Forward (h-going)}}}}  & Si-Tracker Disks & Tracking  &  6 disks of MAPS  &  $1.1 < \eta < 3.75$ & {}\\ \cline{2-6}
{} & Tracking Rings (MPGD) &  Tracking  &  Planar GEMs with annular shape surrounding the Si-disks  &  $1.1 < \eta < 2.0$ & {} \\ \cline{2-6}
{} & dRICH  &  PID  &  Dual RICH with aerogel and gas  &  $1.2 < \eta < 3.7$ & $3 < p < 60$ ($K/\pi$)\newline $0.85 < p < 15$ ($e/\pi$) \\ \cline{2-6}
{} & MPGD Layer  &  Tracking  &  Planar $\mu$RWell disk  &  $1.4 < \eta < 3.75$ & {}\\  \cline{2-6}
{} & pECal  &  e/m Calorimetry  &  W-Powder/SciFi calorimeter   &  $1.2 < \eta < 4.0$ & {} \\ \cline{2-6}
{} & pHCal   &  Hadron Calorimetry  &  Fe/Sci sandwich  &  $1 < \eta < 4.0$ & {}\\ \hline\hline
%
%
\parbox{3mm}{\multirow{11}{*}{\rotatebox{90}{\textbf{Barrel}}}} & Si Vertex-Tracker & Tracking and Vertexing & 3-layer MAPS & $-2.2 < \eta < 2.2$ & {}\\ \cline{2-6}
{} & Si Barrel-Tracker & Tracking & 2-layer MAPS & $-1.05 < \eta < 1.05$ & {} \\ \cline{2-6}

{} & bToF & PID and Tracking & AC-LGAD & $-1.05 < \eta < 1.05$\newline $p_T > 0.23$ GeV/c @ 3T & $p < 1.3$ ($K/\pi$)\newline $p < 0.4$ ($e/\pi$) \\ \cline{2-6}

{} & Barrel Tracker (MPGD) & Tracking & 4 (2+2) layer cylindrical Micromegas & $-1.05 < \eta < 1.05$ & {} \\ \cline{2-6}
{} & hpDIRC & PID & DIRC with focusing elements and fine pixel readout & $-1.64 < \eta < 1.25$\newline $p_T > 0.45$ GeV/c @ 3T & $p < 6.5$ ($K/\pi$)\newline $p < 1.2$ ($e/\pi$) \\ \cline{2-6}
{} & bECal	 & e/m Calorimetry \& Tracking  & 	Hybrid with Astropix imaging layers alternated with Pb/SciFi layers followed by a set of Pb/SciFi layers  & 	$-1.5 < \eta < 1.2$ & {} \\ \cline{2-6}
{} & bHCal	 & Hadron Calorimetry	 & Fe/Sci sandwich  & $-1.0 < \eta < 1.0$ & {} \\ \hline\hline
%
%
\parbox{3mm}{\multirow{8}{*}{\rotatebox{90}{\textbf{Backward (e-going)}}}}  &  Si-Tracker Disks & Tracking & 5 disks of MAPS & $-1.1 > \eta > -3.8$ & {} \\ \cline{2-6}
{} & Tracking Rings (MPGD) & Tracking & Planar GEMs with annular shape surrounding the Si-disks & $-1.1 > \eta > -1.8$ & {} \\ \cline{2-6}
{} & pfRICH & PID & Proximity focusing RICH with aerogel & $-1.5 > \eta > -3.8$ &  $3 < p < 11$ ($K/\pi$)\newline
$0.85 < p < 3$ ($e/\pi$) \\ \cline{2-6}
{} & Inner nECal & e/m Calorimetry & PbWO$_4$ & $-2.3 > \eta > -4.0$ & {} \\ \cline{2-6}
{} & Outer nECal & e/m Calorimetry & SciGlass & $-1.5 > \eta > -2.3$ & {} \\ \cline{2-6}
{} & nHCal  & Hadron Calorimetry & Fe/Sci sandwich & $-1 >  \eta > -4$ & {} \\ \hline
\end{tabular}
\end{minipage}
\end{sidewaystable}

\subsection{ATHENA Capabilities}

The performance benefits of the ATHENA detector design include: high resolution reconstruction of the scattered electron, accompanied by effective electron/pion separation, to help optimize event-by-event kinematics reconstruction; hadron endcap tracking and PID resolution optimized for forward hadron and jet measurements; novel 
barrel electromagnetic calorimetry with superb resolution for electrons and photons, 
providing high precision over a wide $x$ and $Q^2$ range; and an ability to run with lower field to optimize acceptance at different center-of-mass energies. These are
essential for inclusive \gls{dis}, \gls{dvcs}, and \gls{dvmp}.
This ATHENA strategy enables making complementary measurements in each of the key science areas with minimized systematic uncertainties, for example by adding novel jet measurements to the \gls{sidis} studies laid out in the \gls{eic} White Paper.

ATHENA's ability to measure with high resolution over a wide kinematic range will yield significant advances in our knowledge of parton distributions in nucleons and nuclei. This translates to early discovery potential at the \gls{eic} on the origin of spin in polarized \ep\ collisions, in the search for gluon saturation, and measuring \gls{npdfs} at small $x$ in \eA\  collisions. 
Electro- and photo-production of vector mesons in \ep\ and \eA\  collisions are key observables for saturation and origin of the nucleon mass; ATHENA's momentum resolution allows for good separation of resonance  states. Insights into energy loss and transport properties in dense gluonic matter will be enabled by precision measurements over a large $Q^2$ and $x$ range for \gls{sidis} including heavy flavor, jets and their substructure. 

ATHENA will make, with high precision, the challenging measurements required to extract \gls{gpds} and \gls{tmds}, which encode the full structure of the nucleon and nuclei.
Measurements of \gls{gpds} via \gls{dvcs} are enabled by ATHENA's excellent photon measuring capability.
Novel TMD measurements to study the valence region are facilitated by highly effective jet and hadron measurements, identification capabilities, and complementary event reconstruction in the hadron-going direction. 

Spectroscopy measurements are optimized by the superior tracking resolution and \gls{pid} reach in the barrel region. 
Hadronization studies
utilize ATHENA's comprehensive hadron \gls{pid} capabilities for jet fragmentation studies, along with precision measurements of jet substructure. Measurements of long range correlations, of particular interest in \eA\ collisions, require high resolution tracking over a wide range of rapidity and momentum, which is a key attribute of the ATHENA design.

\subsection{Structure of the Proposal}
In Section 2 we present the detailed ATHENA Detector concept and motivate its design and technology choices. 
Section~\ref{chapter:science} contains an assessment of the impact of ATHENA  on \gls{eic} science. 

The results presented in this article 
fall into three categories: 
1) based on Analytical Calculations ({\bf AnaCal}), 2) on full GEANT4 based simulations, 
where we implemented the complete ATHENA baseline geometry, including the material needed for detector services and a realistic reconstruction framework in a newly developed software environment
(Sec.~\ref{sec:software-computing}) leveraging modern developments from the HEP community ({\bf FullSim}), and 3) from fast simulations based on calculations or parameterized smearing of detector resolution validated by full simulations or comparison with full simulation samples ({\bf FastSim}). These acronyms are used in the text to refer to the different approaches.

 \FloatBarrier

\section{The ATHENA Detector}
\label{chapter:detector}

\subsection{Design Considerations}
The ATHENA detector will meet the goals of the EIC physics program over the full range of center-of-mass energies from 20 to 140 GeV with the largest acceptance practically possible. The following principles guided the design:
\begin{description}[leftmargin=0cm]
\item[Overall size:] 
ATHENA benefits from using all the available space for experimental equipment at \gls{ip6} and respects all constraints imposed by the existing detector hall.
Our design accommodates all subsystems without compromising on performance, eases routing of services and leaves room for future upgrades. 
\item[Acceptance:] The central detector will deliver physics in the range of $-3.8 < \eta < 3.75$ (approximately 3$^\circ$ to 177$^\circ$), augmented by far-forward and far-backward detectors for maximum $x-Q^2$  coverage and detection of small angle particles vital for the physics program.
\item[Magnet:] 
A Solenoid with a magnetic field up to 3\,T with ample space to accommodate all ATHENA subsystems and possible future upgrades.
\item[Performance:] ATHENA is designed to deliver the science detailed in the National Academies assessment~\cite{NASRep:2018} and developed in the Yellow Report~\cite{AbdulKhalek:2021gbh}. Careful consideration is given to alternative, cutting-edge technologies 
where these offer substantially improved performance. The proposed detector is designed to be both achievable and low risk. Its performance has been validated through GEANT4 simulations.
\item[Robustness:]  
As the EIC Project has resources for a single general purpose detector, the design must be robust.
Robustness has been a key parameter in the technology choices by including an adequate level of redundancy. 
Accessibility for maintenance has also been carefully considered by the ATHENA integration engineers.
\item[Upgrade capability:]  The detector design foresees room at strategic locations to facilitate future 
upgrades should the first physics results suggest focusing on certain signals or processes that require enhanced performance,
for example expanded particle identification coverage.
\item[Cost effectiveness:] Where possible, cost-effective technologies were chosen without compromising performance. 
\end{description}

\noindent The ATHENA concept arises from almost a year of creative design by physicists and engineers, building upon
prior and ongoing detector research and development~\cite{AbdulKhalek:2021gbh,EIC-CDR,EIC-RD}. This represents progress in several areas
beyond the Yellow Report~\cite{AbdulKhalek:2021gbh}:

\begin{description} [leftmargin=0cm]
\item [Magnet design:]
Design of a large-aperture 3\,T solenoid that satisfies the needs of field projectivity in the forward region posed by the gaseous RICH, but also provides minimal practically achievable material budget ($\sim$1.3 nuclear interaction lengths $\lambda_I$) in the radial direction.
\item [Tracking system:]
Design and performance evaluation of a tracking subsystem, where services are included and accounted for in the simulation.
\item [Particle identification:]
(i) The implementation of \gls{pid} in the forward region of the central detector, where the large-angular-acceptance \gls{drich} has an adequate length of the gas radiator to achieve optimal optical focusing, (ii) the selection of \gls{pid} devices in the forward and in the backward region with a large overlap of technologies, (iii) a time-of-flight system using \gls{aclgad} technology complementing the \gls{hpdirc} in the barrel to ensure \gls{pid} over the complete momentum range of interest, starting below 1 GeV/c.
\item [Calorimetry:]
Hybrid electromagnetic calorimetry in the barrel with imaging layers delivering remarkable $\pi$/e separation at low momenta, a space point following the \gls{hpdirc}, contributing to hadronic calorimetry, and with spatial resolution which improves $\pi^0/\gamma$ separation.
\item [Integration strategy:]
A set of subdetectors matching the physics needs, with a clear strategy for mechanical supports, installation, and initial development of the related engineering model.
\item [Software approach:] An integrated simulation toolkit with a modular structure using software packages such as the DD4hep detector description toolkit, the Gaudi event processing framework, and \gls{acts} suite, which are supported by large international collaborative efforts. The result is a software stack that scales to modern heterogeneous computing architectures, leveraging current and future \gls{htc} and \gls{hpc} capabilities. This is a major step towards a modern software environment for the EIC.
\item [Expertise:]
The expertise of existing and potential future collaborators and collaborating institutions has been considered in the planning of ATHENA.
\end{description}

 \FloatBarrier

\subsection{Magnet Design}
A key feature of 
ATHENA is a large-bore, superconducting solenoid with a maximum field strength of 3\,T. An inner bore diameter of 3.2~m and a coil length of 3.6~m are sufficient to provide the barrel tracking with a uniform high-field region, while also containing the PID devices and the \gls{becal}.

\begin{figure}[htb]
    \centering
    \includegraphics[width=0.8\textwidth]{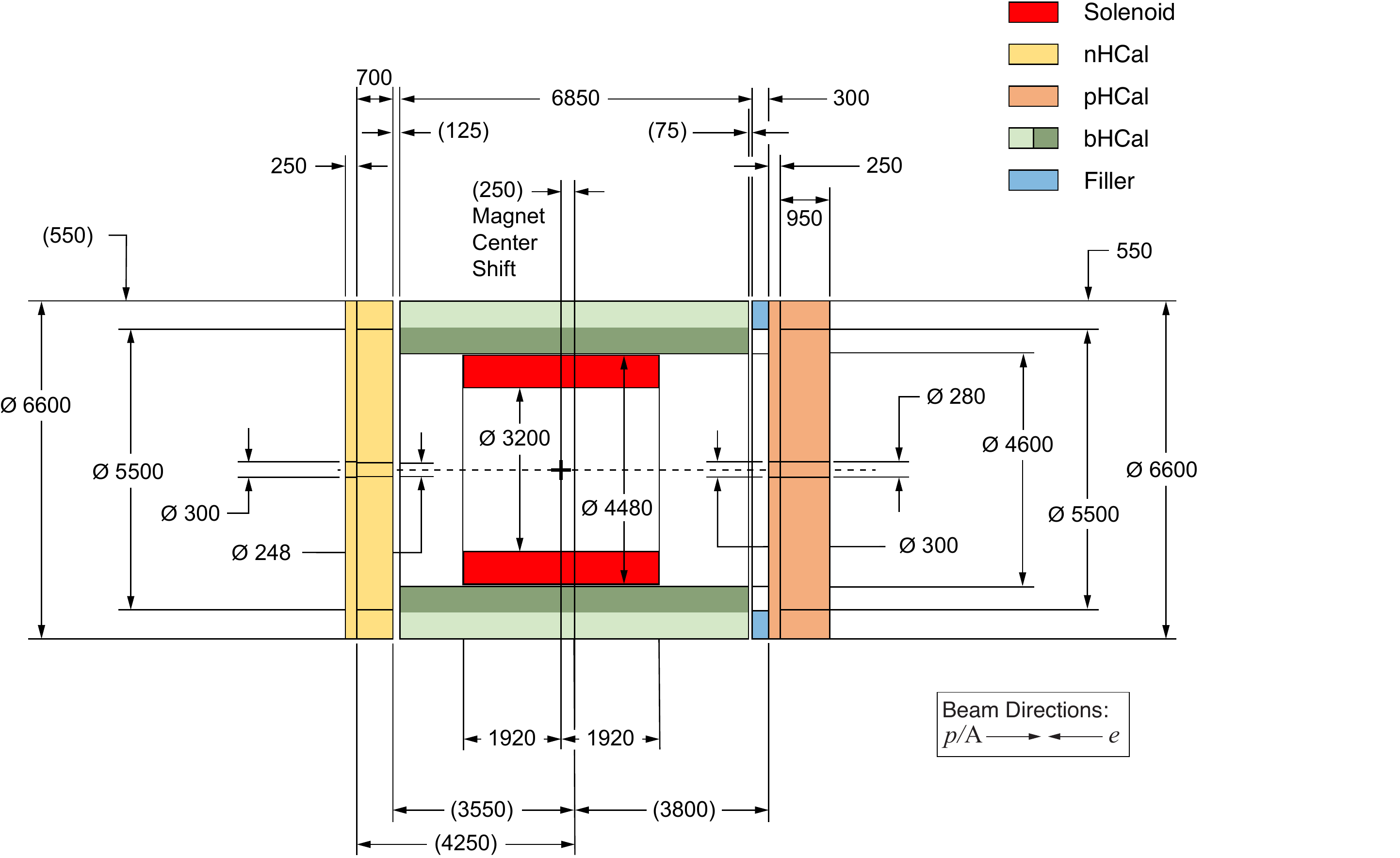}
    \caption{Dimensions of the cryostat and the hadron calorimeters.}
    \label{fig:mag_dimensions}
\end{figure}

To accommodate the magnet coil, the cryostat has an outer diameter of 4.11~m and a total length of 3.84~m. For initial design studies, a simplified barrel was used to reduce the modeling time. For magnetic analysis, only the hadron calorimeters have been considered as these are the only subdetectors constructed from magnetic material. The HCals have three main components: an electron-endcap HCal, a hadron-endcap HCal and a barrel HCal, as illustrated in Fig.~\ref{fig:mag_dimensions}. The magnet center is shifted by 25~cm  with respect to the center of the detector and the nominal interaction point towards the electron endcap. This compensates for the different iron content and location of the two endcap calorimeters, thereby reducing the axial forces on the magnet to enable a
solenoid design with minimum mechanical support.  The current model assumes a generic B-H curve; actual material properties will be incorporated into the model at a later date.

\subsubsection{Field shape}

Key design constrains for the magnet were the flat-field region, the shape of the magnetic field lines in the volume of the \gls{drich} in the forward direction, and limits on the tolerable stray field for the accelerator components.
The space for the dRICH extends from $z=+190$~cm to $z=+330$~cm with an angular range of 3.5$^\circ$ to 25$^\circ$. 
Ideally, the field lines in this region should be projective to minimize distortions due to track bending in the gaseous volume of the \gls{drich} (see Sec.~\ref{dRICH}). This requirement affects the uniformity achievable in the flat-field region and therefore requires careful optimization.
To make the effect of the solenoid on the accelerator tolerable, the requirement is for a stray field smaller than 5G in the region $z = \{7.4,8.0\}$~m and $z = \{-5.3,-7.1\}$~m, respectively. 
The \gls{rcs} is radially 335.2~cm from the magnet central axis where the integral field requirement over the length of the detector is $< 0.007$~Tm. The currently achieved fringe fields will be further improved in the next design phase. 

\begin{figure}[htb]
    \centering
    \includegraphics[width=0.95\textwidth]{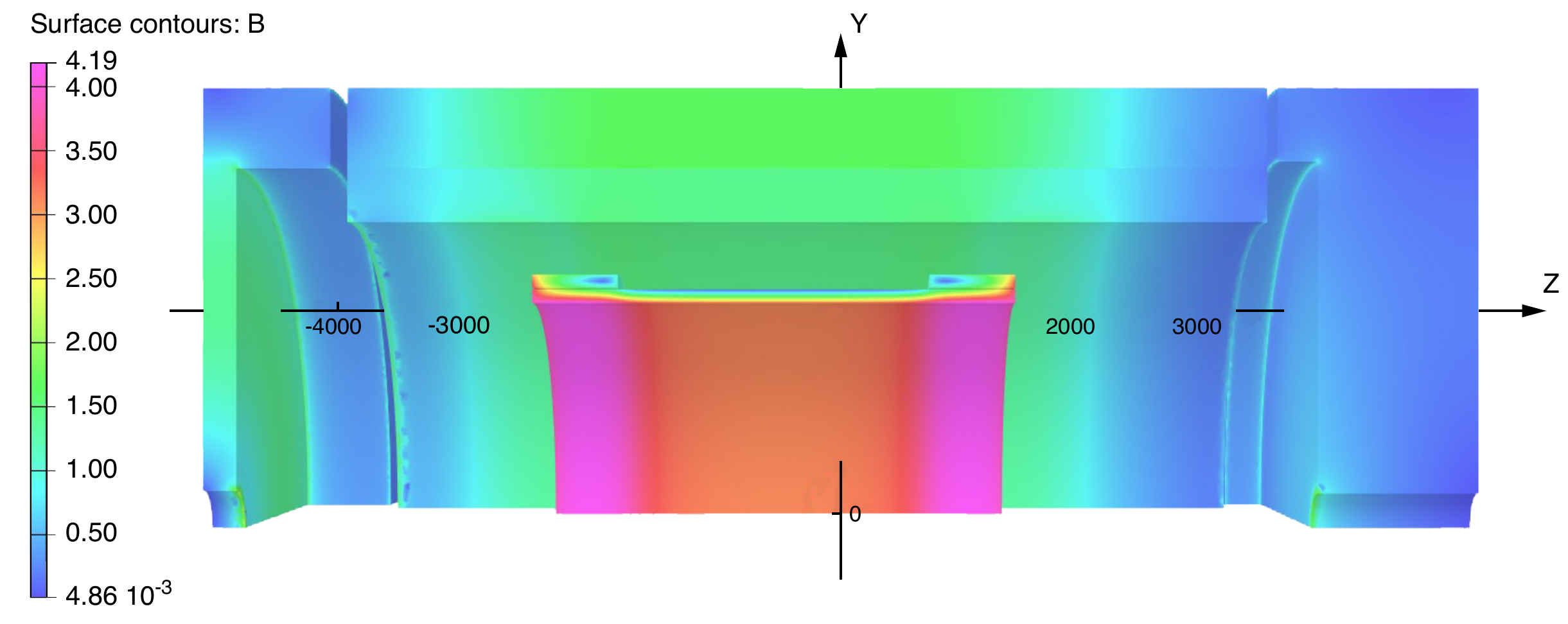}
    \vspace{-5mm}
    \caption{Magnetic field in the coil and barrel and endcap HCals.}
    \label{fig:mag_field}
\end{figure}

The electromagnetic analysis was performed using the SIMULIA Opera simulation package. The field in the coil and in the HCals is shown in Fig.~\ref{fig:mag_field} and the main design parameters are summarized in Tab.~\ref{tab:my_params}.

\begin{table}[htb]
    \centering
    \footnotesize
    \caption{Magnet design parameters.}
    \begin{tabular}{|l|c|l|}
    \hline
      \textbf{Parameter} & \textbf{Value} & \textbf{Units}  \\
    \hline
    \hline
      B in the solenoid centre & 3.02 & T \\
      Peak field in the coil & 4.19 & T \\
      Stored energy & 175.5 & MJ \\
      B @ z = -5.6 m & 10.5 & mT \\
      B @ z = 7.2 m & 5.2 & mT \\
      Homogeneity in flat field area & 27.5 & \% \\
      Projectivity in the forward RICH area & 10.4 & T/Amm$^2$ \\
      Axial force & 37 & kN \\
    \hline
    \end{tabular}
    \label{tab:my_params}
\end{table}

\subsubsection{Choice of conductor}

In order to minimize the material budget in front of the barrel HCal, a NiTi Rutherford cable with a 5N AlNi1\% stabilizer is chosen. The Rutherford cable is made up of 40, 0.84 mm diameter Cu-(NbTi) strands. 
Preliminary calculations show that this conductor is a safe choice with a 1.96~K temperature margin (see Sec.~\ref{Sec.CryoDesign}), a 35\% current margin, and approximately a 50~K hot spot temperature in the event of a quench. 

\subsubsection{Mechanical design}

The preliminary mechanical analysis has been carried out in 2-D. This analysis assumes the orthotropic behavior of the conductor and assumes a sliding contact between the magnet and the support structure. The support structure is 50~mm thick aluminum. Calculations of both the cool down and energization loads have been performed. In both cases, the stresses in the coil and the support structure are well below the design limits. The peak stress at energization (worst case scenario) is 63~MPa in the coils and 73~MPa in the support structure, while the design limits 
are 70~MPa and 135~MPa, respectively. The limits are based on two-thirds of the yield stress of the material.

\subsubsection{Cryogenic design}
\label{Sec.CryoDesign}

The available cryogens for the magnet are supercritical helium at 4.5~K and 3.5~bar for the cold mass and helium gas at 45~K and 15 bar for the shields. The return gas of the cold mass and the shields are expected to be at 4.8~K at 1.28~bar and 80~K at 14 bar, respectively. The allotted power budget for the magnet is 100~W at 4.5~K and around 400~W at 80~K. Based on the magnet size and type, the magnet cooling will be done using the thermosiphon method. Preliminary heat load calculations show that the load is well within the available limits. 

 \FloatBarrier

\subsection{Vertex and Tracking System}
\label{vertex-tracking}
\begin{figure}[htb]
   \center
	\includegraphics[width=0.75\textwidth]{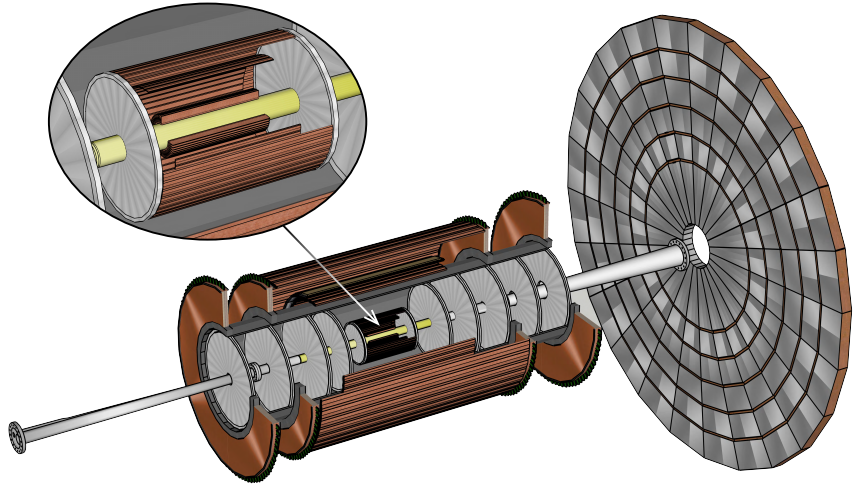}
	\caption{ATHENA baseline hybrid tracking system comprising MAPS vertex and barrel layers and forward/backward \gls{maps} disks complemented by large-area \gls{micromegas} detectors in the outer barrel layers and forward/backward GEM and \gls{murwell} disks.}
    \label{fig:track_baselineconfig}
\end{figure}

ATHENA will utilize silicon \gls{maps} \label{sec:maps} near the interaction point and \glspl{mpgd} \label{sec:mpgd} farther out. This configuration allows for a low material budget 
tracking system with sufficient redundancy over a large lever arm, which is critical to achieve the required momentum resolution.  The layout of the vertex and tracking system is illustrated in Fig.~\ref{fig:track_baselineconfig}. A compact inner silicon barrel consists of three vertex layers and two barrel layers occupying a region that has a maximum radius of $18\cm$ and a total length of $48\cm$. 
The vertex layers are made of large-area, wafer-scale, stitched sensors that are bent around the beam pipe, allowing the first vertex layer to be placed very close to the interaction point at a radius of $33\mm$.  The barrel layers comprise a more traditional stave design that uses smaller stitched sensors. The two outermost barrel layers will each comprise two closely-spaced 2-D layers of Micromegas with mean radii of approximately $49\cm$ and $76\cm$, and maximum total length of approximately $200\cm$.

In the forward and backward directions, the vertex and tracking system consists of silicon disks augmented by large-area \glspl{gem}. The silicon disks will use the same sensor technology as the vertex and barrel layers.  In an effort to minimize material in the backward (electron-going) direction, there are five disks, while in the forward (proton/nucleus-going) direction there are six disks.  They start $25\cm$ either side of the interaction point and extend to $145\cm$ in the backward direction and $165\cm$ in the forward direction. The maximum outer radius of the disks is approximately $43\cm$. The minimum radii are determined by the divergence of the beam pipe. Two triple-\glspl{gem} detectors with an inner and outer radius of about $45\cm$ and $76\cm$, respectively, are implemented near the two silicon disks furthest from the IP to extend the acceptance for tracks and provide additional hit points for track reconstruction in the pseudorapidity interval $1.1 < |\eta| < 2.0$. Finally, a \gls{murwell} detector, with a radius of about $196\cm$ is located behind the \gls{drich} detector in the forward direction. This detector helps seed the \gls{drich} ring finder and improves the momentum resolution in the forward direction.

\subsubsection{Choice of technology}

The silicon detectors will use the latest $65\nm$ \gls{maps} technology that was identified in the Yellow Report \cite{AbdulKhalek:2021gbh} as the best candidate to meet the stringent requirements on vertex and momentum resolution. This technology is currently being developed for an upgrade of the inner tracking system of the ALICE experiment at the LHC at CERN. The upgraded ALICE detector (ITS3) is expected to be ready for installation during the next long LHC shutdown from 2026 to 2028.  

The specifications of the proposed ITS3 sensor already meet most, if not all, of the requirements of the EIC. 
ATHENA plans to use the ITS3 sensor in all parts of its silicon tracker with size optimized for vertex layers with minimal material and for cost effective large area coverage in barrel layers and disks. The overarching goal of ITS3 is to achieve a pixel pitch down to $10\um$ while keeping power dissipation below $20~\mathrm{mW\ cm^{-2}}$ to construct a vertex detector with a space point resolution of better than $5\um$ for a material thickness of just $0.05\%\ X/X_0$ per layer. By comparison, the vertex layers in the current ALICE ITS (ITS2) have a pixel pitch of approximately $30\um$, dissipate $40~\mathrm{mW\ cm^{-2}}$ and each have a material thickness of $0.35\%\ X/X_0$.

For the three innermost vertex layers of the tracking system, ATHENA will adopt the ALICE ITS3 concept of large-area, wafer-scale stitched sensors, thinned to below $50\um$, bent around the beam pipe and held in place using low mass carbon fiber support structures \cite{ITS3det}.  The low power dissipation of the sensor will enable air cooling of the vertex layers, which is a key factor in reducing the material thickness of the innermost tracking layers, crucial to achieve the required vertex reconstruction resolution.  
The barrel layers and forward/backward disks will use more conventional flat sensors, also stitched but not to wafer scale, mounted on flat support structures: staves and half-disks. This leads to an estimate of $0.55\%\ X/X_0$ in each of the two barrel layers and $0.24\%\ X/X_0$ for the disks, based on the ALICE ITS2 \cite{ITS2} experience and the anticipated  power dissipation of the new $65\nm$ sensor. 

The EIC Silicon Consortium has grown out of the EIC generic detector R\&D program (eRD16/eRD18/eRD25) and its leadership is made up of members of the ATHENA collaboration.  The EIC Silicon Consortium will co-develop with ALICE-ITS3 the wafer-scale sensor for the vertex layers, while also developing an EIC-specific, stitched but not wafer-scale version of the same sensor for the barrel layers and disks, together with support structures and services (see Sec.~\ref{RDNeeds}).

\gls{mpgd} technologies such as \glspl{gem}, \gls{micromegas} and \gls{murwell} detectors are a cost-effective solution for large-area tracking systems requiring a low material budget. \glspl{gem} and \gls{micromegas} are both mature technologies and have been used in many nuclear and high-energy physics experiments including COMPASS and the upgrades of ATLAS, CMS, ALICE, and LHCb at CERN; SBS, CLAS12, and PRad at Jefferson Lab; as well as the STAR Forward \gls{gem} Tracker and PHENIX Hadron Blind Detector at BNL. In general, \glspl{mpgd} are gaseous devices for electron amplification with a high granularity strip or pad anode readout to provide good 2-D space point resolution ($< 100\um$), fast signals ($\sim 10\ns$), high rate capability (up to 1~$\mathrm{MHz\ cm^{-2}}$), low material budget, radiation hardness and large area coverage. Through the EIC generic detector R\&D program (eRD3/eRD6~\cite{Posik:2018oat,Posik:2015gha,Gnanvo:2014hpa,Gnanvo:2015xda,Hohlmann:2017sqj,Zhang:2017dqw,Zhang:2016vbk,Zhang:2015kxy,Zhang:2015pqa,Vandenbroucke2018,azmoun:2020tns,azmoun:2021tns}) advancements towards low material and large-area \gls{mpgd} detectors with low channel counts and high granularity readout structures have been made. Each \gls{micromegas} layer in the barrel and each of the forward/backward triple-\gls{gem} disks has a material thickness well below $1\%\ X/X_0$. 
The addition of a \gls{tof}-layer based on \gls{aclgad} in the barrel will improve the pattern recognition and tracking performance at higher $p_T$. In the evaluation of the tracker performance discussed in the following this device was not taken into account.

\subsubsection{Requirements and subsystem performance}

The combined silicon and gaseous detector technology design, illustrated in Fig.~\ref{fig:track_baselineconfig}, has been chosen after careful consideration of tracking and vertex reconstruction performance, cost, ease of integration, and with the overarching aim of minimizing material.  The solution allows the services to the central silicon barrel to be routed along the conical/cylindrical support structure that encapsulates the silicon disks. 

\begin{figure} [htb]
   \centering
	\includegraphics[width=\textwidth]{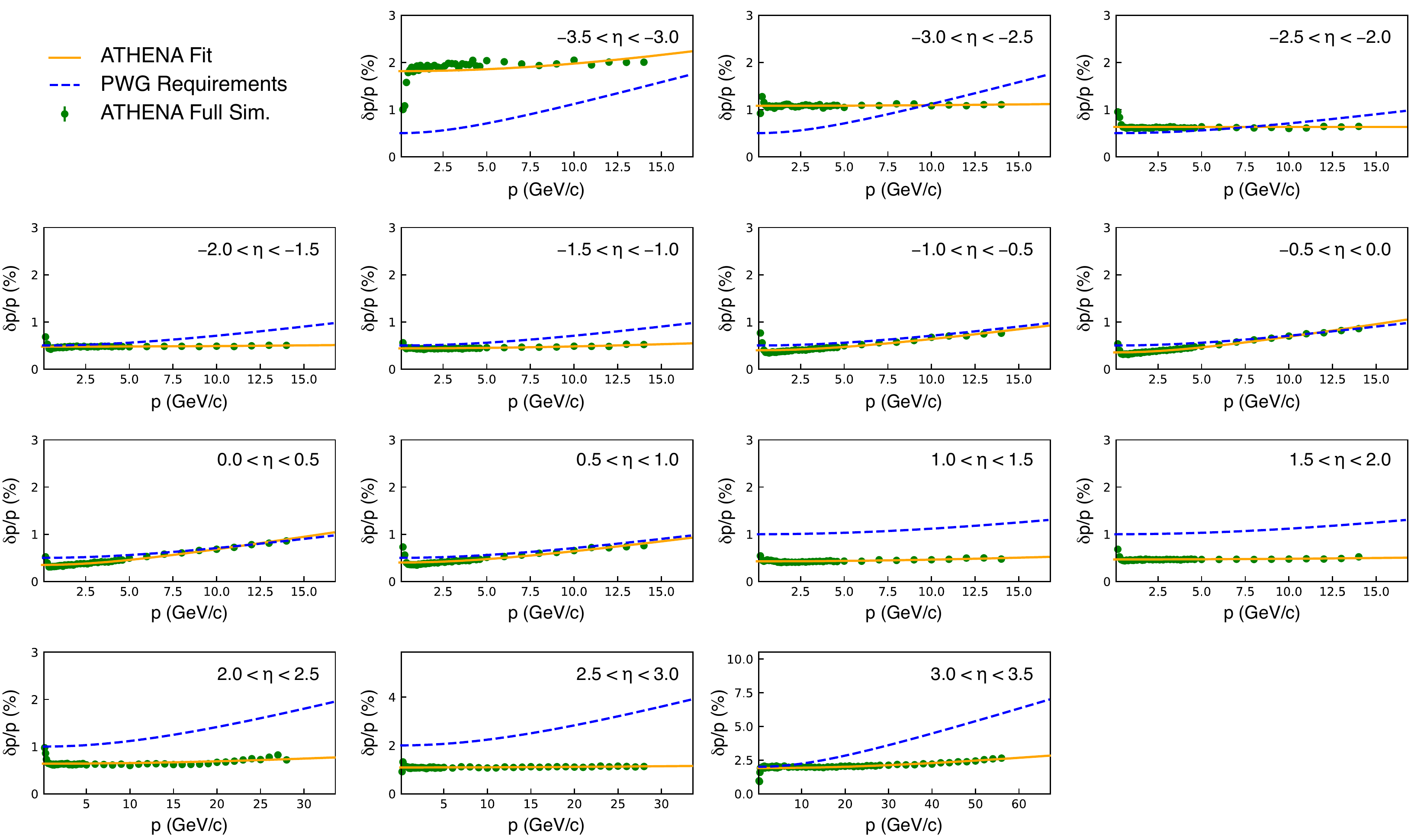}
	\caption{ATHENA momentum resolutions versus momentum of generated pions compared to the Yellow Report requirements (dashed lines) for selected $\eta$ bins. (FullSim).}
    \label{fig:track_performance}
\end{figure}

\begin{figure} [htb]
   \centering
	\includegraphics[width=\textwidth]{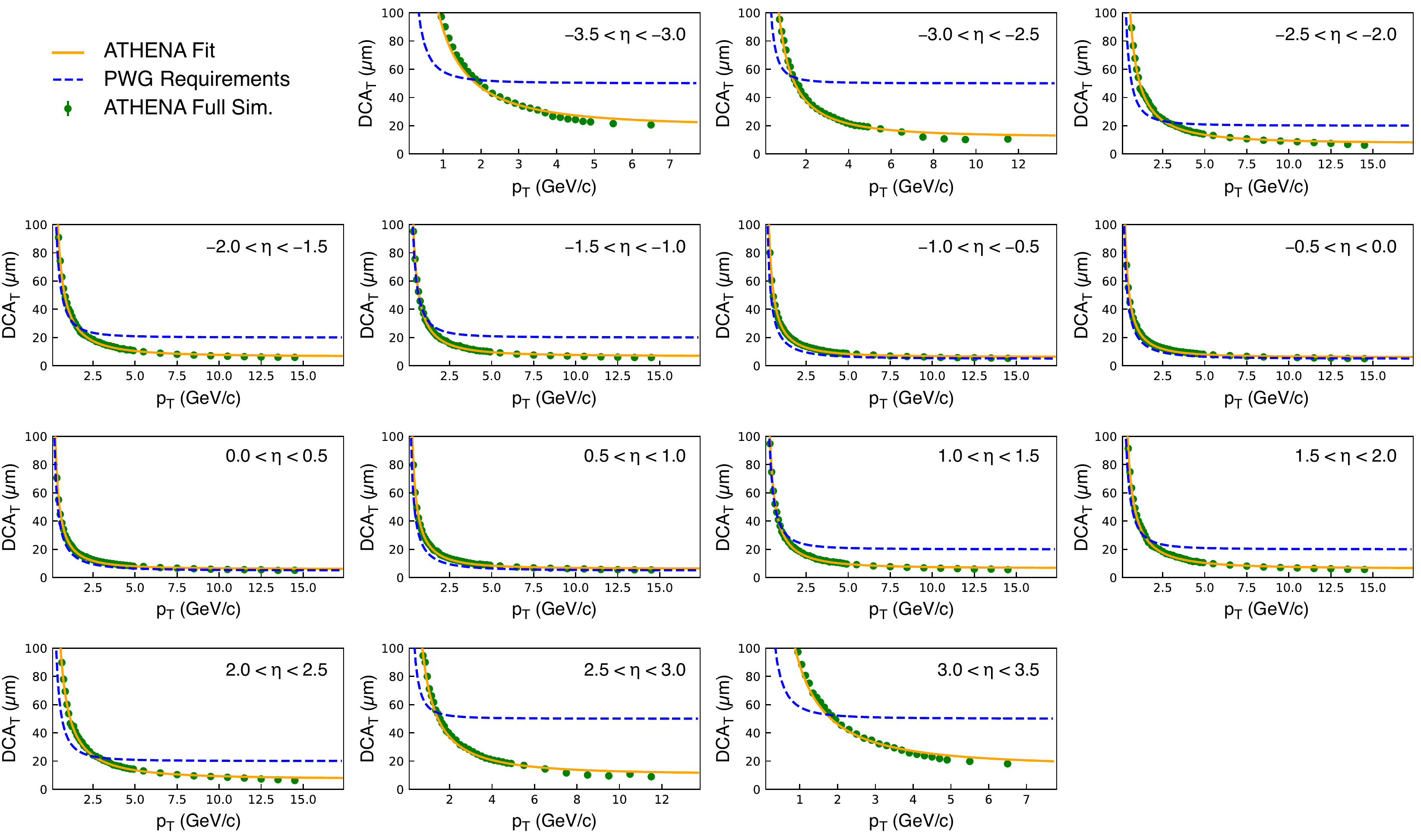}
	\caption{Transverse DCA resolution versus transverse momentum of generated pions compared to the Yellow Report requirements (dashed lines) for selected $\eta$ bins. (FullSim).}
    \label{fig:dca_performance}
\end{figure}

The overall performance of the chosen vertex and tracking system is illustrated in Fig.~\ref{fig:track_performance} and Fig.~~\ref{fig:dca_performance}, which shows the reconstructed relative momentum resolution ($dp/p$) as a function of momentum (Fig.~\ref{fig:track_performance}) and the transverse distance of closest approach ($\mathrm{DCA_T}$) to the primary vertex (pointing resolution) as a function of transverse momentum (Fig.~\ref{fig:dca_performance}) for primary pions generated in three bins of pseudorapidity. The dashed lines represent the corresponding performance requirements from the EIC Yellow Report.  
The relative momentum and transverse pointing resolutions were obtained from full GEANT4 simulations and have been parameterized by fitting the same functional forms as the Yellow Report requirements. 
The fits were performed in bins $\Delta \eta = 0.5$ 
and then combined to match the binning found in the Yellow Report. The results are summarized in Table~\ref{tab:track_performance_table} presenting a side-by-side comparison of the achieved performance and detector requirements as a function of pseudorapidity.


\begin{table} [htb!]
\footnotesize
\centering
\caption{\label{tab:track_performance_table}
Comparison of performance and Yellow Report requirement parameterizations for relative momentum and transverse pointing resolutions as a function of momentum for the ATHENA baseline tracking system.}
\begin{tabular}{|c|c|c|c|c|}
\hline
{} & \multicolumn{2}{|c|}{\textbf{Momentum resolution $\sigma(p)/p$}} & \multicolumn{2}{c|}{\textbf{Transverse pointing resolution $\sigma(\mathrm{DCA_T})$}}\\
\hline
{} & Performance & Requirements & Performance & Requirements\\
\hline
{-3.5 $<$ $\eta$ $<$ -2.5} & 
{$\sim 0.04 \% \times p \oplus 1.5 \%$} & 
{$\sim 0.1\% \times p \oplus 0.5\%$} & 
{$\sim 80/p_T \oplus 10$ $\mu{m}$} & 
{$\sim 30/p_T \oplus 50$ $\mu{m}$}\\
{-2.5 $<$ $\eta$ $<$ -1.0} & 
{$\sim 0.01 \% \times p \oplus 0.5\%$} & 
{$\sim 0.05\% \times p \oplus 0.5\%$} & 
{$\sim 50/p_T \oplus 5$ $\mu{m}$} & 
{$\sim 30/p_T \oplus 20$ $\mu{m}$}\\
{-1.0 $<$ $\eta$ $<$ 1.0} & 
{$\sim 0.05\% \times p \oplus 0.4\%$} & 
{$\sim 0.05\% \times p \oplus 0.5\%$} & 
{$\sim 30/p_T \oplus 5$ $\mu{m}$} & 
{$\sim 20/p_T \oplus 5$ $\mu{m}$}\\
{1.0 $<$ $\eta$ $<$ 2.5} & 
{$\sim 0.01\% \times p \oplus 0.5\%$} & 
{$\sim 0.05\% \times p \oplus 1\%$} & 
{$\sim 50/p_T \oplus 5$ $\mu{m}$} & 
{$\sim 30/p_T \oplus 20$ $\mu{m}$}\\
{2.5 $<$ $\eta$ $<$ 3.5} & 
{$\sim 0.02\% \times p \oplus 1.5\%$} & 
{$\sim 0.1\% \times p \oplus 2\%$} & 
{$\sim 80/p_T \oplus 10$ $\mu{m}$} & 
{$\sim 30/p_T \oplus 50$ $\mu{m}$}\\
\hline
\end{tabular}
\end{table}

The tracking performance meets or exceeds the momentum resolution requirements stated in the Yellow Report, except for the most backward pseudorapidities. 
One way to improve this would be to further increase the $B \cdot dl$ of the tracking system.
However, a significantly larger field value is impractical, the tracking lever arm cannot be extended further due to spatial constraints in the current ATHENA detector configuration, and adding silicon disks to the proposed array would worsen the momentum performance because of the additional material they would introduce.
The backward momentum resolution requirement in the Yellow Report can thus not be met by current detector technology.
Achieving the science will require the combination of tracking information with that from the high resolution crystal \gls{necal} to improve the electron measurement, further minimization of the material inside the backward disk array, a different trade-off with the PID subsystem in the associated acceptance region, an alternative analysis approach, or a combination of these factors.

 \FloatBarrier
   
\subsection{Calorimetry}
\subsubsection{Electron endcap electromagnetic calorimeter}
The \acrfull{necal} is a high-resolution electromagnetic calorimeter designed for precision measurements of the energy of scattered electrons and final-state photons in the region $-4 < \eta < -1.5$. Based on the Yellow Report \cite{AbdulKhalek:2021gbh}, the required high energy resolution is driven by inclusive DIS where precise measurement of scattered electrons is critical to determine the event kinematics. 
The inner part of the nECal consists of 1976 PbWO$_4$ crystals, each of size $20 \times 20 \times 200~\mathrm{mm}^3$ $(\sim22~X_0)$~\cite{Horn:2019beh, Asaturyan:2021ese}. The expected energy resolution for PbWO$_4$ crystals is $2\%/\sqrt{E} \oplus 1\%$ \cite{AbdulKhalek:2021gbh}.
The outer part of the nECal consists of 1104  \gls{sciglass} blocks, each of size $40 \times 40 \times 550 \mathrm{mm}^3$ $(\sim 20~X_0)~$\cite{Buchner:1988fu,E705:1993imr}, with expected energy resolution 
of $2.5\%/\sqrt{E} \oplus 2.0\%$. Both the PbWO$_4$ and SciGlass blocks will be read out with arrays of \glspl{sipm}. \label{sec:cal}

The technology choice and overall
design concept of the \gls{necal} is the same as in the Yellow Report.
Since the Yellow Report, the design has been further developed by the EEEMCAL consortium~\cite{Horn:2020eoi}. 

\begin{figure}[ht]
\centering
\includegraphics[width=0.9\textwidth]{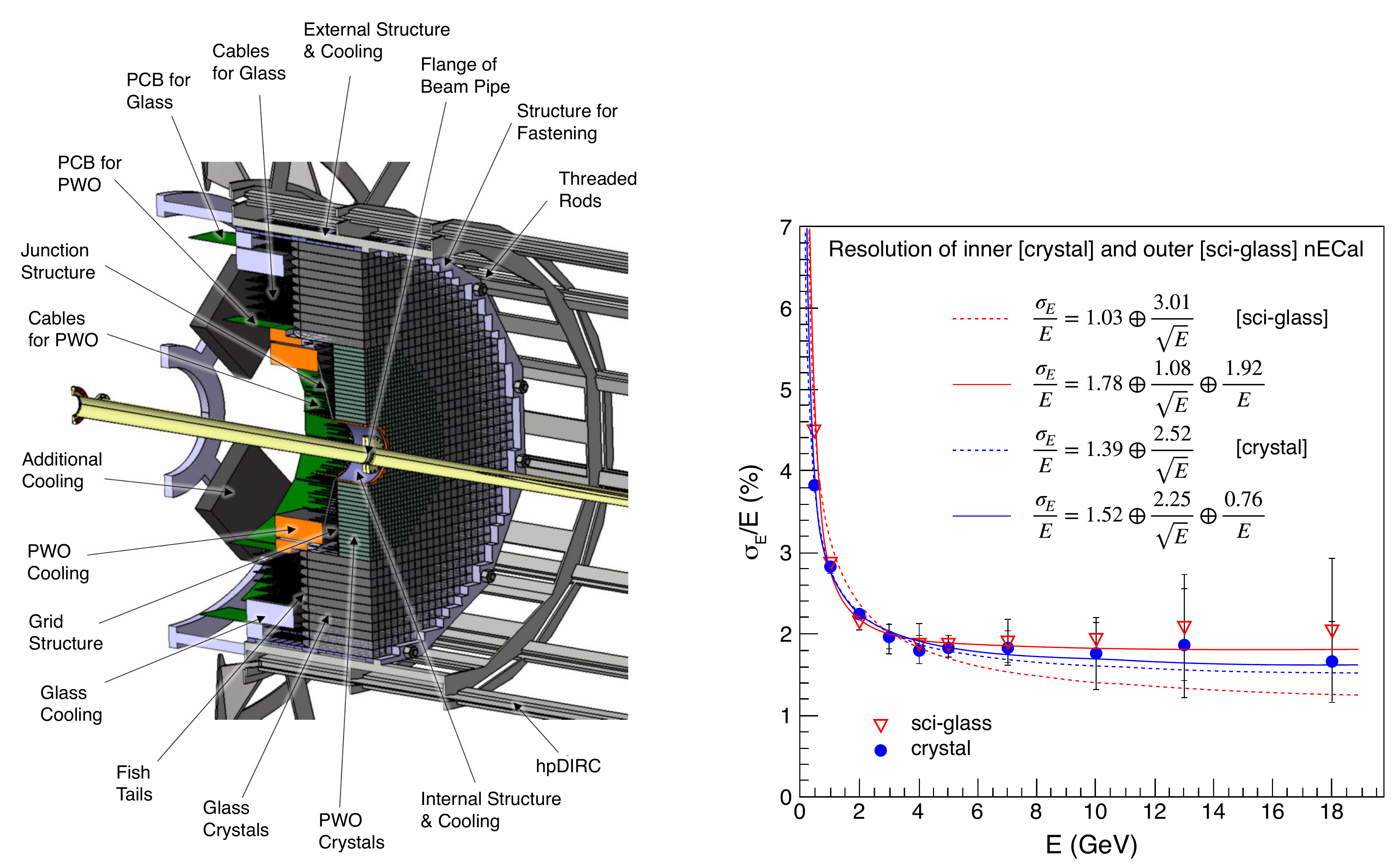}
\caption{Left: The  mechanical design of hybrid crystal/glass calorimeter nECal. Right: Expected nECal performance for the stand alone calorimeter, the energy resolution curves for inner PbWO$_4$ $(\sim22~X_0)$ and outer SciGlass $(\sim20~X_0)$ regions (FullSim).}
\label{fig:EEEMCAL}
\label{fig:EEEMCAL_mechanics.png}
\label{fig:EEEMCAL_performance.png}
\end{figure}

The \gls{necal} calorimeter concept was developed as part of the EIC generic detector R\&D program~\cite{eRD1:2017aaa}. The team 
collaborated closely with producers of PbWO$_4$ crystals and SciGlass to establish robust QA protocols at all stages of production, ensuring the highest quality of blocks. 
R\&D for SciGlass will continue under the auspices of the EIC Project to show the feasibility of production scale up. 
In the event that SciGlass R\&D is delayed, the fallback technology is lead glass. 

A detailed design of the nECal is underway among the collaborating institutions of EEEMCAL, focussing on
mechanical design, scintillator, readout, and software/simulation development. 
Pre-design activities, in particular for the support structure have started in 2021. The mechanical integration of this detector is shown in Fig.~\ref{fig:EEEMCAL_mechanics.png}. 
This concept is based upon existing detectors the team has constructed, and in particular, the Neutral Particle Spectrometer at Jefferson Lab~\cite{Horn:2015yma}. The final assembly of the detector will be performed at BNL. 
 
\subsubsection{Barrel electromagnetic calorimeter}

\begin{figure}[hbt]
\centering
\includegraphics[width=0.9\textwidth]{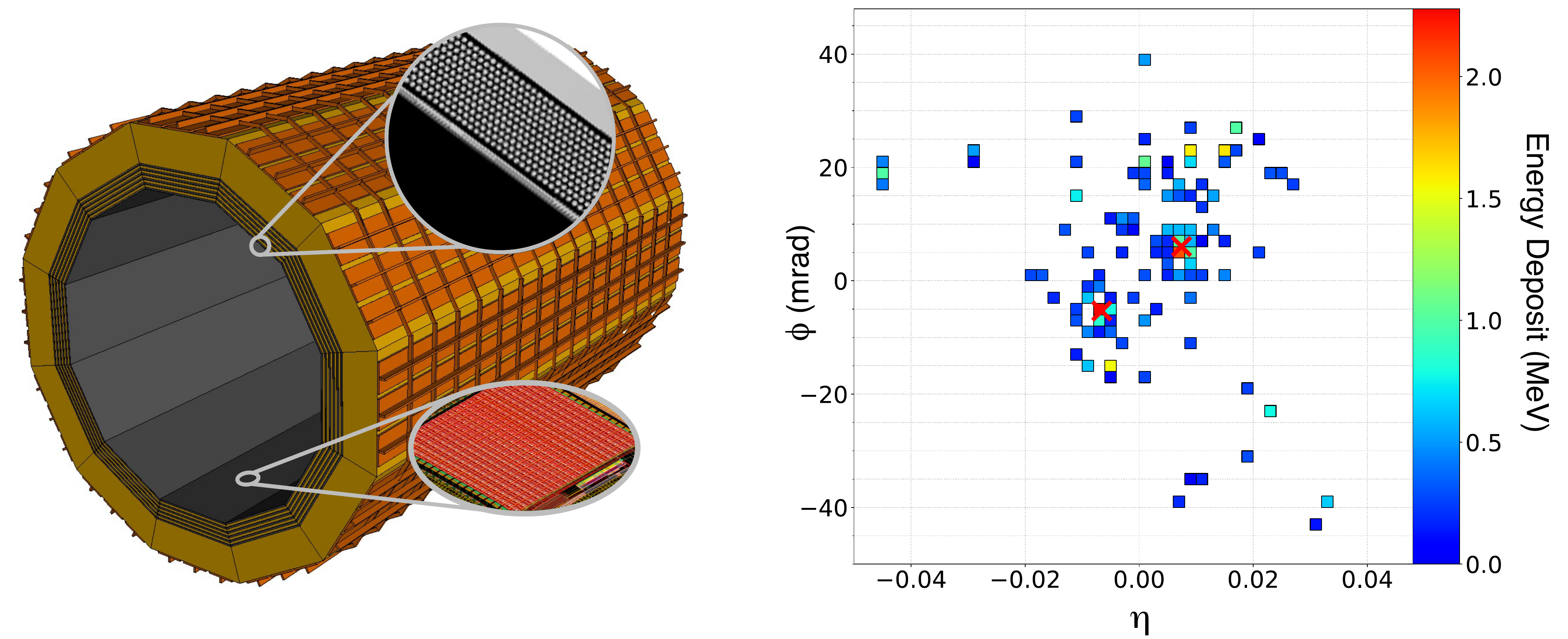}
\caption{Left: The \gls{becal} as included in detector simulations.
The insets show the structure of the Si imaging layers (bottom) and the 1.59 cm thick Pb/SciFi (top).  
The imaging layers are followed by a thicker Pb/SciFi section, for a total thickness 
(not including the support structure) of 40 cm.  Right: Energy deposited in pixels in the imaging calorimeter demonstrating clean separation of the two clusters for a 15 GeV $\pi^0$.  The red crosses mark the reconstructed clusters centers (FullSim). }
\label{fig:barrel-ecal}
\end{figure}

The \acrfull{becal} 
will detect scattered and secondary electrons and separate them from pions, detect and reconstruct full kinematic information for photons, and provide sufficient spatial resolution to identify neutral pions from $\pi^0\rightarrow\gamma\gamma$ decay at high momenta.  The proposed design is hybrid, using light-collecting calorimetry based on \gls{scifi} embedded in Pb and imaging calorimetry based on AstroPix monolithic silicon sensors ~\cite{brewer2021astropix}. The imaging of particle showers is achieved by six layers of silicon sensors interleaved with five Pb/SciFi layers, followed by a thick layer of Pb/SciFi calorimeter resulting in a total radiation thickness of about 20~X$_0$.
The barrel is composed of 12 staves, as shown in Fig.~\ref{fig:barrel-ecal} presenting the geometry of the barrel calorimeter used in the current ATHENA simulations. The inner radius of the barrel is 103 cm. The first (closest to the beam) six layers 
are imaging layers with a width of 55.2 cm and length of 405 cm. Each imaging layer is separated by a Pb/SciFi layer  that is about 1.59 cm thick. Because of ATHENA's geometry, the \gls{becal} 
not only functions 
as the barrel calorimeter, but also provides significant coverage in the electron-going direction for the overall ECal system.  

The Yellow Report stipulates
that the \gls{becal} should have energy resolution of approximately 
(10--12)\%/$\sqrt{E} \oplus$ (1--3)\%, 
electron-pion separation up to $10^4$, 
spatial resolution to separate $\gamma$ from $\pi^0$ decay for momenta up to 15~GeV/c, and the capability 
to detect photons down to 100~MeV.

\begin{figure}[ht]
\centering
\includegraphics[width=0.9\textwidth]{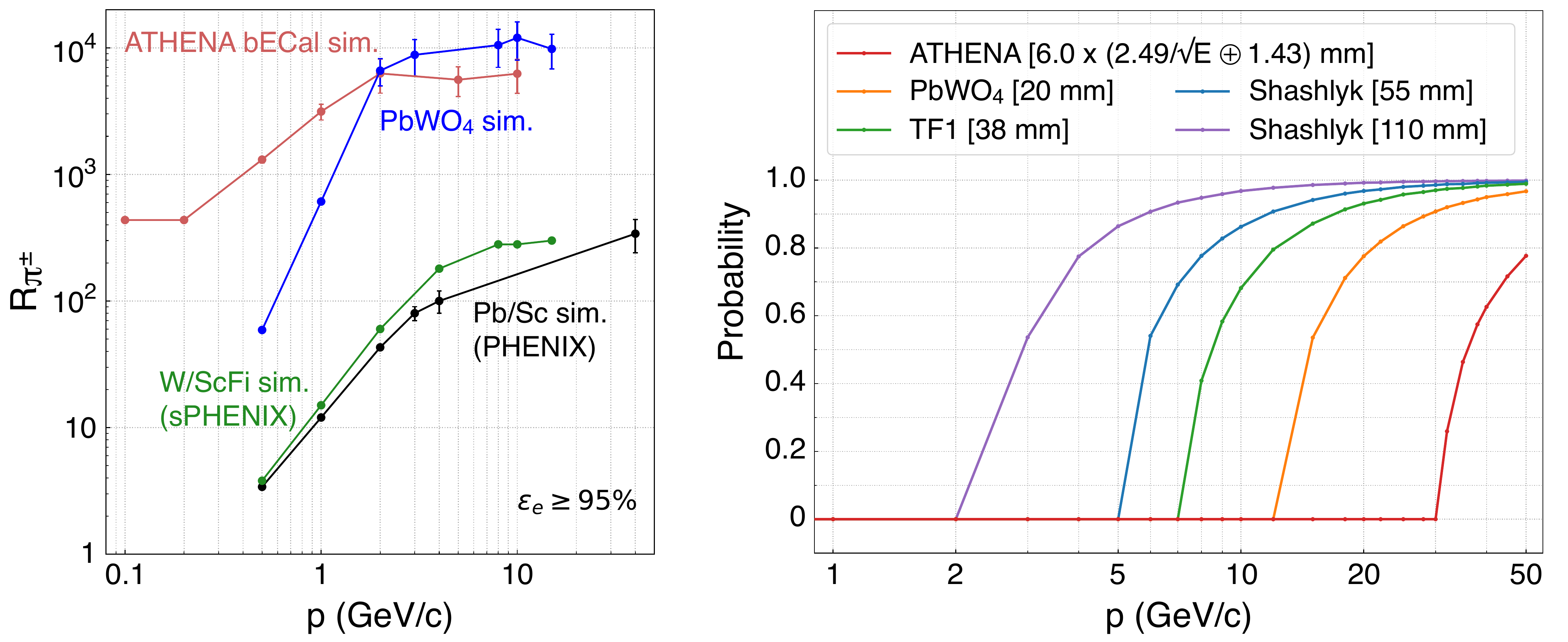}
\caption{Left: The pion rejection power of the \gls{becal} (red solid line) compared with other technologies listed in the EIC Yellow Report. The rejection power of the \gls{becal} is 
obtained from
the $E/p$ method and pattern matching, while the other rejection powers are determined solely from  the $E/p$ method \cite{AbdulKhalek:2021gbh}. All the curves, including simulations and data, are obtained for the standalone calorimeter, i.e., no other materials are placed in front of the calorimeter and no magnetic field is involved. 
Right: The merging probability for two $\gamma$s from $\pi^0$ decay for the \gls{becal} is determined by $6\sigma$ of the spatial resolution ($2.4/\sqrt{E} \oplus 1.3$ mm), since its pixel size (0.5mm) is much smaller than the cluster profile. For the other technologies, the cell size is used to estimate the probability\cite{AbdulKhalek:2021gbh}.
\label{fig:RPie}}
\end{figure}

This hybrid design 
provides precise measurements of both the energy and position of the incident particle's cascade in 3-D. Utilizing \gls{ai} techniques for pattern recognition of the 3-D shower images, this calorimeter will provide pion-electron rejection 
better than that achievable with traditional sampling calorimetry, especially at lower particle energies $(\lesssim 5\-\mathrm{GeV})$; see Fig.~\ref{fig:RPie}. 
The layers of the Pb/SciFi sampling calorimeter improve the overall sampling fraction and hence, the energy resolution of the calorimeter. Table~\ref{tab:bEcal-perf} summarizes the expected detector performance based on simulations with AstroPix sensor digitization and reconstruction algorithms for both 3-D and 2-D shower clustering. 

The imaging layers are based on off-the-shelf AstroPix sensors, the successor of ATLASPix~\cite{schoening2020mupix}, a low-power pixel sensor developed for ATLAS, and further optimized for the NASA AMEGO-X~\cite{fleischhack2021amegox} mission. These sensors have 
excellent energy resolution at low energy ($\sim$~7\% at 30~keV) and do not have stringent power and cooling requirements. This technology was discussed in the Yellow Report as an alternative to light-collecting calorimeters. The imaging capabilities replace the need for the projective geometry of the Yellow Report W/SciFi calorimeter. 
The proposed Pb/SciFi design is based on the existing GlueX barrel calorimeter with an energy resolution of $\text{5.2}\%/\sqrt{E} \oplus \text{3.6}\%$~\cite{Beattie:2018xsk} and $z$-position resolution $\sigma_z=1.1\,\text{cm}/\sqrt{E}$ at normal incident angle \cite{leverington:2008zz}, exceeding the Yellow Report requirements. 
These numbers were obtained from fits to low-energy data ($E<2.5$~GeV) that do not fully constrain the constant term.

The fine pixelation of the \gls{becal} allows for tagging of final state radiative photons that may be extremely close to the scattered electron. 
This is important for radiative corrections, enabling the benchmarking of \eA\ Monte Carlo generators.  Furthermore, the first %
Si layer
provides a space point 
for the \gls{hpdirc} reconstruction, 
obviating the need for additional large-radius tracking.
The outer, thick layer of Pb/SciFi %
contributes to neutral hadron identification since 70\% or more of produced neutrons will begin showering within the \gls{becal}.  This partially compensates for losses in the magnet material
and improves the overall hadronic reconstruction.
The tracking layers lower
the threshold of $\pi/e$ separation, 
expanding the available phase space for all physics objectives. 
Furthermore, the 3-D shower profiles measured by the \gls{becal} enable effective $\mu$ identification.
 
The overall concept for a tracking calorimeter is novel, but it relies on two well-developed technologies. The tracking layers rely on the technology of CMOS pixel sensors;  
a first version of the AstroPix sensor has already been delivered, and the second version is expected soon. Pb/SciFi technology is very mature, having been used for over 30 years; this design will scale up the design for the KLOE and GlueX experiments~\cite{Adinolfi:2002zx,Beattie:2018xsk}.

\begin{table}[htb]
\centering
\caption{Expected \gls{becal} detector performance.}
\label{tab:bEcal-perf}
\begin{tabular}{|p{0.2\textwidth}|p{0.7\textwidth}|}
\hline
Energy Resolution & $5.5\%/\sqrt{E} \oplus 1\%^a$ \\ \hline
$e/\pi$ separation &  $>$ 99.8\% pion rejection with 95\% electron efficiency at $p \ge 0.1$ GeV/c$^b$. \\ \hline
$E^\gamma_{\text{min}}$ & $<100$\,MeV$^c$ \\ \hline
Spatial Resolution & Cluster position resolution for 5 GeV photons at normal incident angle is below $\sigma = 2$\,mm (at the surface of the stave $r=103$\,cm) or $0.12^{\circ}$. For comparison, the minimal opening angle of photons from $\pi^0 \rightarrow \gamma \gamma$ at 15\,GeV is $\sim 1.05^{\circ}$ (about 19 mm -- 37 pixels -- of separation at $r=103$\,cm).
 \\ \hline
\end{tabular}
\begin{footnotesize}
\begin{flushleft}
$^a$Based on the photon simulations with $-0.5 < \eta < 0.5$ and $0<\phi< 2\pi$. The constant term does not include calibration effects. \\
$^b$Based on simulation for a standalone \gls{becal}, see Fig. \ref{fig:RPie} for detailed results. \\
$^c$Based on simulations, 100~MeV photons leave an energy deposit of $\sim 15$\ MeV in SciFi layers and of $\sim 1$\,MeV in the imaging layers. This simulation includes digitization with electronics noise and a noise suppression cut.
\end{flushleft}
\end{footnotesize}
\end{table}

\subsubsection{Hadron endcap electromagnetic and hadronic calorimeters}
					
ATHENA's high resolution, high granularity, compact hadron endcap calorimeter system consists of a compensated \acrfull{pecal} and Fe/Scint (20~mm / 3~mm) sandwich \acrfull{phcal}. The expected electromagnetic resolution is $\sim 11\%/\sqrt{E}\oplus2\%$, while the hadronic resolution 
is $\sim 32\%/\sqrt{E} \oplus 2\%$, including a tail-catcher cut. 
This is based on GEANT4 simulations and requires experimental validation. 
The experimentally achieved energy resolution with a similar but thinner (only 4.5 $\lambda_I$) HCal and shashlyk ECal system built for STAR was $\sim 60\%/\sqrt{E}\oplus 6\%$.
The e/h ratio for the \gls{pecal} is tuned to be $\sim1$ above 10~GeV, while for the \gls{phcal} it is $\sim1.2$. The \gls{phcal} will have four longitudinal sections for software compensation, i.e., re-weighting energy deposition in sections with localized high electromagnetic fraction~\cite{FeHCal}, and 3-D shower imaging. The transverse tower size is 2.5~$\times$~2.5~cm$^2$ for the \gls{pecal} and 10~$\times$~10~cm$^{2}$ for the \gls{phcal}, respectively.  The longitudinal space required for about seven interaction lengths is 150~cm.  The \gls{phcal} absorber structure serves as a support for the \gls{pecal}. Both calorimeters will be read out with \glspl{sipm}. 
The choice of technology is the same as for the reference detector described in the EIC CDR~\cite{EIC-CDR}.

The \gls{pecal} is made of W powder with embedded scintillating fibers~\cite{Tsai:2012cpa}. This technology was pioneered as part of the generic detector R\&D program for EIC, adopted by sPHENIX. A novel, efficient, and scalable construction method~\cite{Tsai:2015bna} developed for the STAR Forward Calorimeter System (FCS), will be used to build the \gls{phcal}. 

Longitudinal segmentation in the \gls{phcal} will be achieved by using scintillation tiles with two different time constants, similar to the CALEIDO2 Prototype for the ILC \cite{CALEIDO:2001vdz}.  Thus, with just two independent readout channels per tower, longitudinal information from four segments of the tower becomes available. The longitudinal segmentation of the \gls{phcal} and the achievable hadronic energy resolution will be verified in the near future with beam tests of a large scale prototype. A fallback solution, in case the two-channel readout technique would not be optimal, is to use an optical readout scheme with additional independent readout of the three last scintillation tiles in the \gls{phcal} towers. This scheme is similar to that used in the STAR FCS, and was studied in EIC generic detector R\&D.

The hadron endcap will be assembled in situ, as was done for the STAR FCS.  The STAR forward HCal will be re-used for the ATHENA \gls{phcal}. 
a region of the \gls{phcal} close to the beam pipe will have a potential upgrade path  to replace scintillation tiles with Si sensors.
   
\subsubsection{Hadronic calorimeters in electron endcap and barrel}

In combination with information from 
the electromagnetic calorimeters, tracking and \gls{pid} detectors, the hadron calorimeters in electron endcap and barrel assist with the detection of neutral hadrons \cite{Page_2020}. 
The \gls{bhcal} with passive magnet steel provides mechanical support for all the detector systems of ATHENA. The steel structures of all the hadron calorimeters provide the return flux path for the magnetic field.

The thickness of the ATHENA Magnet ($\sim$~1.3~$\lambda_I$) between the \gls{becal} and \gls{bhcal} precludes good calorimetric energy measurements of hadrons. 
The \gls{becal} in front of the magnet cryostat will be $1-1.7$ $\lambda_I$ deep, thus it is sufficient to instrument only about two $\lambda_I$ (tail-catcher) after the magnet cryostat to contain about 95\% of hadronic showers. 
The remaining steel required for the flux return and mechanical support of the barrel detectors will 
consist of re-used flux return steel bars of the STAR Magnet. A barrel made of these bars will weigh about 540~tons. 

The \gls{bhcal} is a five layer steel and scintillator sandwich  (4~cm/5~mm layer structure). ATHENA will re-use existing scintillation mega-tiles from the STAR barrel ECal \cite{STAR:2002ymp}.
These 
tiles have lost less than 5\% of their initial light yield after 20 years of operation. 
Existing scintillation mega-tiles consist of 80 optically isolated scintillation tiles with sigma grooves 
and wavelength-shifting fibers for light collection. 
The tiles are arranged in a projective geometry with excellent granularity 0.05 x 0.05 in $\eta$ and azimuthal angle $\phi$. 
Each individual tile will be read out with $1.3 \times 1.3~\mathrm{mm}^2$ \glspl{sipm}. 
The expected light yield will exceed 10 photoelectrons
for \glspl{mip} providing very good efficiency for low energy hits. Pre-assembled and calibrated with cosmic muons, mega-tile cassettes will be inserted into
the \gls{bhcal}.  
Moderate energy resolution for hadrons in the barrel region can be obtained from  the \gls{becal} Pb/SciFi layers. 

A similar approach will be used for the \gls{nhcal} by utilizing scintillation mega-tiles from the STAR endcap ECal \cite{STAR:2002eml}. 
The structure for the \gls{nhcal} is similar to the \gls{bhcal}, consisting of approximately 10 layers. The exact number of instrumented layers in both detectors is under optimization.
 
Re-using components from the STAR detector (magnet steel, cradles, scintillation mega-tiles) significantly reduces the cost of these two subsystems. 

 \FloatBarrier

\subsection{Particle Identification}
Particle identification in ATHENA requires multiple technologies to address the physics goals.  Measurements of Cherenkov radiation provide the 
greatest reach at higher momentum, but are limited in their low-momentum reach.
Furthermore, the $\eta$-dependence of the momentum spectrum along with space constraints require distinct technologies in the forward, backward, and barrel regions.  Our solution to this challenge involves:
\begin{itemize}
    \item A dual radiator RICH (\acrshort{drich}) in the forward region utilizing aerogel and gas radiators focused by mirrors onto a focal plane instrumented by \glspl{sipm} with built-in capability to "anneal-in-place" to combat the inevitable dark current generated by radiation damage.
    \item A 100-cm radius high-performance DIRC (\acrshort{hpdirc}) very close to the design of the Yellow Report in the central region.  The hpDIRC is complemented by an AC-LGAD 
    ToF detector at a 52.5 cm radius.  
    The AC-LGAD layer dramatically improves the PID reach to low momentum by catching particles that do not reach the hpDIRC, while sidestepping the limitations of Cherenkov detectors at lower momenta.
    \item A proximity-focusing aerogel RICH (\acrshort{pfrich}) with 40 cm proximity gap.  This deviates from the mRICH technology used in the Yellow Report. This design features minimal material budget, easy pattern recognition, large acceptance, and can simultaneously function as a threshold gas Cherenkov detector.
\end{itemize}
\subsubsection{Forward direction}
\label{dRICH}

\noindent\textbf{dRICH:}
The highest momentum particles produced at the EIC are emitted in the forward direction.  Gas-based Cherenkov ring imaging is the only presently-known technology that addresses $\pi$-K-p separation up to the required 50 GeV/c.  
Optimization of any Cherenkov detector design effectively boils down to minimizing the Cherenkov angular resolution of a single photon's contribution to the ring (linear dependence) and maximizing the number of detected photons per ring ($\sqrt{N}$ dependence).  The single photon resolution is affected by internal factors (radiator chromaticity, optical aberration, pixel size) as well as external factors including stray magnetic field and track pointing resolution.
The ATHENA design is the result of an intensive optimization on all fronts using full GEANT simulations at the level of optical photon propagation.

ATHENA underwent extensive studies of the magnet design with 
simulations demonstrating that the most significant positive impact involved distancing the dRICH from the collision point.  
The optics of the dRICH also represent a solved challenge with 
broad implications.  
The focal plane location entails an optimization involving three factors: maximizing the radiator length (prefers longer focal length), shielding the photon sensors from radiation (prefers tilted mirrors), and fitting the detector into the available space.  
This is realized by careful positioning of the dRICH detector with respect to the ATHENA magnet.
The optimization is a practical implementation of the dRICH concept from the Yellow Report that fully realizes the device's potential in the face of the compromises that must be made in a realistic detector design.

\begin{figure}[bht]
	\centering
    \includegraphics[width=\textwidth]{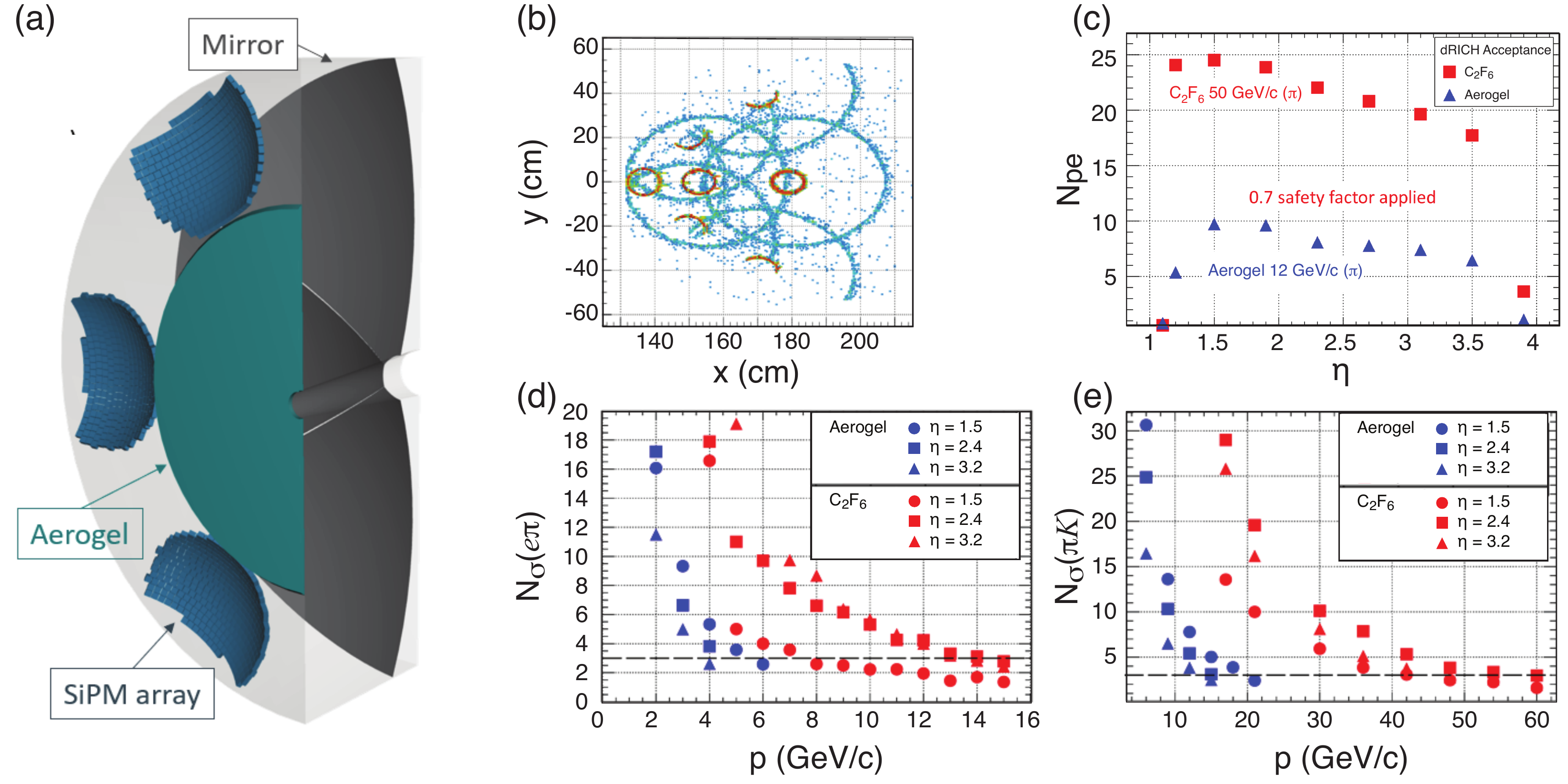}
\caption{Panel a) shows the layout of the dRICH radiators, mirror, and \gls{sipm} arrays located in the focal planes. Panel b) shows the superposition of hits from 1000 events with identical primary particles.  This effectively captures ring shape (aerogel-large, gas-small), Rayleigh scattering, optical aberration, multiple scattering, tracking resolution, chromaticity, and signal-to-noise effects in one image.  Panel c) shows the number of photoelectrons per single ring vs $\eta$, and thereby illustrates the acceptance range. Panels d) and e) demonstrate the PID separation for $e/\pi$ and $\pi/K$. The performance meets the Yellow Report specification. Aerogel performance is indicated in blue and $C_2F_6$ in red (FullSim).}
\label{Fig:PID-dRICH}
\end{figure}

The sensor choice for a dRICH is 
quite challenging.  
After careful consideration, the only 
viable choice is the 
well known \label{sec:drich}\gls{sipm} technology.  These devices are ideal in terms of quantum efficiency, sensitive wavelength range, and single-photon signal size.  
The difficult aspect regards dark currents that grow with radiation exposure.  
It has been demonstrated by studies at INFN \cite{Calvi:2018ulw} that this damage can be repaired by thermal annealing, while R\&D efforts towards in-situ annealing are ongoing.
The ATHENA design thereby features \gls{sipm}-based photon sensors in both the dRICH and the backward \gls{pfrich}.  
Detailed calculations of the worst-case dark current rates have been used as the basis of estimates for the DAQ needs of the dRICH.
Figure~\ref{Fig:PID-dRICH} shows the layout and performance of the ATHENA dRICH.  
As described previously, the optimization of the dRICH location (panel a) and optics (panel b) have already achieved the 
performance requirements in the Yellow Report.

\subsubsection{Barrel region}

\begin{figure}[hbt]
	\centering
    \includegraphics[width=\textwidth]{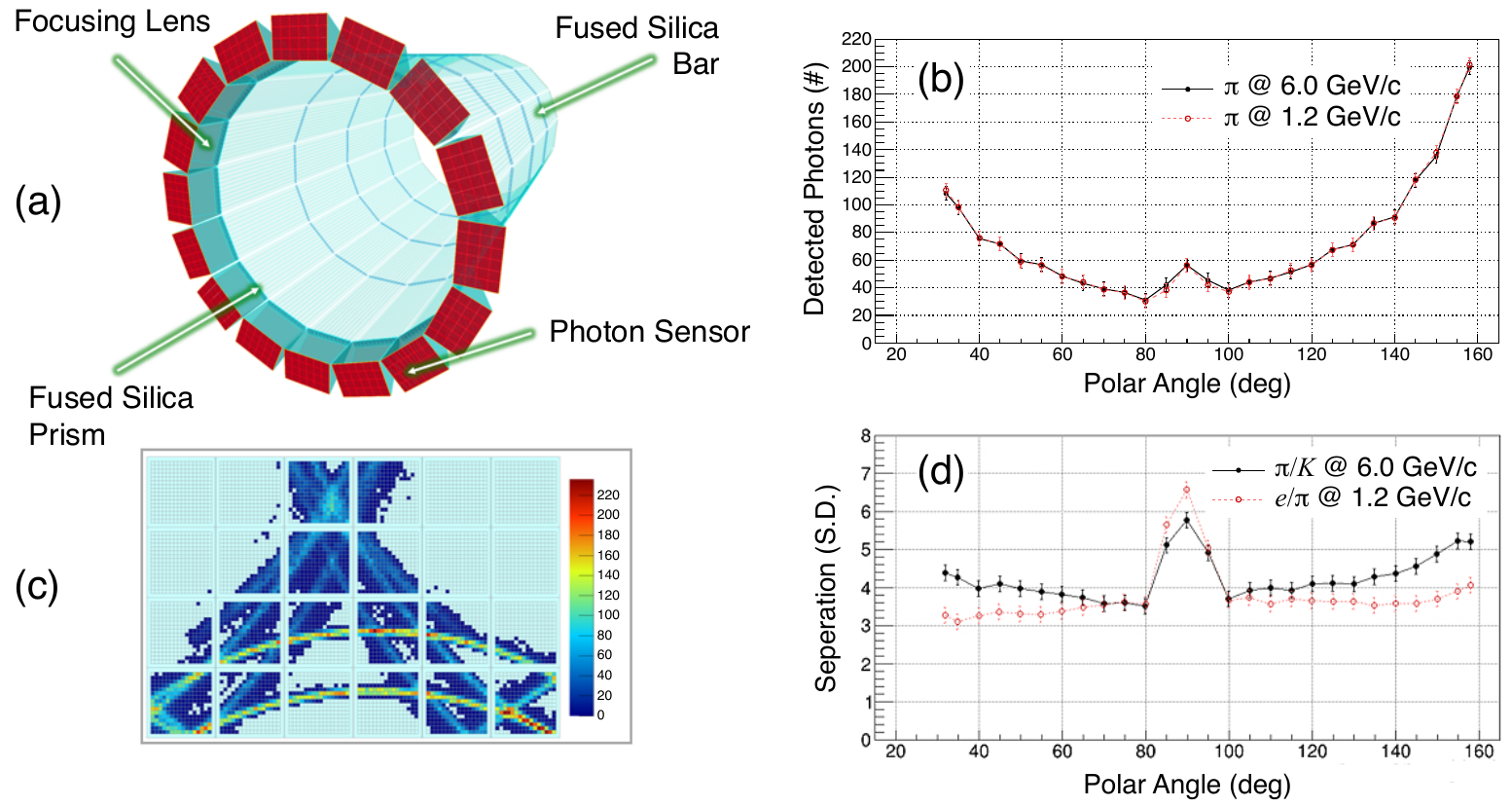}
    \vspace{-2mm}
\caption{a) Configuration of the DIRC.  b) Number of detected photoelectrons as a function of polar angle. c) Superposition of the distribution of photon hits from 6 GeV/c identical pions. d) Separation power at the maximum momentum requirement stated in the Yellow Report. Results from a stand-alone GEANT4 simulation.}
	\label{Fig:PID-DIRC}
\end{figure}

\noindent\textbf{hpDIRC:}
The hpDIRC closely resembles the concept described in the Yellow Report.
This presents an evolution of the original BaBar design.
The ATHENA barrel is surrounded by a series of radiator bars made of synthetic fused silica. Rings are imaged in an ”expansion volume”,  onto a focal plane containing photon detectors oriented normal to the magnetic field. 
The detector and its performance are shown in Fig.~\ref{Fig:PID-DIRC}.  
The hpDIRC focuses the ring onto a sensor plane. 
This will improves the performance of the device compared to the BaBar DIRC (doubling the 3-$\sigma$ separation limits), while making the readout system significantly more compact. 

\begin{figure}[hbt]
\centering
\includegraphics[width=0.85\textwidth]{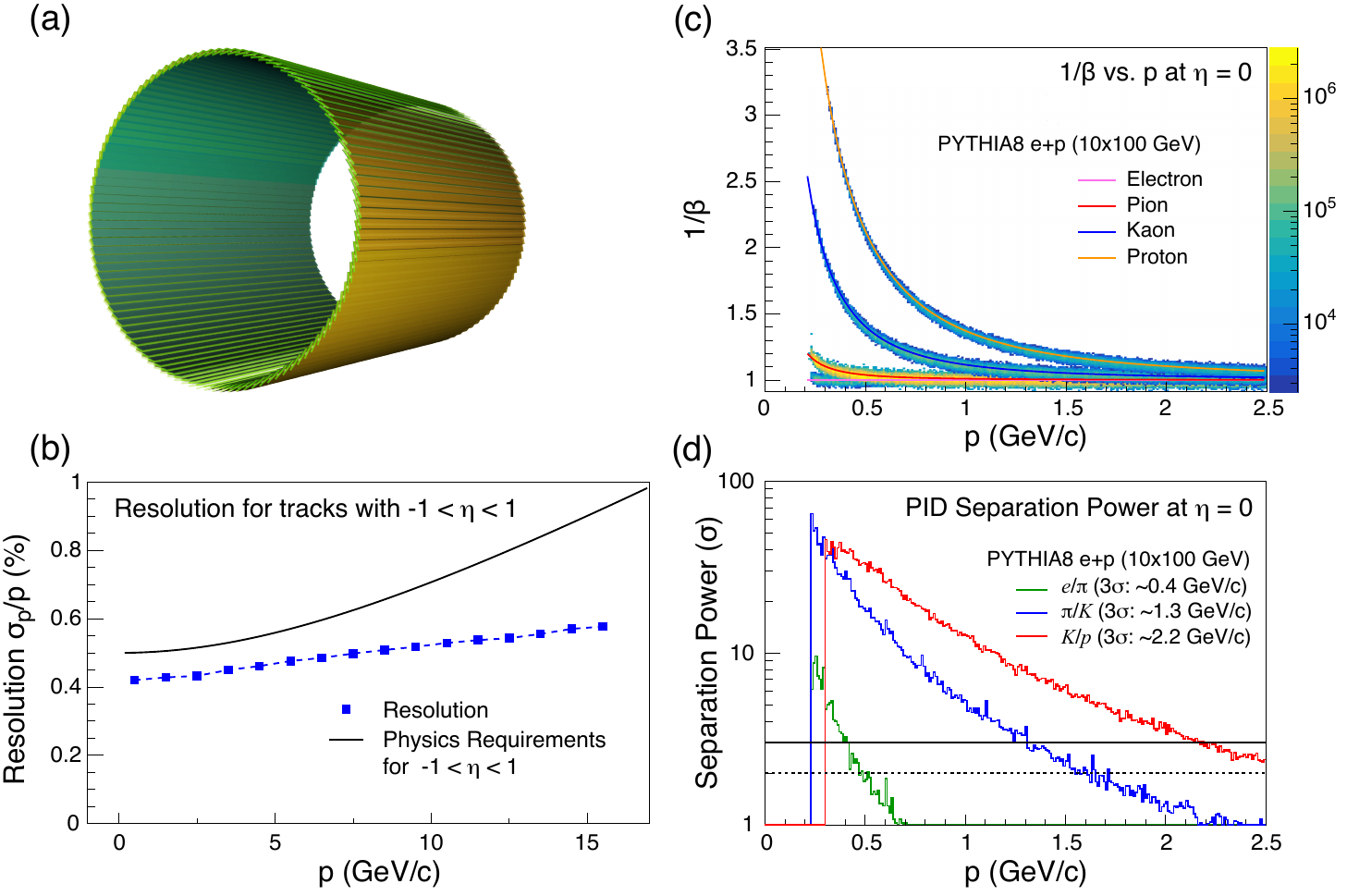}
\caption{a) Configuration of the \gls{btof}. 
b) Anticipated impact from the spatial hit from the ToF system on the tracking performance, with improved baseline performance at higher momenta. 
c) Illustrates that the \gls{btof} satisfies the PID requirements in the momentum range below the DIRC kaon threshold (0.47 GeV/c), thereby filling in the PID to 230 MeV/c. 
d) Separation power in number of $\sigma$ separation (FullSim).}
\label{Fig:PID-TOF}
\end{figure}

\noindent\textbf{AC-LGAD ToF:}
ATHENA defines positive PID as a signal beyond the Cherenkov threshold for $K$-radiation.
As such, the positive PID range for the DIRC begins at a momentum higher than 0.47 GeV/c (the $K$ threshold in synthetic fused silica).
This nicely matches the momentum cutoff of the ATHENA magnetic field.
Extending the range downward requires a dedicated detector located at a smaller radius. \label{sec:tof}  AC-LGAD devices deliver an excellent time resolution ($<30$ ps/layer) and spatial resolution.  
A single-layer array of AC-LGAD devices at a radius of 52.5 cm in the ATHENA design provides an effective PID coverage between 0.23 and 1.3 GeV/c for a 3 T field, and an additional spatial hit to supplement the tracking system in the barrel region.
Figure~\ref{Fig:PID-TOF} shows the configuration of the \gls{btof}.

One should note that state-of-the-art timing detectors present many technology challenges beyond just the performance of the sensor.  
These include, but are not limited to, the determination of a start time or $t_0$, and timing jitter which is invariably present in a large electronic system sending signals (such as the crossing clock) over long distances.  Because we emphasize the identification of particles at momenta below 1 GeV/c, conservative estimates of the uncertainty in $t_0$ using only the crossing clock and time distribution jitter show that this system will easily  perform beyond its requirements for $\pi$-K-p separation.

\subsubsection{Backward direction}

\begin{figure}[!hbt]
	\centering
    \includegraphics[width=0.85\textwidth]{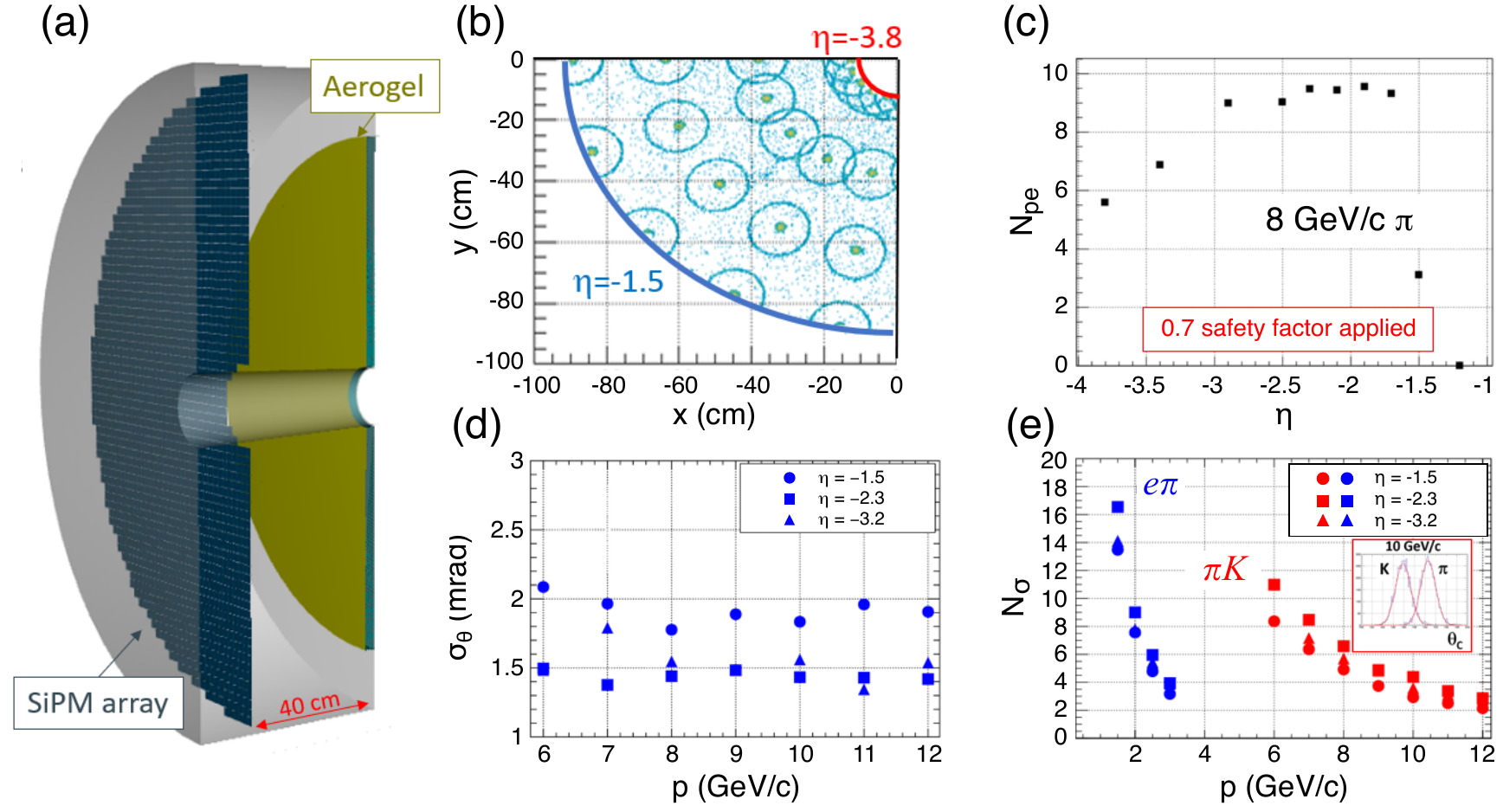}
	\caption{a) Configuration of the \gls{pfrich}. b) Response in $\frac{1}{4}$ of the detector to 1000 simulated events with regularly spaced 8 GeV/c pions.  This highlights the aerogel rings, gas-blobs, and indicates the properly scaled (rather small) background. c) Number of detected photoelectrons for single 8 GeV/c pions vs. $\eta$.  d) Cherenkov-angle resolution as a function of momentum in three regions of $\eta$.  e) Shows the separation power, demonstrating full realization the Yellow Report performance goal (FullSim).}
	\label{Fig:PID-pfRICH}
\end{figure}

\noindent\textbf{pfRICH:}
In the backward direction, \gls{pid} is provided by an aerogel radiator proximity-focusing RICH with a 40~cm proximity gap, shown in Fig.~\ref{Fig:PID-pfRICH}.
This technology choice maximizes acceptance while minimizing material in front of the \gls{necal}, with a uniform performance across its entire aperture.
The simplicity of the design makes the pattern recognition much simpler than for typical RICH detectors.
The baseline aerogel and photon detector technologies for the pfRICH are identical to those for the \gls{drich}. This minimizes the number of PID technologies used in ATHENA.

One additional feature emerges from this design:  
The Yellow Report identifies the need for additional electron identification (beyond $e/\pi$ from calorimetry and tracking) in the backward direction at momenta up to 4 GeV/c.
While this is not possible with an aerogel-based design, this can be achieved with a traditional gas threshold Cherenkov detector.
The 40 cm gap between the aerogel and the sensor plane
is sufficiently long for threshold-based electron identification, further improving ATHENA's PID performance in the backward direction.

 \FloatBarrier
  
\subsection{Far Forward Detectors}

\begin{figure}[htb]
\centering
\includegraphics[width=0.8\textwidth]
{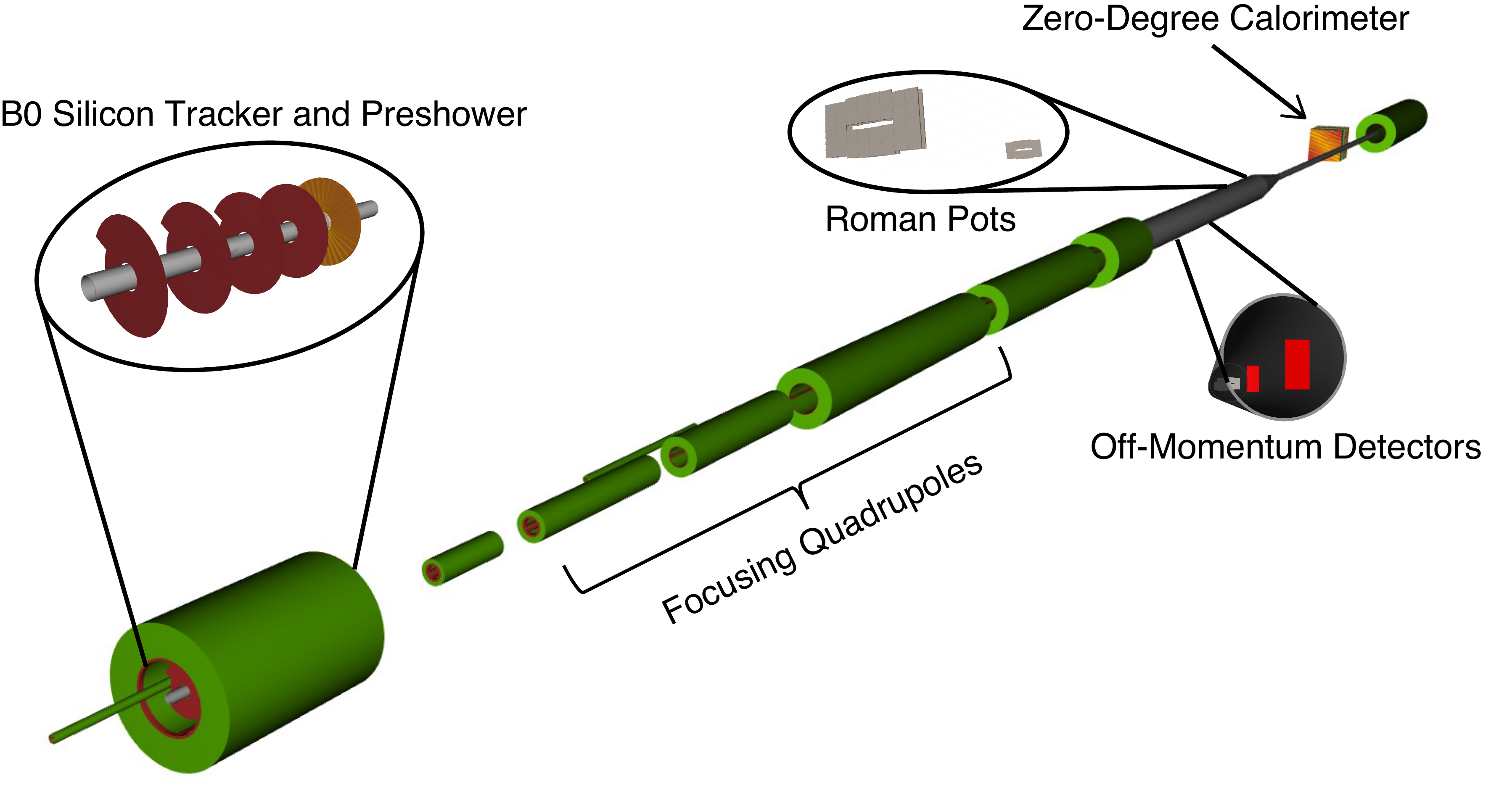}
\caption{\label{fig:FF_ATHENA_DD4HEP}3-D rendering of the far-forward region of IP6 with the proposed ATHENA detector instrumentation from the DD4HEP geometry implementation.}
\end{figure}

EIC collisions include many final-states 
with charged or neutral particles 
with $\eta > 4.0$.
These particles 
are outside the acceptance of the central detector and therefore require detectors integrated with the accelerator beamline.
Maximum acceptance across all beam energies and species requires multiple subsystems, whose acceptance is dictated by the \gls{ir} design. 
This is summarized in Tab.~\ref{tab:FFDetectors_acceptance}. 
A 3-D layout of the far-forward region is shown in Fig.~\ref{fig:FF_ATHENA_DD4HEP}.

\begin{table}[ht]
\footnotesize
\centering
\caption{\label{tab:FFDetectors_acceptance} 
Summary of the geometric acceptance for far-forward protons and neutrons
in polar angle $\theta$ and magnetic rigidity percentage provided by the baseline EIC far-forward detector design \cite{AbdulKhalek:2021gbh}. 
$^*$The Roman Pots acceptance at high values of rigidity depends on the optics choice for the machine.} 
\begin{tabular}{|l|c|c|c|c|}
\hline
\textbf{Detector} & 
$\mathbf{\theta}$~\textbf{accep. [mrad]} & 
\textbf{Rigidity accep.} & 
\textbf{Particles} & 
\textbf{Technology} \\
\hline
 B0 tracker    &  5.5--20.0  & N/A & \begin{tabular}{c}
 Charged particles \\  Tagged photons
 \end{tabular} & 
 \begin{tabular}{c}
 MAPS \\ AC-LGAD
 \end{tabular} \\
 \hline
 Off-Momentum Detector &  0.0--5.0   & 45\%--65\% & Charged particles & AC-LGAD \\
 \hline
 Roman Pots    &  0.0--5.0   & 60\%--95\%$^*$ & \begin{tabular}{c}
 Protons \\ Light nuclei
 \end{tabular} & AC-LGAD \\
 \hline
 Zero-Degree Calorimeter & 0.0--4.0 & N/A & \begin{tabular}{c}
 Neutrons \\ Photons
 \end{tabular} & 
 \begin{tabular}{c}
 W/SciFi (ECal) \\ Pb/Sci (HCal)
 \end{tabular} \\
\hline
\end{tabular}
\end{table}

\subsubsection{Technology choices}

The B0 spectrometer requires approximately 20 $\mu$m  position resolution to provide the required $p_T$ resolution for high-momentum hadrons near the beam line and good timing resolution to aid in background rejection, and to correct for effective vertex smearing from the crab cavity rotation.
Our design consists of three silicon MAPS disks serving as the first, second, and fourth layers of the detector, complemented by a single \label{sec:ff}AC-LGAD layer with 500~$\mu$m pixel pitch and $20-30$~ps timing resolution. Each tracking layer is separated by 30~cm. The silicon preshower following the tracker has two radiation lengths of lead as a photon converter, and a layer of silicon to tag the produced lepton pair. 
We envision use of \gls{aclgad} sensors for the preshower. These sensors enable the required spatial resolution to measure lepton pairs while providing excellent timing resolution to reduce background contamination.
While some observables could benefit from an expanded electromagnetic calorimeter, this is precluded by  engineering constraints (size, weight).

The \gls{omds} and \gls{rps} each consist of two stations separated by 2~m, with each station consisting of two active layers. Both detectors require spatial resolution better than $\sim$150$~\mu$m, and $\sim$30~ps timing resolution , making \glspl{aclgad} an ideal choice for both detectors.
 
We 
will insert the silicon detector packages for the RP and OMD directly into the beam pipe vacuum (i.e., without the usual ``pot" vessels) with thin foils surrounding the detector packages  to maximize acceptance. The RP detectors, in particular, need to be inserted (vertically only due to spatial constraints) into the beam line as close to the beam as possible (usually a few mm from the beam, depending on the beam optics).


The \gls{zdc} consists of the following major systems: a) silicon charged particle veto layer, 
b) W/SciFi sampling calorimeter with 2.5 $\times$ 2.5~cm$^2$ towers, 17~cm long (identical to the \gls{pecal}), and
c) Pb/Scint imaging hadronic calorimeter composed of a total of 120 layers of 1~cm Pb and 0.25~cm scintillator, corresponding to seven $\lambda_I$. 
The full \gls{zdc} delivers approximately eight $\lambda_I$ in total.

\subsubsection{Detector performance}

\begin{figure}[htb]
\centering
\includegraphics[width=\textwidth]{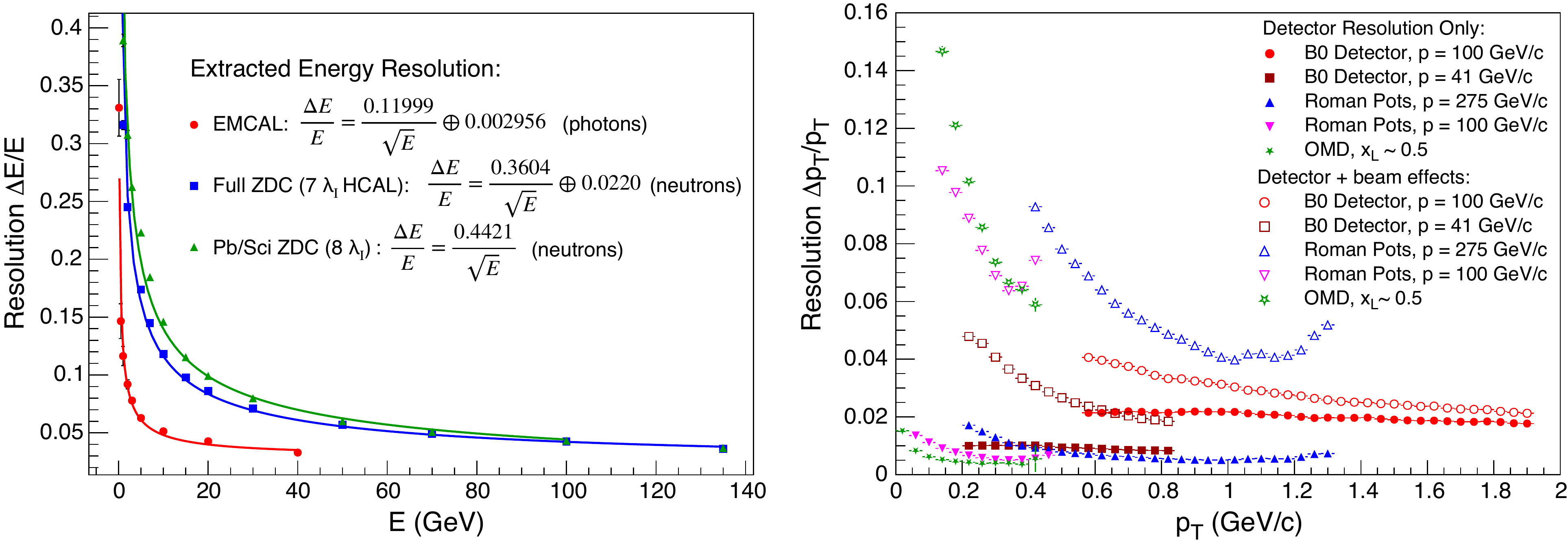}
\vspace{-5mm}
\caption{\label{fig:neutron_pt_res} Left: Energy resolution for the \gls{zdc} ECal for photons and combined ECal+HCal resolution for neutrons. Right: Transverse momentum resolution for protons in the 3 silicon tracking far-forward subsystems for various energies and rigidities, with and without beam effects 
(FullSim).}
\end{figure}

Resolutions for the far forward detectors are shown in Fig.~\ref{fig:neutron_pt_res}. 
The RPs and OMDs
utilize a transfer matrix which specifies the transport of protons from the IP through the magnetic elements 
of the far-forward lattice. However, the OMDs 
require a more sophisticated approach to handle the non-linear transport of protons with low rigidity ($\sim50$\% or less); these trajectories are at the edges of the quadrupole fields, which cause additional bending not captured by the linear transport matrices.

The performance demonstrated by the full GEANT4 simulations of the baseline \gls{zdc} is shown in the left panel of Fig.~\ref{fig:neutron_pt_res}. 
Preliminary results obtained from the W/SciFi + Pb/Scint \gls{zdc} simulations when compared to test beam data indicate that the performance meets the physics requirements.
In addition, an 8 $\lambda_I$ version of the detector was benchmarked against ZEUS test beam data. 
For photons, the measured performance of the W/SciFi ECal is consistent with previous studies. Simulations were made for photon energies down to 100~MeV for photons, the results indicating that this W/SciFi ECal can measure low energy photons.

 \FloatBarrier

\subsection{Luminosity Measurements and Low-\texorpdfstring{$Q^2$}{Q2} Tagging} 
\begin{figure}[!ht]
  \centering
    \includegraphics[width=\textwidth]{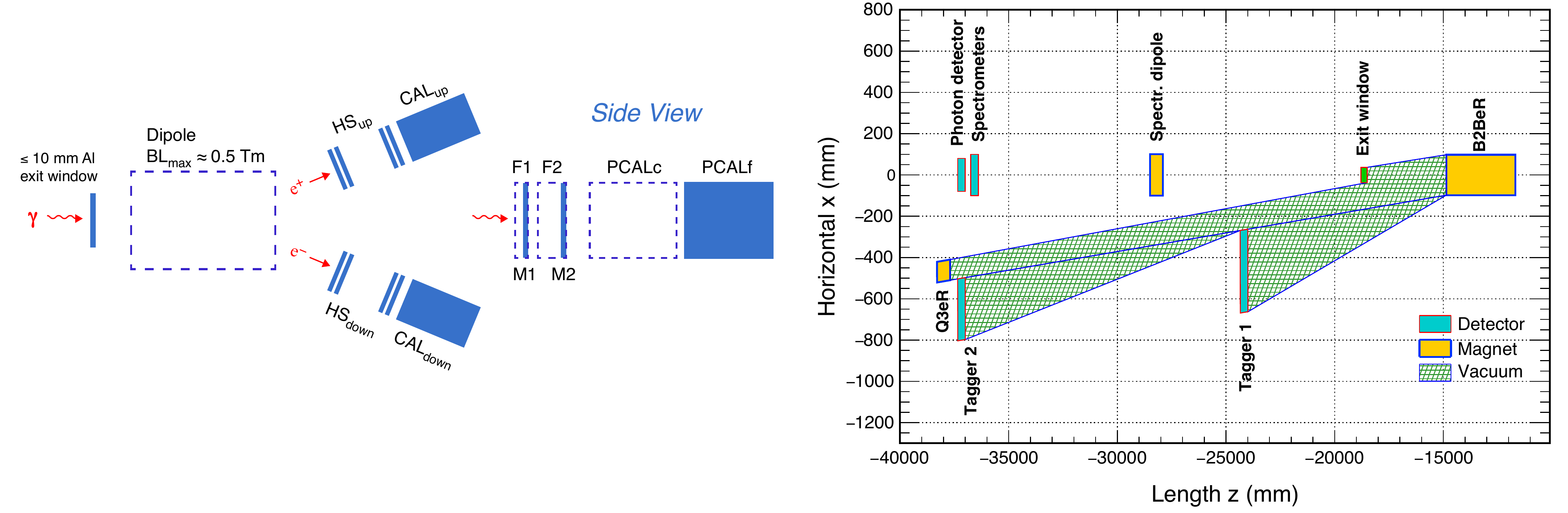}
  \caption{Left: Layout of the luminosity bremsstrahlung photon detectors. Right: Top view of far-backward region as simulated in Geant4, including two low-Q$^2$ taggers.}
  \label{fig:lumi-layout}
\end{figure}

As 
described in the Yellow Report, we 
will measure the EIC luminosity using the bremsstrahlung process~\cite{AbdulKhalek:2021gbh}.
To meet the
requirements of 1\% precision in the absolute luminosity measurement and 0.01\% precision in relative luminosity, we 
add 
further instrumentation
compared to the Yellow Report. 
This will enable 
data-driven corrections and systematic checks, 
and facilitate three largely independent and complementary measurement methods~\cite{Piotrzkowski:2021cwj}. The first method is based on counting photons converted in a thin exit window by applying a horizontal magnetic field and measuring $e^+ e^-$ pairs with two small calorimeters, CAL$_\mathrm{up/down}$ (left panel of Fig.~\ref{fig:lumi-layout}). This method is not sensitive to direct \gls{sr}, but at the nominal \ep\ luminosity the effects of event pileup 
are mitigated by two small hodoscopes, HS$_\mathrm{up/down}$. In the second method, the total energy carried by unconverted photons will be measured by the (movable) calorimeter, PCALf. By construction, it is not affected by the event pileup but direct SR 
must be suppressed using a set of filters, F1 and F2. This simple and robust measurement will also enable the online luminosity monitoring.
The third method is based on counting the unconverted photons using the movable calorimeter, PCALc, which will be used at small electron bunch current, when the event pileup and SR levels are low.
This is also essential to validate special corrections to the bremsstrahlung cross section~\cite{Piotrzkowski:2020uya}. 
All three measurements will be performed on a bunch-by-bunch basis, with negligible deadtime.

The detectors CAL$_\mathrm{up/down}$ and PCALc will be made using the same technology: Spaghetti W-calorimeter with radiation-hard scintillating fiber, read out with fast PMTs \label{sec:fb}.
Due to the very large irradiation levels in this location, PCALf and the SR monitors M1 and M2, will use Cherenkov-radiating quartz fibers read out by SiPMs.
Each of the hodoscopes HS$_\mathrm{up/down}$ will comprise four front (back) planes made by straight 1(2)~mm square scintillating fibers, also read out by SiPMs. 
Half of the fibers will be horizontal and half will be rotated by a small angle to allow for the determination of the horizontal coordinates of the $e^+ e^-$ hits. 
Signals from all detectors will be sampled with 100 MHz custom readout chips.

\begin{figure}[!ht]
  \centering
    \includegraphics[width=\textwidth]{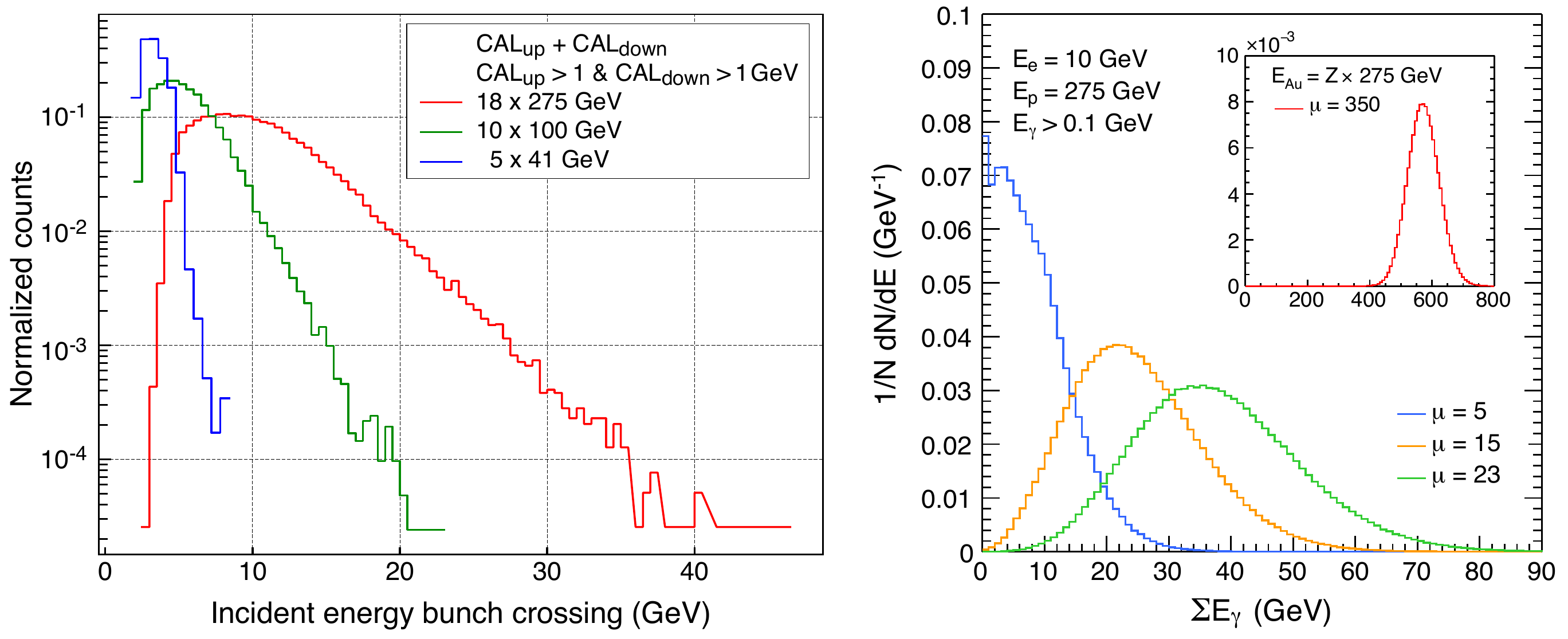}
  \caption{
  Left: Distributions of total energy deposited in CAL$_\mathrm{up}$ and CAL$_\mathrm{down}$ at the nominal EIC luminosities, for events where at least 1 GeV is deposited in each calorimeter. The high energy tails are due to significant event pileup (FullSim). Right: Distributions of total photon energy reaching PCALf for luminosities of L = 2.2, 6.5 and $10 \times 10^{33}$ cm$^{-2}$s$^{-1}$, resulting in the average photon multiplicities, $\mu$, as indicated.  In the insert, such a distribution is shown for \eAu\ collisions at the nominal EIC luminosity (FastSim).}
  \label{fig:lumi-phot-spectra}
\end{figure}

Bremsstrahlung electrons will be measured by two small detectors, Tagger~1 and Tagger~2, placed behind thin exit windows about 24~m and 37~m from IP6, respectively (right panel of Fig.~\ref{fig:lumi-layout}). They will be very similar in design to CAL$_\mathrm{up/down}$ and will have similar hodoscopes in front.  The major difference is that half of their fibers are horizontal and half are vertical.

While all detectors will be built using existing technologies, the huge bremsstrahlung event rates at the EIC, sometimes in excess of 10 GHz (Fig.~\ref{fig:lumi-phot-spectra}), make the design of the readout electronics and of radiation hard detectors challenging. 
Consequently, optimization of both the detector and electronics designs will require extensive studies and verification using test beams. 
This will include the development of suitable techniques to control, in-situ, all relevant systematic effects, as well as providing the \gls{daq} system for the luminosity measurement.

Our design will permit efficient tagging of very low-$Q^2$ events during the first two years of EIC running, when the event pileup is expected to be relatively low. 
An upgrade path after the initial phase consists of replacing the tagger hodoscopes 
with thin, high-resolution pixel detectors operated in the primary beam vacuum and read out by Timepix4 sensors~\cite{Ballabriga:2020nbo}. This will enable a very effective separation between the bremsstrahlung and low-$Q^2$ electrons at nominal EIC luminosities~\cite{Adam:2021hei}.

 \FloatBarrier

\subsection{DAQ and Readout Electronics}
\label{DAQ.and.Elec}
The EIC will provide \ep\ and \eA\ collisions at rates up to 0.5~MHz with 
beam bunches separated by 10~ns.
ATHENA will digitize hit position, charge and timing signals originating from nearly 30 distinct subdetectors using a variety 
of technologies including
\glspl{sipm}, \gls{maps}, \glspl{aclgad}, \glspl{mcppmt}, \glspl{pmt}, and \glspl{mpgd}.
The signals will be zero-suppressed where possible, digitized and aggregated using front-end boards containing a variety of \glspl{asic} and \glspl{fpga}. 
The primary function of the \gls{daq} system will be to aggregate data and record all collision related hits.
The system must also control, configure, and monitor the acquisition of data and ensure data quality.

\begin{figure} [htb]
   \centering
	\includegraphics[width=0.9\textwidth,trim=0 0 0 50,clip]{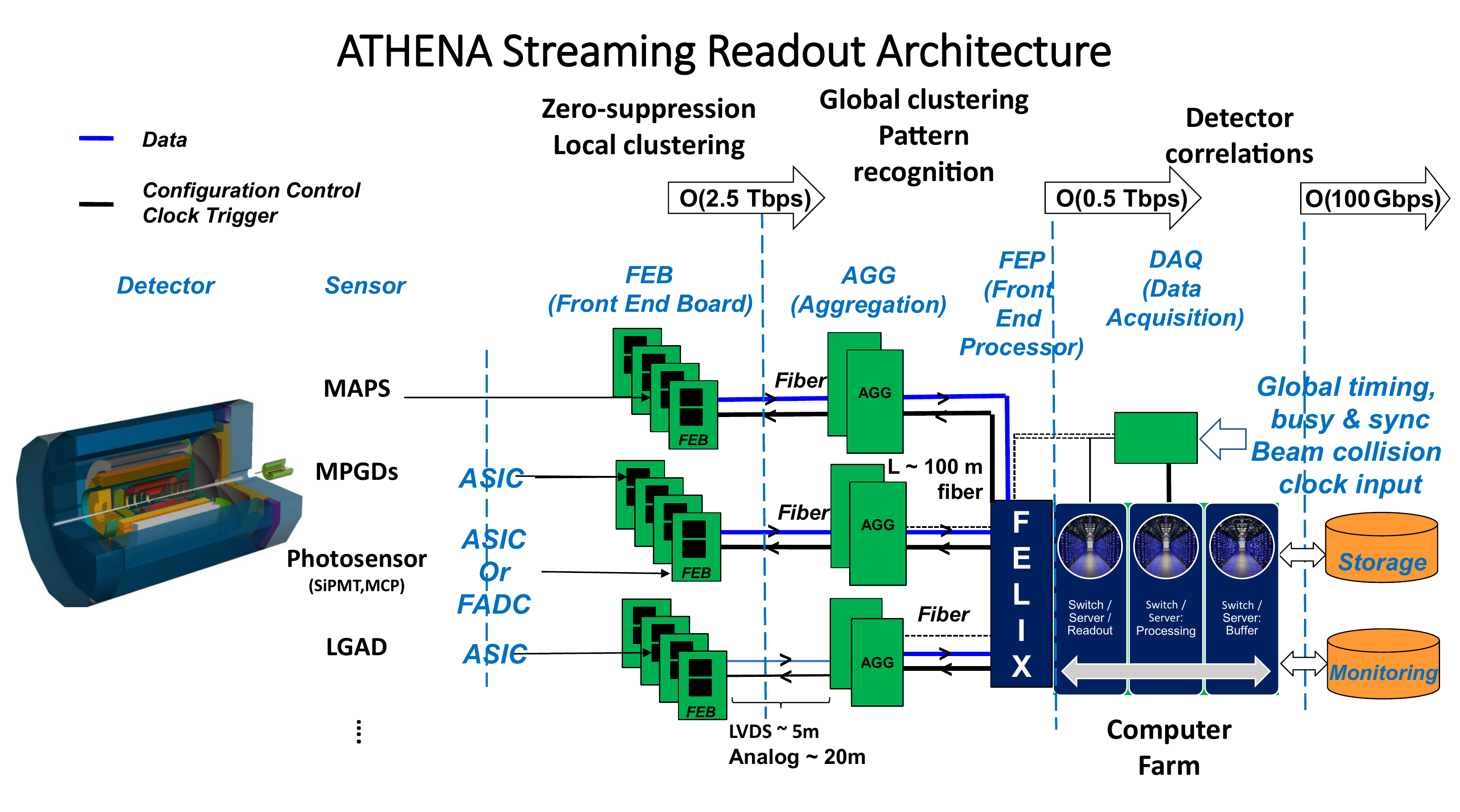}
	\caption{Overview of the ATHENA DAQ architecture.}
    \label{fig:daq_architecture}
\end{figure}

The chosen architecture is illustrated in Fig.~\ref{fig:daq_architecture}. It will be a streaming \gls{daq} system following the scheme outlined in the Yellow Report. A global timing system is needed to synchronize the system with the bunch structure of the EIC to a resolution of 10~ps. \gls{felix} boards, implementing GBTx links towards \gls{fee} with 10 Gbps/link bandwidth, are used as a basis to provide common interface between the \glspl{feb} and the commodity DAQ computers.
The \gls{felix} boards are capable of transferring data to the \gls{daq} computers and of transmitting clock and configuration information to the \glspl{feb}. 
A farm of 50 computers on a 100~Gbps Ethernet network is needed to read out the FELIX boards.
An additional 50 computers will be required to perform further data reduction, 
ensure data integrity and monitoring, buffer, and transfer data to tape.

An estimated maximum data volume by detector subsystem is shown in Tab.~\ref{fig:daq_rate}.
The 2.5~Tbps stream of data transmitted from the \glspl{feb} needs to be reduced to 100~Gbps for long-term storage without losing detector hits that arise from beam collisions. 

\begin{table}[!htb]
    \renewcommand{\arraystretch}{1.2}
    \footnotesize
    \centering
        \caption{Maximum data volume by detector.} 
        \label{fig:daq_rate}
        \begin{tabular}{|l|c|c|c|}
            \hline
            \textbf{Detector} & \textbf{Channels} & \textbf{DAQ Input~(Gbps)} & \textbf{DAQ Output~(Gbps) }\\
            \hline
            B0 Si & 400M & \textless 1 & \textless 1\\
            B0 AC-LGAD & 500k & \textless 1 & \textless 1 \\
            RP+OMD+ZDC & 700k & \textless 1 & \textless 1 \\
            FB Cal & 4k & 80 & 1 \\
            ECal & 34k & 5 & 5 \\
            HCal & 39k & 5.5 & 5.5 \\
            Imaging bECal & 619M & 4 & 4 \\
            Si Tracking & 60B & 5 & 5 \\
            Micromegas Tracking& 66k & 2.6 & .6 \\
            GEM Tracking  & 28k & 2.4 & .5 \\
            \textmu RWELL Tracking & 50k & 2.4 & .5 \\
            dRICH & 300k & 1830 & 14 \\
            pfRICH & 225k & 1380 & 12\\
            DIRC & 100k & 11 & 11 \\
            TOF & 332k & 3 & .8 \\
            \hline
            Total &  & 3334 & 62.9 \\
            \hline
        \end{tabular}
\end{table}

The biggest challenge to the goal of fully reading out the ATHENA detector with no deadtime will be the dark currents from the \gls{sipm} readout, expected to gradually increase with accumulated radiation dose.
The current estimates assume an average rate of up to 300~kHz/sensor over the full detector.
This dark current is indistinguishable from signals from single photoelectrons.
We hope to reduce this by a factor of three to five in the \glspl{feb} using sample cuts relative to the bunch crossing time.
Further reduction can be obtained by a software trigger applied in the \gls{daq} computers.
Requiring a collision to be present will provide a data reduction by a factor of at least 200 allowing the ATHENA \gls{daq} to write all collision data to tape.
Another option for data reduction is by machine learning techniques implemented in the \glspl{fpga} of the FELIX boards; dedicated development and feedback from initial data are needed.

Unexpectedly high noise in any detector, as well as the high rate from the \glspl{sipm} used as sensors in the PID devices, represent a potential unknown challenge.  The most obvious and immediate mitigation strategy is a \gls{daq} architecture that preserves the possibility of providing hardware trigger signals to specific detectors resulting in a hybrid triggered-streaming DAQ system. 
The potential avenues for data volume reduction will be the main R\&D required.
Significant development needs are expected to integrate each detector FEE with the FELIX board.   

 \FloatBarrier

\subsection{Software and Computing}
\label{sec:software-computing}
Software and computing will be critical to the success of any EIC experiment.
ATHENA chose to already now lay the foundations for  a long-term software strategy for the EIC.
To accomplish this, we focused on modern scientific computing practices: We developed a toolkit
of modular, orthogonal components designed 
for performance
in heterogeneous computing environments in the context of both
\gls{htc} and \gls{hpc}.
Furthermore, we emphasized modern development practices built around the use of a dedicated
GitLab server with a continuous-integration backend for reproducible containerization and automated tests and
benchmarks.

We 
leverage mature, well-supported, and actively developed software components
allowing us to focus our limited resources on those parts of the toolkit requiring custom development work.
ATHENA benefits from cutting-edge CERN-supported software developed for the (HL-)LHC.

\begin{figure}[htb]
  \centering
    \includegraphics[width=0.9\textwidth]{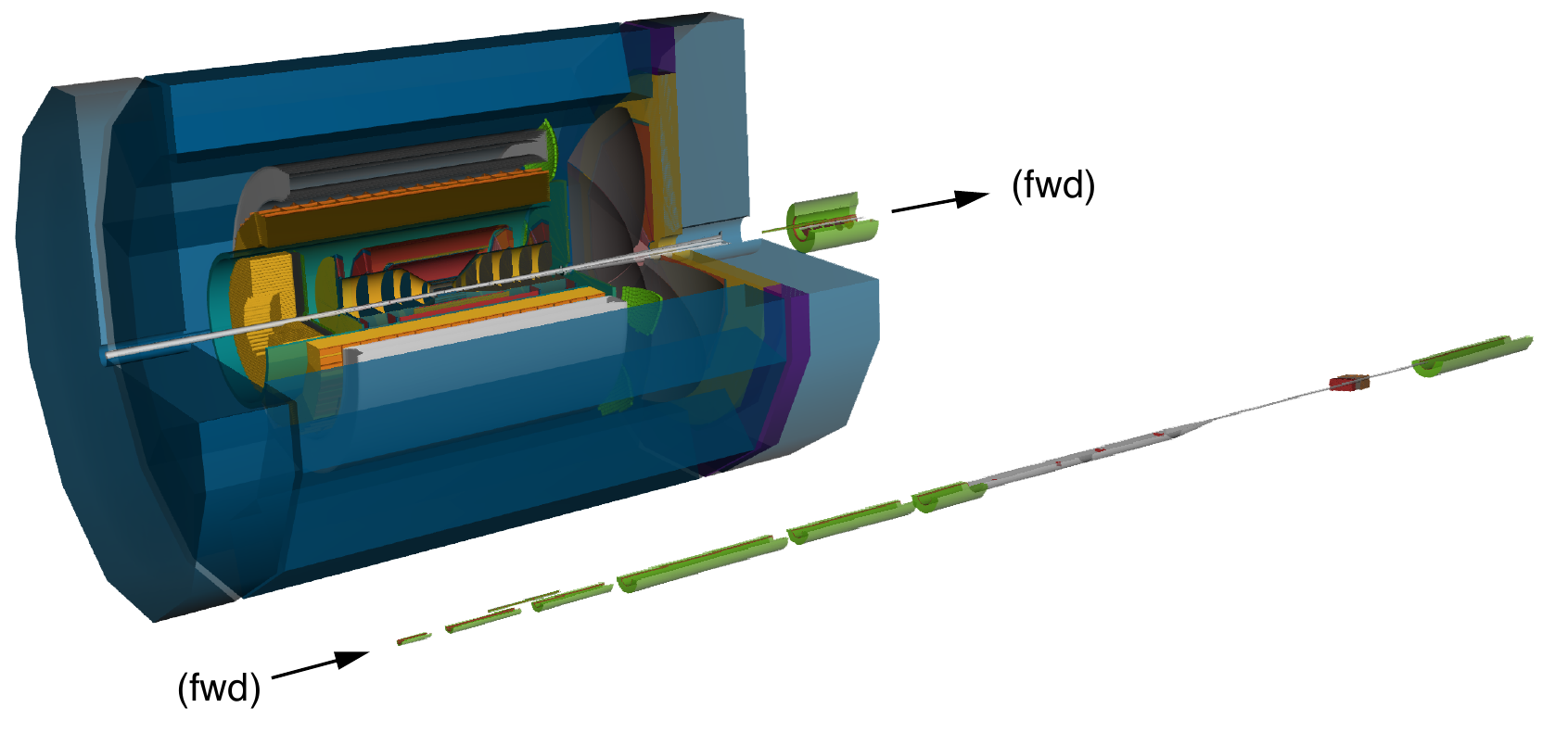}
  \caption{
  DD4hep implementation of the ATHENA central detector (top) and far-forward region (bottom). This geometry description is used in the GEANT4 simulation and the reconstruction.}
  \label{fig:ATHENA-dd4hep}
\end{figure}

We implemented our detailed detector geometries~\cite{Software:athena,Software:ip6,Software:npdet} in DD4hep~\cite{Software:DD4hep}, which provides geometry services for both the full GEANT4 simulation and our reconstruction algorithms (Fig.~\ref{fig:ATHENA-dd4hep}).
For the reconstruction framework, we chose Gaudi~\cite{Software:Gaudi}, as it supports modern task-based concurrency ideally suited for heterogeneous computing environments. On top of Gaudi, we built Juggler~\cite{Software:Juggler}, our library of digitization, reconstruction, and analysis routines, where we used \gls{acts}~\cite{Ai:2021ghi} for highly
performant tracking, and Tensorflow~\cite{Software:Tensorflow} for \gls{ai}.
These modular components communicate through a robust flat data model, EICD~\cite{Software:eicd},  implemented using the PODIO data model library~\cite{Software:podio}.

The maturity and robustness of the software components in the toolkit enabled us to build out, from scratch, a performant simulation and reconstruction software stack
over the short timeline since the call for detector proposals.
This toolkit implements the ATHENA detector in all its details, including  the far-forward and far-backward, accelerator, magnet, and detector components.
This setup allowed us to conduct effective detector optimizations for the proposal and prepare ATHENA for the road towards the \gls{tdr}. The simulation results in this proposal were obtained using our new software environment, deployed on an extensive range of systems, including the \gls{osg}, Jefferson Lab, BNL (including S3 storage), Compute Canada, ALCF (ANL), LCRC (ANL), NERSC (LBNL), INFN-CNAF, and a dedicated continuous-integration cluster at ANL.
We strongly believe that this innovative approach, introduced within ATHENA, represents a significant step forward for the EIC community.

 \FloatBarrier

\subsection{Integration and Installation}
As a new EIC detector to be installed in the existing experimental hall at 
\gls{ip6} at the RHIC collider complex, there are many space constraints impacting the ATHENA design:
\begin{itemize}[leftmargin=0.5cm]
\item{The layout of the IR magnets provides an accelerator element free region for the detector of overall 9.5~m, resulting in 
-450~cm to 500~cm around the IP in the outgoing electron and hadron-beam direction, respectively. This constrained space makes any assembly and most inner detector maintenance impossible in-situ.}
\item{Approximately 50~cm space between both endcaps and the first IR magnets is occupied with valves and vacuum pumps.}
\item{The height of the IP above the floor is 432~cm, influencing the design of the cradle and the integration of the detector with the cradle.}
\item{The detector solenoid needs to be aligned with the electron beam direction to minimize the generation of synchrotron radiation, therefore the entire detector must be rotated by 8~mrad in the horizontal plane away from the central axis.}
\item{The door size between the assembly and the collider hall is 823~$\times$~823~cm$^2$. This directly
limits  the size of the detector, as it will need to be rolled to the assembly hall for installation and maintenance.}
\item{The \gls{rcs} runs at 335.2~cm radial distance from the IP at a height of 372~cm above the floor level. This limits the outer radius of the detector to 330~cm.}
\item{The detector solenoid fringe field should stay below $5 \times 10^{-4}$~T past the endcaps in the vicinity of the IR magnets, and the field integral should not exceed 0.007~Tm along the \gls{rcs} beam line. This has a direct impact 
on the design of the flux return of the detector.}
\item{The beam pipe widens towards the endcaps to accommodate the synchrotron radiation fan and the cone of protons, neutrons, and particles from nuclear breakup. This has a direct impact on how the detectors need to be installed, either in a clamshell configuration or in sectors around the beam pipe.}
\item{Due to the compactness of the IR and the detector design, the first valve on the outgoing hadron side can only be integrated after the first bending magnet in the hadron beamline downstream of the IP at approximately 7~m. In the outgoing electron side, such a valve can be placed after the nHCal at approximately 4.5~m from the IP.}
\end{itemize}

\begin{figure}[htb]
    \centering
    \begin{minipage}[c]{0.65\linewidth}
    \includegraphics[width=0.9\linewidth]{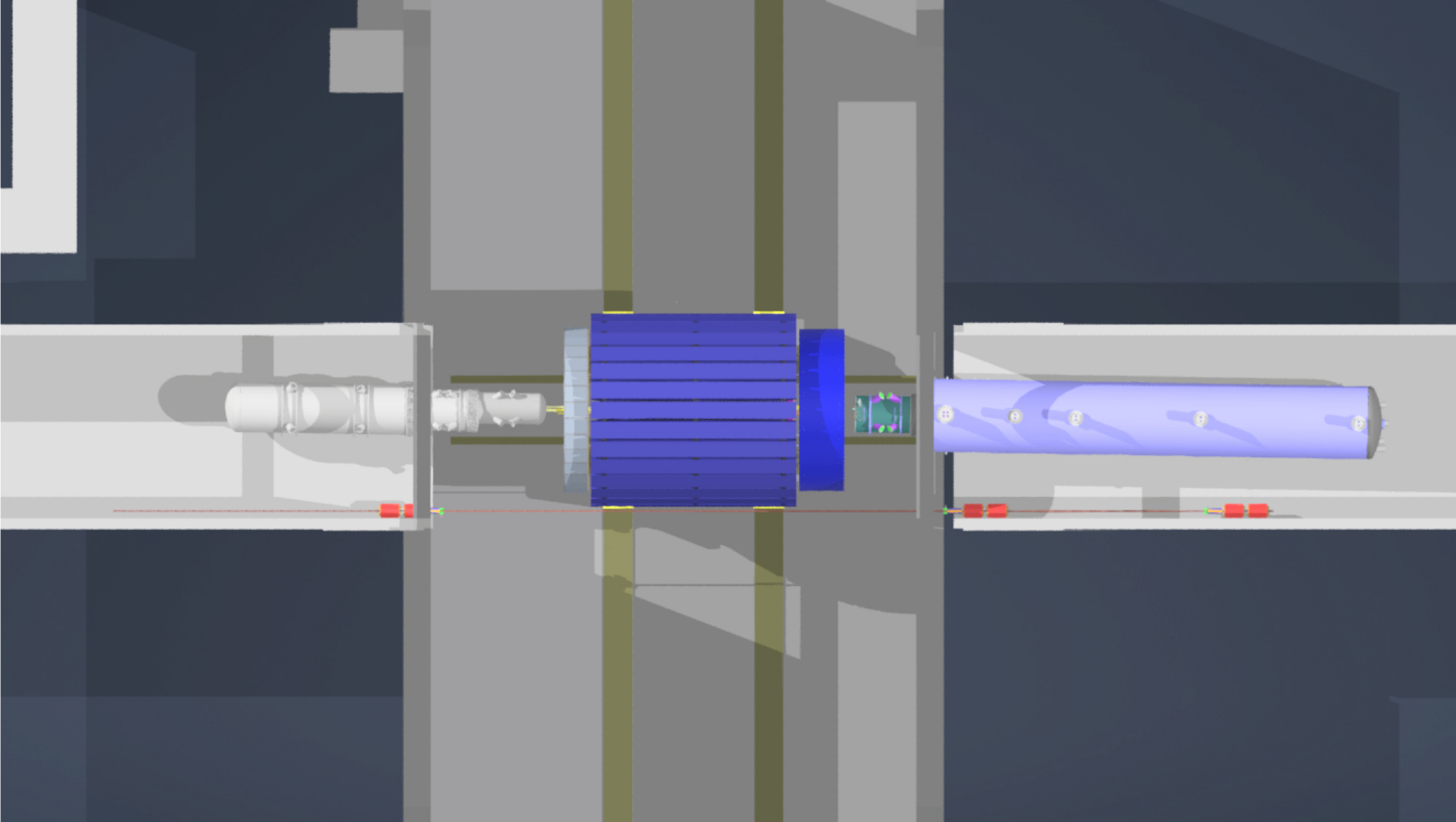}
    \end{minipage} 
    \begin{minipage}[c]{0.3\linewidth}
    \caption{Integration of the ATHENA detector in the experimental hall at IP6 and the interaction region (top view).}
    \label{fig:ATHENA-IP6-Layout}
    \end{minipage}
\end{figure}

\begin{figure}[!hbt]
    \centering
    \includegraphics[width=0.94\linewidth]{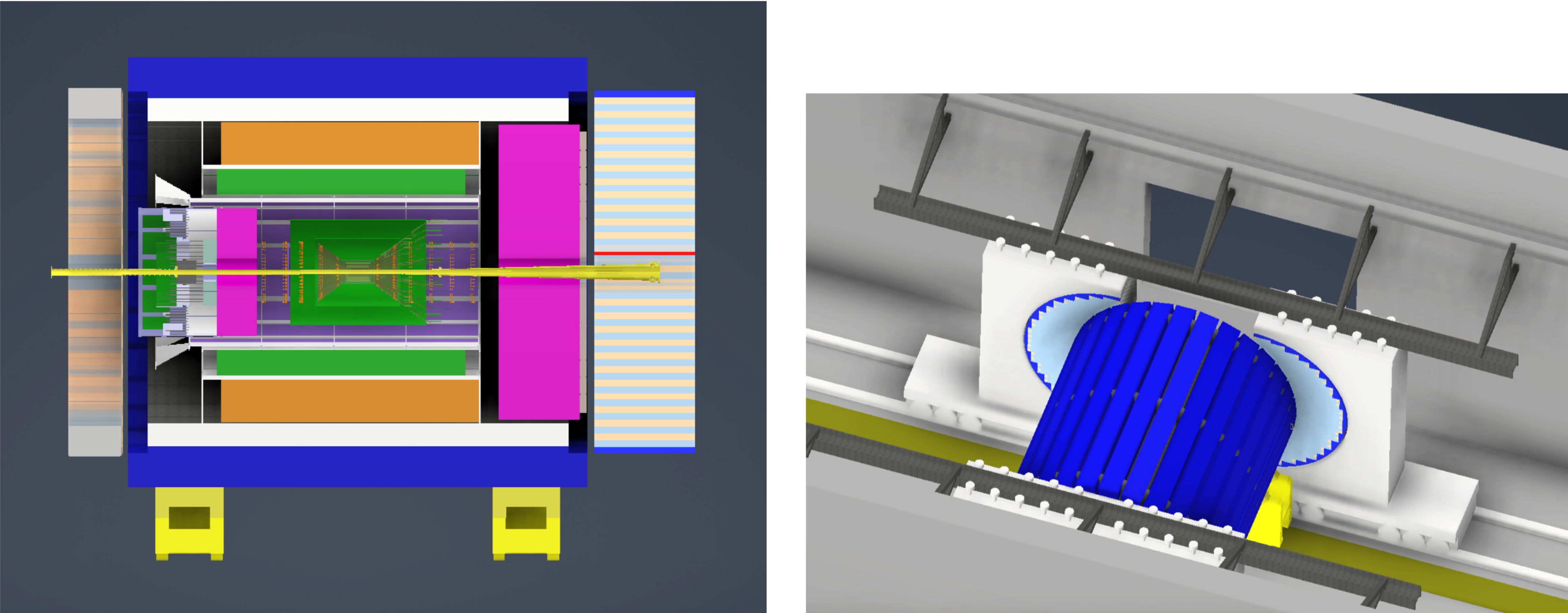} 
    \caption{Left: Conceptual design of the ATHENA detector in ProE-Creo, the computer-aided design program used for the EIC Project. Right: Conceptual design of the pECal and pHCal opening perpendicular to the beam axis.}
    \label{fig:ATHENA-Layout}
\end{figure}

Figure \ref{fig:ATHENA-IP6-Layout} shows ATHENA fully integrated into the interaction region and the experimental hall at IP6, obeying all the above listed constraints.
Because of the stringent space constraints, it is important to keep the assembly procedure of the detector as a design consideration. As a consequence, the central part of the detector is installed from the outside in and the electron and hadron endcaps from the inside out. Figure \ref{fig:ATHENA-Layout} left shows the overall concept for the ATHENA detector as implemented in ProE-Creo, the computer-aided design program used for the EIC Project.
The installation sequence is as follows:
\begin{description}[leftmargin=0cm]
    \item[Barrel Detector:]
    The detector cradle is followed by the lower half of the flux return and the hadron calorimeter. Then the solenoid and the upper half of the flux return and the hadron calorimeter are installed. The bECal and its support structure are then installed, followed by the hpDIRC support structure, the DIRC bars, and barrel MPGD and MAPS trackers. The MAPS tracker has a clamshell design around the beam pipe. 
    \item[Electron Endcap:] The first detectors to be installed are the MAPS disks, followed by the MPGD disks, the aerogel pfRICH, the nECal, and finally the DIRC readout. The nECal is supported like the pfRICH from the DIRC support structure.
    \item[Hadron endcap:] As on the electron side the first detectors to be installed are the MAPS disks, followed by the MPGD rings. The dRICH is installed next followed by the MPGD tracker. Both are supported from the pHCal.
    \item[Endcap calorimeters:] The nHCal, pECal, and pHCal are installed independently on their own cradles in the collider hall. These calorimeters can be opened perpendicular to the beam axis to disconnect the beam pipe and roll the central part of the detector to 
    the assembly hall (Fig.~\ref{fig:ATHENA-Layout} right). It is noted that the \gls{rcs} beam pipe needs to be separated to allow the endcap HCals to open.
\end{description}

To route the cables to the outside of the detector, $\sim$10~cm service gaps are foreseen between the barrel and the endcap calorimeters. In addition, there is a service gap between the end of the solenoid, bECal, DIRC and the dRICH to route the barrel and hadron endcap tracker services to the outside. On the side of the electron endcap, the services are routed between the pfRICH, the nECal, and the hpDIRC towards the service gap.

 \FloatBarrier

\subsection{R\&D Needs}
\label{sec:rd_needs}
\label{RDNeeds}
R\&D on many ATHENA subsystems was initially conducted through the \href{https://wiki.bnl.gov/athena/index.php/Detector_R\%26D}{EIC Generic Detector R\&D Program}~\cite{EIC-RD} that ran from 2011 until September 2021. 
Several subsystems matured  to levels where little basic R\&D remains and the focus is shifting to the construction and testing of full chain prototypes (detector, electronics, supplies, and DAQ) in test beams. These developments will be supported by \href{https://wiki.bnl.gov/conferences/index.php/General_Info}{EIC project funded R\&D} \cite{EIC-Project-RD} that will start in FY22. 
Project funds
will also support R\&D of components that have not reached their full potential yet and need further optimization. Only a limited number of technologies still need substantial R\&D and have been selected because of the performance benefits they offer. Here we give a brief summary of ATHENA's R\&D needs.

\begin{description}[leftmargin=0cm]
	\item[MAPS:]
	The EIC Silicon Consortium is partnering with the ALICE collaboration to develop the ITS3 sensor and to modify it, as necessary, for use at the EIC.
	This partnership will help to reduce the risk inherent in developing a detector solution in 65~nm technology. The 
	plan is to use
	the ITS3 wafer-scale sensor for the vertex layers and to develop a smaller version for the silicon tracker barrel layers and disks. A full description of the planned development can be found in the eRD25 proposal~\cite{eRD25prop}. 
	A full detector infrastructure will be developed in parallel with the sensor, to modify the vertex layers to fit the geometry of the beam pipe at the EIC and to develop stave and disk configurations of sensors, which are outside of the scope of ITS3. 
	There are two EIC Project R\&D activities associated with 
	these developments:
	eRD111 deals with
	R\&D towards forming modules from sensors, developing the EIC-specific infrastructure required to produce staves and disks, mechanical support structures, and cooling. eRD104 carries out R\&D into services reduction, powering, and readout, which has the potential to further enhance the performance of the system.
	
	\item[GEMs:]
	While GEM technology is very mature and has been used in a wide variety of experiments, the EIC has a unique requirement of demanding very low-mass trackers. Work carried out by the eRD6 Consortium, within the EIC generic detector R\&D program, was able to successfully build and test two 1~m long triple-GEM detectors.
	These detectors achieved a material budget in the active area of $0.4\%~X/X_0$.
	With the material in the active area of the GEMs minimized, efforts are now focused on reducing the GEM support frames, which sit in the $\phi$-$\eta$ tracking acceptance.
	This R\&D effort is being addressed within the eRD108 EIC R\&D project, along with the optimization of the readout structure.
	
	\item[$\boldsymbol{\mu\mathrm{RWell}}$:]
	Using $\mu\mathrm{RWell}$ detectors instead of GEM detectors in the endcap could reduce material budget, simplify construction, and lower cost for the overall endcap tracker. However, $\mu\mathrm{RWell}$ detectors have not yet been used in any major nuclear or high energy physics experiment. R\&D would therefore be needed to build and test a large $\mu$RWell detector and is one of the goals of the eRD108 EIC R\&D project. If $\mu\mathrm{RWell}$ does not develop enough within the EIC time frame, GEM technology can be used as a fallback to the large-area $\mu\mathrm{RWell}$ detector located behind the dRICH.
	
	\item[Micromegas:]
	The proposed, curved Micromegas technology has been successfully installed in the central tracker of the CLAS12 experiment at Jefferson Lab. The low material budget  of less than $0.4\%~\mathrm{X/X_0}$ in the active region and the ability to shape the detectors in cylindrical structures are crucial features for the ATHENA central detector. 
	The ongoing R\&D focuses on the optimization of the Micromegas 2-D readout and it will also be addressed within the eRD108 EIC R\&D project .
	This R\&D is also of interest for GEM and \gls{murwell} detectors.
	
	\item[hpDIRC:]
	The objective of the R\&D program, eRD103, is to validate the PID performance of a
	cost-optimized hpDIRC design with a vertical-slice prototype in a particle beam
	by FY24. Key topics include: the usability of BaBar DIRC bars, the development of compact readout electronics for the fast detection of single photons with high-density sensors, and the validation of the PID performance of a cost-optimized design.
	
	\item[dRICH:]
	The main technical goals for near term R\&D is the preparation of the basic version of the dRICH prototype and
	first test beams at CERN. The tests are organized in synergy with ALICE and will have, as complementary targets, the study of the single-photon response of \gls{sipm} coupled to the ALCOR readout electronics, and the comparative use of Russian and Japanese aerogel. The goal of these initial test-beams is to commission the
	dRICH prototype and to define what has to be improved to reach the design performance. FY23/FY24 R\&D will be targeted to the definition of technical specifications to meet the EIC requirements, matching to EIC-driven (developed by other EIC R\&D) photosensors and readout electronics, and validation of cost-effective component technologies to mitigate the construction risk.
	
	\item[AC-LGAD bToF:]
	The  eRD112 R\&D efforts for a bToF with AC-LGAD sensors can be categorized in 3 main areas: sensors, electronics, and system design (including cooling, engineering, and construction). On the sensor side extensive R\&D is already ongoing at several institutes in the LGAD Consortium, e.g., reducing the active volume thickness and optimizing implantation parameters so that a time resolution below 20 ps can be achieved. One important effort specific to the bToF is to develop long strip AC-LGAD sensors for a lower material budget. The needs for fast timing performance and finer granularity also pose significant challenges to the readout electronics and specifically to the \gls{asic} readout chips. ATHENA  will collaborate on addressing these challenges. For example, the IJCLab (Orsay), \'Ecole Polytechnique/Omega and CEA (Saclay) groups are currently developing a new ASIC that meets the requirements set by the EIC Roman Pot detector, and are thus providing in-kind contributions. First discussions among the institutes of the LGAD Consortium and ATHENA have started to develop an ASIC that serves both the pixelated design (RPs) and the strip design (bToF).
	
	\item[Photosensors:]
	The objective of the eRD110 R\&D effort is to mitigate technical, cost,
	and schedule risk related to readout sensors of EIC Cherenkov detectors and calorimeters. The effort should allow for a  well-informed decision for a baseline sensor solution for each PID detector in FY23, taking into account the impact of B-field strength and relative sensor orientation with respect to the field direction, as well as the expected radiation levels.
	The proposed R\&D activities related to the characterization of (i) \textit{Photek/Photonis} \glspl{mcppmt}, (ii) the pixelization and improvement of the field resistance  of \glspl{lappd} and \glspl{hrppd}, and (iii) the improvement of the radiation hardness of \glspl{sipm} and the optimization of the sustainability of their proper temperature treatment in collaboration with various manufacturers.
	
	\item[Calorimetry:]
	There are four project R\&D programs related to ATHENA calorimetery: eRD105 for development of high-resolution electromagnetic calorimeters based on scintillating glass; eRD106 and eRD107 for development of a very compact, high-resolution hadron endcap system; and eRD110 for studies of the impact on calorimeter performances due to radiation damage of the \glspl{sipm}. The hadron calorimeters for the barrel and negative endcap do not require R\&D. For the novel barrel electromagnetic tracking calorimeter, which is not covered by any of the R\&D projects, the construction of a smaller prototype to validate the Monte Carlo simulations of the device is imperative. The manufacture and assembly of the prototype will be used to streamline the later manufacture and assembly of the full calorimeter. An estimated 80\% of the AstroPix sensors used in the prototype should be recovered for use in the full scale electromagnetic calorimeter.
	
	\item[Auxilliary Detectors:]
	For the \gls{rps} and \gls{omds}, use of AC-LGADs is envisioned, which relies on successful completion of R\&D eRD112 efforts for the design of a new ASIC to be used with the AC-LGAD sensors available from various vendors, and optimal choice of pixel size (which impacts the ASIC design). The charge sharing capability of the AC-LGAD sensor allows for improvements of spatial resolution up to 20 times that of conventional silicon, which could enhance the applicability for other subsystems (e.g., the B0 tracker). Given the large community interest to develop this technology, it should be a safe option for these subsystems. If needed, it could be substituted with other silicon technologies with the added need for a separate timing layer.
	
	\item[Electronics \& \glspl{asic}:]
	Much of the R\&D efforts on electronics and especially \glspl{asic} will be coordinated through the EIC project through eRD109.  Substantial R\&D will be needed for the development of \glspl{fee}. Here we define \glspl{fee} as \glspl{asic}, \glspl{feb}, and \glspl{fep}. \glspl{fep} may be required if the \glspl{asic} do not provide all required features. 
	The choice of using streaming readout for ATHENA excludes several existing \gls{asic} chips. We estimate the need for up to four \glspl{asic} used for the readout of (i) \gls{sipm}, (ii) \gls{mcppmt}, (iii) \gls{micromegas}/\gls{gem}/\gls{murwell}, and (iv) \gls{aclgad}.  Development of a new \gls{asic} takes four to five years, while an update or modification of an existing design requires less time. Considering the project timelines, developments will likely have to occur concurrently, requiring the involvement of multiple groups.
	
\end{description}

 \FloatBarrier

\subsection{Challenges and Mitigation Policies}
\label{sec:challenges}
A detailed assessment of the challenges associated with the technologies selected by ATHENA is presented in Tables  \ref{tab:trk}, \ref{tab:cal}, \ref{tab:pidA}, \ref{tab:pidB}, and \ref{tab:aux}. The technological maturity listed is evaluated on a scale of 0 to 10, where 10 corresponds to a fully developed technology. The  mitigation strategies to minimize the impact of the challenges are also presented.

\begin{sidewaystable}
\centering
\footnotesize
\captionof{table}{Challenges and mitigation policies for tracking technologies.}
\label{tab:trk}
\renewcommand{\arraystretch}{1.4}
\begin{tabular}[h]{|p{1.8cm}|p{4cm}|p{8.5cm}|p{7cm}|}
%
%
\hline
\multicolumn{4}{|c|}{\textbf{Tracking}} \\ \hline
\rowcolor[gray]{.9}
\multicolumn{1}{|c}{Component} &
\multicolumn{1}{|c}{Technology} &
\multicolumn{1}{|c}{Challenge considerations} &
\multicolumn{1}{|c|}{Alternatives \& Mitigation} \\
\hline \hline
%
%
Si-trackers, vertex layers &
Extremely thin (0.05\% $X/X_0$ per layer), wafer-scale, curved MAPS in 65 nm technology and pixel size $\cal{O}$(10 $\mu$m). &  
Development mainly for ALICE ITS3 upgrade; large international and inter-laboratory synergies; synergies within the EIC community via the Silicon Consortium; highly innovative: \textbf{Maturity Level of the Technology: 6} &
ITS3 fallback solution: new sensor in 180~nm technology, with pixel pitch and power specifications close to the 65 nm sensor. In case of delays in the development of both the ITS3 sensor and the fallback solution, the experiment will start with ALPIDE and upgrade to a new sensor as soon as possible. It is understood that higher power consumption of the ALPIDE will translate in more material and thus some degradation in the performance of the tracking detector and electromagnetic calorimeter.
\\
\hline
Si-trackers, barrel layers and disks &
Same sensor technology as for vertex, size optimized for cost-effective, large area coverage, mounted on staves and disks. With this arrangement: 0.55\% $X/X_0$ per barrel layer, 0.24\% $X/X_0$ per disk.
&
Same as for vertex. &
Same as for vertex.  \\
\hline
Cylindrical Micromegas  &
Curved Micromegas. &  
Curved Micromegas in operation at CLAS12 at Jefferson Lab. 2-D readout by large strips $\cal{O}$(1.5~mm)  with fine resolution $\cal{O}$(0.15~mm) has to be established, even if supported by MICROMEGAS with 2-D readout used in ASUCUSA at CERN and for muography.  \textbf{Maturity Level of the Technology: 8}   &
Fallback options: use of MAPS trackers in an all silicon tracking system; Micromegas with smaller pitch readout strips resulting in an increased number of readout channels; small flat GEMs combined to approximate the cylindrical shape.\\
\hline
Planar GEMs  with annular shape surrounding the Si disks &
Large-size GEMs with reduced support material. &  
Large GEMs of comparable size are under construction for the upgrade of the CMS muon system and large-size chambers have been validated with test beam studies. Large GEMs with reduced material in the support and readout planes have been developed for SBS and PRad at Jefferson Lab. \textbf{Maturity Level of the Technology: 9}  &
Fallback options: detector segmentation; different technology using large-size Micromegas (ATLAS NSW).  \\
\hline
$\mu$RWELL  &
large-size $\mu$RWELL.  &  
$\mu$RWELL never used in an experiment; foreseen in LHCb upgrade; considered also for SOLID and CLAS12 upgrade at Jefferson Lab; \textbf{Maturity Level of the Technology: 7}  & 
Alternative is using different gaseous detector technologies: 
GEM, Micromegas, sTGC. \\
\hline
\end{tabular}
\end{sidewaystable}

\begin{sidewaystable}
\centering
\footnotesize
\renewcommand{\arraystretch}{1.4}
\caption{Challenges and mitigation policies for calorimeter technologies.}
\label{tab:cal}
\begin{tabular}{|p{2.5cm}|p{4cm}|p{8cm}|p{6cm}|}
%
\hline
\multicolumn{4}{|c|}{\textbf{Calorimetry}} \\ \hline
\rowcolor[gray]{.9}
\multicolumn{1}{|c}{Component} &
\multicolumn{1}{|c}{Technology} &
\multicolumn{1}{|c}{Maturity} &
\multicolumn{1}{|c|}{Alternatives \& Mitigation} \\
\hline \hline
%
%
HCal (Forward, Barrel, Backward) &
HCal Fe/Scint sandwich. &
Well established technology with recent up-to-date implementations as STAR forward HCal. The longitudinal segmentation has to be implemented. \textbf{Maturity Level of the Technology: 9}  & Further studies to optimize the details of the layout are needed. Fallback option: implement the same technology as used for the STAR forward HCal. \\
\hline
EMCal, forward &
W-Powder/SciFi calorimeter. &
W-Powder/SciFi calorimeter with SiPM sensors, going to be used in sPHENIX. The optimization of the uniform light collection is still to be established. \textbf{Maturity Level of the Technology: 9} &
Further studies to optimize the light collection. \\
\hline
EMCal barrel &
Hybrid calorimeter with front imaging layers using AstroPix sensors alternated with Pb/SciFi layer followed by a set of Pb/SciFi layers. & 
Innovative design. The performance  
and interplay between the ECal device and the HCal located outside of the  solenoid has to be understood and established. The required technologies, namely Pb/SciFi calorimetry (most recently: GlueX) and AstroPix, are established. \textbf{Maturity Level of the Technology: 9}&
Further studies to optimize performance are needed.\\
\hline
EMCal backward &
Center equipped with PbWO$_4$ crystals, peripheral area by novel scintillating glass.  &
PbWO$_4$ crystals: procurement uncertainties only.  Scintillating glass: the novel development specific for EIC is already well advanced.  Synergies within the EIC community exploited by the EEEMCAL consortium. \textbf{Maturity Level of the Scintillating Glass Technology: 7} &
For crystals: anticipate the market survey and purchasing. For scintillating glass, consider alternative technologies as  lead-glass (possibility of material re-use). \\
\hline
\end{tabular}
\end{sidewaystable}

\begin{subtables}
\begin{sidewaystable}
\centering
\footnotesize
\renewcommand{\arraystretch}{1.4}
\caption{Challenges and mitigation policies for particle identification technologies.}
\label{tab:pidA}
\begin{tabular}{|p{1.5cm}|p{4cm}|p{8cm}|p{7cm}|}
%
%
\hline
\multicolumn{4}{|c|}{\textbf{Particle Identification}} \\ \hline
\rowcolor[gray]{.9}
\multicolumn{1}{|c}{Component} &
\multicolumn{1}{|c}{Technology} &
\multicolumn{1}{|c}{Maturity} &
\multicolumn{1}{|c|}{Alternatives \& Mitigation} \\
\hline \hline
%
%
dRICH (forward), global design &
Combination of gas and aerogel in large acceptance focusing  RICH in magnetic field. &
Two-radiator RICHs already operated in experiments (HERMES, LHCb). The specific radiator combination in relation with the expected environment at EIC to be validated. \textbf{Maturity Level of the Technology: 9}  &
Completion of the ongoing studies by prototyping and test beam runs.  \\
\hline
hpDIRC (barrel), global design &
DIRC concept empowered by focusing element and fine-pixel readout. &
Focusing DIRC extensively developed and confirmed by test beam for PANDA at GSI. Principle of hpDIRC (more refined focusing by lenses) demonstrated with optical studies. Confirmation of the performance of the hpDIRC needed. \textbf{Maturity Level of the Technology: 9}  &
Completion of the ongoing studies by prototyping and test beam runs.  \\
\hline
pfRICH (backward), global design &
Proximity focusing RICH with aerogel radiator and large proximity gap for fine resolution. &
Proximity focusing demonstrated in various experiments (ALICE, BELLE II); aerogel as RICH radiator demonstrated (HERMES, LHCb, BELLE II, CLAS12). Confirmation of the performance of the pfRICH needed. \textbf{Maturity Level of the Technology: 9} &
Completion of the simulation studies specific to EIC. \\
\hline
Photosensors by SiPMs (for dRICH and pfRICH) &
SiPMs at low temperature ($\sim -40^\circ$~C). &
The validation of the approach is via a dedicated  R\&D program including the study of the dark current versus irradiation dose and versus repeated thermal annealing cycles; SiPM selection by characterizing devices by different providers; highly innovative approach. \textbf{Maturity Level of the Technology: 6}  &
Continue pursuing the development of an alternative approach by large-size MCP devices: LAPPDs; as both the baseline choice (SiPM)  and the alternative option (LAPPD) are not established, both items require management attention, adequate support, and investment.  \\
\hline
\end{tabular}
\end{sidewaystable}
\begin{sidewaystable}
\centering
\footnotesize
\renewcommand{\arraystretch}{1.4}
\caption{Challenges and mitigation policies for particle identification technologies.}
\label{tab:pidB}
\begin{tabular}{|p{1.5cm}|p{4cm}|p{8cm}|p{7cm}|}
%
%
\hline
\multicolumn{4}{|c|}{\textbf{Particle Identification}} \\ \hline
\rowcolor[gray]{.9}
\multicolumn{1}{|c}{Component} &
\multicolumn{1}{|c}{Technology} &
\multicolumn{1}{|c}{Maturity} &
\multicolumn{1}{|c|}{Alternatives \& Mitigation} \\
\hline \hline
%
%
Photosensors by MCP-PMTs (for hpDIRC) &
Commercial 1-inch MPC-PMTs (most likely by \it{Photonis}). & Challenges arise from the
production rate and cost. The use of the technology in high magnetic field that can reduce the gain and, therefore, the time resolution needs to be confirmed. \textbf{Maturity Level of the Technology: 9} &
Early purchasing procedure. Further studies of performance in realistic magnetic field.
Fallback option:
continue pursuing the development of an alternative approach by large-size MCP devices (LAPPD).   \\
\hline
Radiator gas (for dRICH) &
C$_2$F$_6$. &
Procurement difficulties and increasing cost because of increasing usage restrictions (worldwide) related to the extremely high \gls{gwp}).  Recirculation plant required. \textbf{Maturity Level of the Technology: 7} &
Fallback option: Develop the approach by pressurized Argon to replace the use of fluorocarbon gas.   \\
\hline
Aerogel (for dRICH and pfRICH) &
Low refractive index (1.02) aerogel &
The homogeneity and yield in the production of hydrophobic or hydriphilic low refractive index aerogel has to be demonstrated as well as the production rate.  \textbf{Maturity Level of the Technology: 7} &
Develop the low refractive index aerogel together with more than a single producer
(University of Chiba, Japan, Budker Institute, Russia, and  ASPEN AEROGELS, INC., USA where aerogel development is supported by an SBIR grant); early purchasing.    \\
\hline
Synthetic fused silica bars (for hpDIRC) &
Synthetic fused silica bars with high precision mechanical parameters and very fine surface polishing; re-use of BaBar material expected. &
A detailed protocol for BaBar silica bar disassembly to be established; a protocol for barrel assembly in ATHENA needs to be put in place. \textbf{Maturity Level of the Technology: 9} &
Guidance from BaBar and PANDA experience.  \\
\hline
Barrel TOF  &
Sensors: \gls{aclgad}. &
Dedicated development, already advanced at the present time; synergies with other proposed applications: ATLAS roman pots for HL-LHC, LHCb upgrade, ALICE3, NA62 (CERN), PIENUX  (TRIUMF), and PAN space missions. \textbf{Maturity Level of the Technology: 8}  &
Continue pursuing the development of an alternative approach for PID at low momenta to complement PID by DIRC in the barrel: R\&D dedicated to a miniTPC with sensors by GridPix technology.  \\
\hline
\end{tabular}
\end{sidewaystable}
\end{subtables}

\begin{sidewaystable}
\centering
\footnotesize
\renewcommand{\arraystretch}{1.4}
\caption{Challenges and mitigation policies for technologies deployed in far-forward and far-backward detectors.}
\label{tab:aux}
\begin{tabular}{|p{1.5cm}|p{4cm}|p{6cm}|p{6cm}|}
%
%
\hline
\multicolumn{4}{|c|}{\textbf{Far-Forward Detectors}} \\ \hline
\rowcolor[gray]{.9}
\multicolumn{1}{|c}{Component} &
\multicolumn{1}{|c}{Technology} &
\multicolumn{1}{|c}{Maturity} &
\multicolumn{1}{|c|}{Alternatives \& Mitigation} \\
\hline \hline
%
%
B0 &
Sensors: MAPS. &
ALPIDE in operation at ALICE, CERN  is foreseen. \textbf{Maturity Level of the Technology: 10}  &
No need.  \\
\hline
Roman Pots, Off-momentum detectors &
Sensors: AC-LGAD. &
See comments for AC-LGAD in Tab.~\ref{tab:pidB}.  &
Consider alternative technologies as a combination of pixel (MAPS) and timing layers (DC-LGAD).  \\
\hline
ZDC &
Electromagnetic component: W-powder/SciFi; hadronic component: Pb/scintillator with imaging layers by Pb/Si. &
Established technologies. A cost issue can arise if the need for a longer hadronic calorimeter is demonstrated. \textbf{Maturity Level of the Technology: 10} &
Pb/SciFi is considered as alternative to Pb/Si.\\ \hline
%
%
\hline
\multicolumn{4}{|c|}{\textbf{Far-Backward Detectors}} \\ \hline
%
%
EMCal  &
Radiation-hard scintillating fibers, with quartz fibers to SiPM sensors &
Established technologies. \textbf{Maturity Level of the Technology: 10}  & 
No need. \\ \hline
Hodoscopes  &
SciFi with SiPM sensors.&
Established technologies. \textbf{Maturity Level of the Technology: 10}  &
No need. \\ \hline

\end{tabular}
\makeatother
\end{sidewaystable}

\FloatBarrier

\subsection{Upgrade Path}
\label{sec:upgrade_path}
Two 
upgrade paths are considered 
by ATHENA, both strictly related to ongoing R\&D activities (Sec.~\ref{sec:rd_needs}): 
(i) improvements of the baseline detector described in this proposal by alternative techniques offering performance or cost advantages that can become mature in the near future, and
(ii) upgrades that can be implemented after a few years of data taking, informed by the experience gained with the initial detector configuration.

\subsection*{Potential Upgrades to the Baseline Detector}
\begin{description}[leftmargin=0cm]
\item[$\boldsymbol{\mu\mathrm{RWell}}$:]{ The use of \gls{murwell} \glspl{mpgd} instead of \glspl{gem} for gaseous detectors in the endcap regions of the central detectors is an option that can provide easier detector construction, lower material budget, and potentially save around 25\% in material cost. \gls{gem} technology, including large-size chambers, is well established, e.g., the upgrade of the CMS muon system. Since \gls{murwell} technology is a more recent development, it has not yet been adopted in any experiment. 
This upgrade option during the ATHENA design phase requires finalizing the R\&D of this novel technology.
}

\item[Photosensors:]{A major challenge are the photosensors for the Cherenkov-imaging  
\gls{pid} devices. 
The baseline design assumes commercial \glspl{mcppmt} for the \gls{hpdirc}, with their associated issue of cost. \glspl{sipm} operated at low temperature are foreseen for the \gls{drich} and the \gls{pfrich}.  So far, no experiments deploy RICH detector readout using these photon sensors; therefore, a systematic and detailed R\&D is ongoing to establish them as single photon detectors by mitigating the challenges posed by the high dark count rate, which increases with irradiation damage.
An alternative option offering reduced cost for the \gls{hpdirc} and a fallback solution for the \gls{drich} and \gls{pfrich} are \glspl{lappd}. In spite of  significant progress, pixelized versions of
these detectors are not sufficiently mature, with their field performance needing further improvement.
If the ongoing efforts at universities,
laboratories and industry  converge in the next few years, \glspl{lappd} can become
an attractive  option for the \gls{hpdirc}, reducing the detector cost. 
They also present an upgrade option for the \gls{pfrich} in the backward endcap. 
Equipped with \glspl{lappd}, this detector could provide \gls{tof} information through the photons produced in the sensor window by the charged particles. 
Furthermore, they represent a fallback option for the \gls{drich} sensors.}

\item[Radiator Gases:]{A further challenge is related to the choice of radiator gas in the RICH detectors (\gls{drich} and \gls{pfrich}).
In the baseline design the radiator gasses are fluorocarbons, 
gasses that exhibit extremely high \gls{gwp}. These gasses are increasingly prohibited all across the world. 
Where used, complex and expensive close circulation systems are imposed and increasing procurement issues are expected. 
However, RICH performance is preserved when 
fluorocarbons at atmospheric pressure are replaced with argon pressurized at a few bar.
The challenge is to design a vessel that allows safe high-pressure operation while minimizing its impact on the overall detector material budget.
A solid pressurized vessel constructed with light materials would make  this innovative approach possible. It would enable eco-friendly operation of the detector, while enabling substantial decreases in cost due to a cheaper gas and largely simplified gas system.}

\item[\gls{daq}:]{In the present design of the \gls{daq} system,  \gls{felix} boards are used both as interface between the \gls{feb} and the commodity \gls{daq} computers and 
for data aggregation. The development of a dedicated generic aggregation board consisting of a simplified \gls{felix}-like design can save cost by avoiding radiation hard components, using a less expensive \gls{fpga}, and replacing the PCIx interface with fiber output, while maintaining the data interface, timing interface, and potential trigger capabilities of the board.}
\end{description}

\subsection*{Longer-term Upgrade Options}
Possible upgrades to be introduced after the initial data taking period are briefly mentioned here. 
\begin{description}[leftmargin=0cm]
\item[\gls{gem}-based \acrshort{trd}:]{
Space is available in the ATHENA design to introduce a \gls{gem}-based \gls{trd} in front of the \gls{drich} to provide enhanced $e/\pi$ separation power in the forward direction, if required.
R\&D is ongoing to establish the technology needed to construct a large-scale detector~\cite{BARBOSA2019162356,eRD22}.}
\item[\gls{aclgad} \gls{tof}:]{
A possible increase of the $e/\pi$ separation power and $h$-\gls{pid} capabilities at low momenta in the endcaps, can be achieved by introducing \gls{tof} layers by downstream of the \gls{pfrich} and upstream or downstream of the \gls{drich}, using the same \gls{aclgad} technology adopted for the barrel.}
\item[GridPIX \acrshort{minitpc}:]{
ATHENA has undertaken detailed studies of a GridPIX-based \gls{minitpc} that could be installed in the barrel region at radii between 20--45~cm. GridPIX provides spatial and energy measurements with unprecedented fine-space granularity. This detector would extend the identified particle reach down to 100~MeV/c through dE/dx and provide tracking information.}
\item[\gls{zdc}:]{
The \gls{zdc} hadronic calorimeter could be improved by adding another interaction length (an additional 17 layers), and a tail-catcher (independent readout of the final few layers) to further improve the energy resolution for neutrons at high energies.}
\item[\gls{phcal}:]{A region of the \gls{phcal} close to the beam pipe has a potential upgrade path to replace scintillation tiles with Si sensors.}
\item[Nanowire-based RPs:] {
Superconducting nanowire particle detectors are under development for the EIC (eRD28) and might provide an upgrade for the RP silicon sensor technology. Such detectors will provide excellent position and time resolution, are radiation hard, and can operate in high magnetic fields of up to 5T.}
\end{description}

 \FloatBarrier

\section{EIC Science with ATHENA}
\label{chapter:science}

\subsection{Acceptance and Performance}
The primary motivation in designing ATHENA is maximizing the acceptance and performance so that we can deliver the entire suite of EIC physics goals. In this chapter, we demonstrate ATHENA's acceptance and kinematic reach, reconstruction of physics observables, and the quality of the physics measurements. 

\begin{figure}[htb]
\centering
\includegraphics[width=0.65\textwidth]{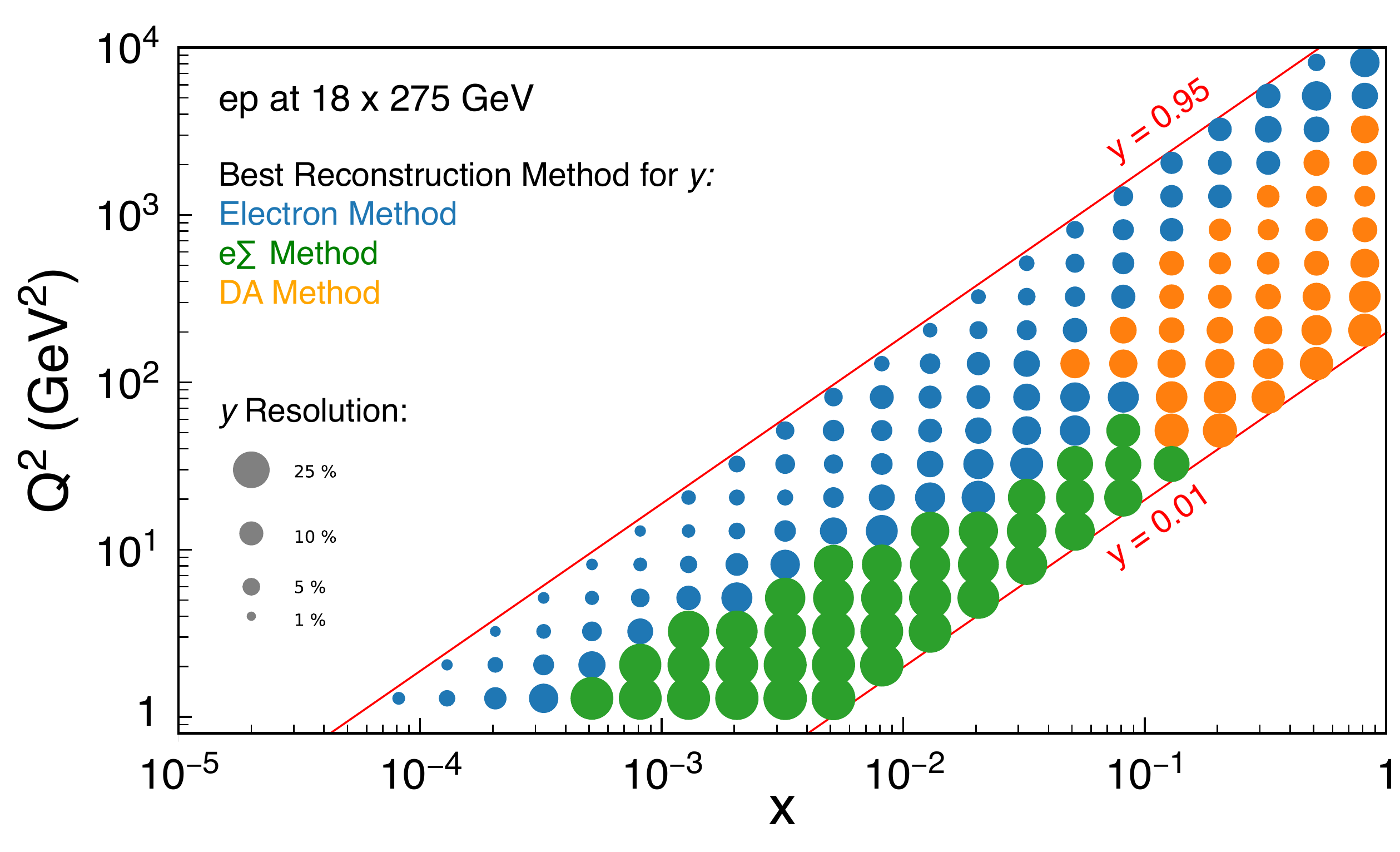}
\caption{Variation of the estimated ATHENA resolution on the kinematic variable, $y$, with $x$ and $Q^2$, for the case of $18 \ {\rm GeV}$ electrons colliding with $275 \ {\rm GeV}$ protons. At each point in the kinematic plane, the best performing reconstruction method is chosen and indicated by the color of the corresponding marker, while the size of the marker indicates the magnitude of the resolution obtained (FullSim).}
\label{fig:kinrec}
\end{figure}

The kinematics of inclusive DIS processes are usually discussed in terms of $x$ and $Q^2$. 
These quantities are reconstructed using either the scattered electron, the inclusive hadronic final state (defined as all remaining final state particles after excluding the electron), or a mixture of the two. 
The kinematic variables $Q^2$, $x$, and $y$ are related by the center-of-mass energy squared $s$: 
$Q^2\approx sxy$. The variable $y$ is closely related to the scattering angle in the lepton-quark center-of-mass frame, and thus also to the direction of the hadronic final state in the laboratory frame. 
Figure~\ref{fig:kinrec} shows
the $y$ resolution throughout the accessible kinematic plane in $x$ and $Q^2$ 
for $18 \ {\rm GeV}$ electrons 
on $275 \ {\rm GeV}$ protons. 
The electron-only method performs best over most of the kinematic phase space and will be used for \gls{nc} DIS measurements at all but the lowest $y$ values. The ATHENA ECal and tracker provide excellent energy and angular resolution for the scattered electron. 
At the lowest $y$ values, the electron method resolution degrades like $1/y$. 
Here, the ability of ATHENA to reconstruct the overall hadronic final state with good resolution can be exploited using the $e-\Sigma$  or \gls{da} methods~\cite{Bassler:1994uq}, to ensure high quality measurements for $y \lesssim 0.1$, leading to high quality reconstruction at 
high $x$.
In \gls{cc} DIS, the only available method is \gls{jb}, which relies entirely on the hadronic final state. Once again, the high quality response of ATHENA to hadrons over a wide range of $\eta$ and $p_T$
is key. The resolution with the \gls{jb} method is at the 20--30\% level throughout the kinematic range. 

\begin{figure}[htb]
\centering
\includegraphics[width=0.65\textwidth]{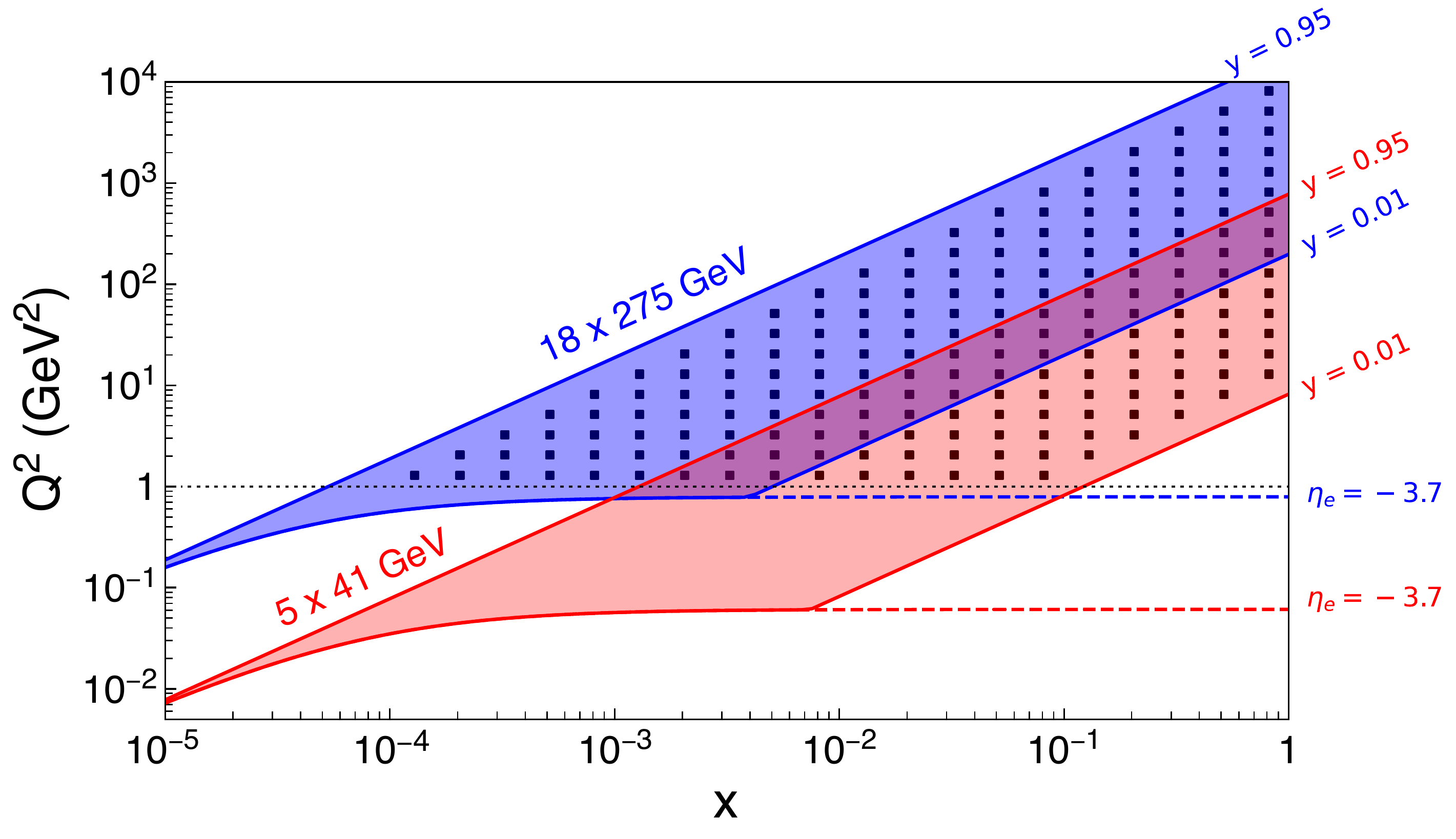}    
\caption{Kinematic coverage of simulated ATHENA data for EIC running at the largest and smallest center-of-mass energies. The positions in the kinematic plane of
simulated measurements in the deep-inelastic region
with selection requirements: $Q^2 > 1 \ {\rm GeV^2}$ and $0.01 < y < 0.95$ (FullSim).}
\label{fig:ncdata}
\end{figure}

Figure~\ref{fig:ncdata} summarizes the kinematic range and binning used in 
simulated ATHENA data for unpolarized \ep\ collisions. The kinematic range is restricted to $Q^2 > 1 \ {\rm GeV^2}$ corresponding to the deep-inelastic regime and a region in which the ATHENA ECal and tracking detectors provide full acceptance across the accessible $x$ range. The requirement $y < 0.95$ is applied to ensure sufficiently large scattered electron energies and clean electron identification conditions (Fig.~\ref{fig:elecpurity}) and a further cut $y > 0.01$ 
is made to ensure that the kinematic variables can be sufficiently well reconstructed (Fig.~\ref{fig:kinrec}). 
The resolution 
allows for five logarithmically spaced bins per decade in
$x$ and $Q^2$. A running time of one year is assumed for each beam energy combination. 
Statistical uncertainties for 
inclusive \ep\ cross sections are negligible at all but the very highest $x$ and $Q^2$ values. These uncertainties, however, become important for the asymmetry measurements 
sensitive to spin dynamics. The systematic uncertainties are taken to be the 
average of the optimistic and pessimistic scenarios 
in the Yellow Report, which are compatible with simulations carried out to date 
for the ATHENA detector. 
HERA experience has shown that point-to-point systematic uncertainties 
vary from 1.5\% to 2.5\% depending on $y$, whilst there are overall normalisation uncertainties for each beam energy pairing at the level of 2.5\%.
The inclusive NC cross section is the fundamental ingredient of all studies of collinear parton densities at EIC, as well as underlying semi-inclusive, exclusive, and hadronic final-state cross-section measurements. 

\begin{figure}[ht!]
  \begin{minipage}[c]{0.65\textwidth}
    \centering
    \includegraphics[width=0.55\textwidth]{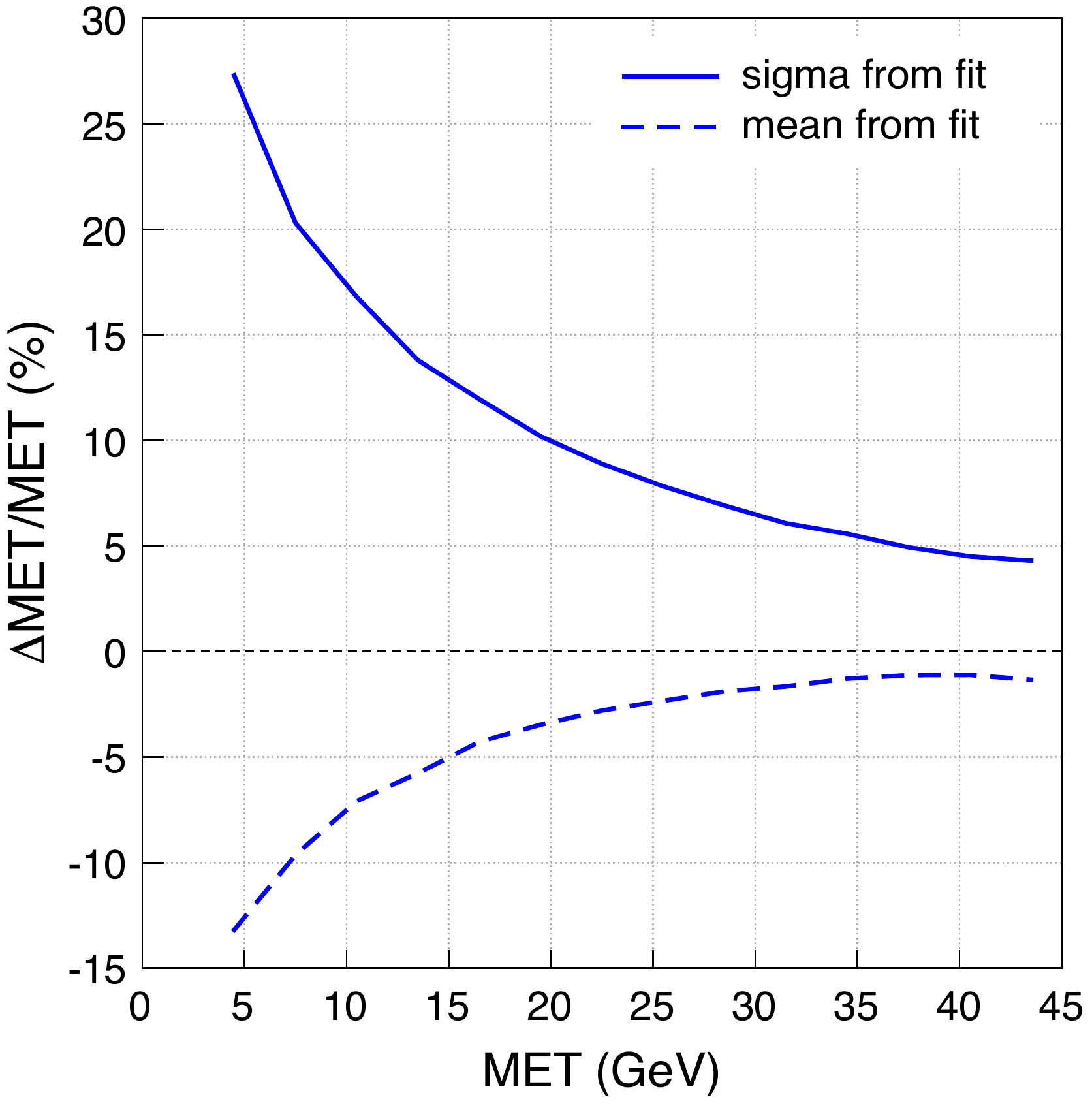}
  \end{minipage}
  \begin{minipage}[c]{0.3\textwidth}
    \caption{MET resolution (solid line) and bias (dashed line) as a function of generated MET. The reconstructed MET is defined as the magnitude of the vector sum of the transverse momentum of all energy-flow objects (FastSim).}
    \label{fig:METperformance}
  \end{minipage}
\end{figure}

The \gls{met} is defined as the magnitude of the vector sum of the momentum of all final-state particles in the event. It is reconstructed using energy-flow candidates. This quantity is used in the hadronic reconstruction methods in CC DIS (for example in the \gls{jb} method: $Q^{2}$ = $\mathrm{MET}^{2}/(1-y)$). Unbiased MET reconstruction with good resolution enables CC DIS measurements with  $Q^{2} >$ 100 GeV$^{2}$.  Good missing energy performance is also required for kinematic reconstruction in NC DIS at low $y$.
Figure~\ref{fig:METperformance} shows the MET resolution and bias. At MET = 10 GeV, the resolution is 15$\%$ and bias less than 10$\%$.
The hermeticity of ATHENA for all particles (both charged and neutral hadrons, photons, and leptons) will be crucial for measurements of low-MET events. 

\subsubsection{Electron identification}
\label{preamble}

The acceptance and performance of the ATHENA backward  endcap and barrel calorimeters leads to near 100\% reconstruction of electrons in most of the  deep-inelastic domain. The backward beamline instrumentation continues the coverage, albeit with lower acceptance, into the photoproduction domain.

The kinematic range over which precision measurements can be made in NC DIS
also depends on the ability to cleanly identify scattered electrons.
Assuming a selection based on 
a calorimeter electromagnetic cluster associated with a charged particle track, the dominant source of misidentification is 
from residual $\pi^-$. 
Suppression factors have been estimated  to vary 
from $10^2$ to  $10^4$ depending on the electron energy and pseudorapidity. The $\pi^-$ rejection factors have been
convoluted with the predicted yields of electrons and pions in a PYTHIA6 
simulation of NC DIS to estimate  misidentification rates. 
Isolation and coplanarity requirements have been applied to the scattered electron candidates. 
The summed $E-p_z$ of all detected final state particles  must be compatible with the expected
value of twice the electron beam energy. Together, these requirements provide 
another order of magnitude misidentification suppression. 
Event kinematics and topology together with the PID subsystems can be used to further reduce the misidentification.

\begin{figure}[!htb]
\centering
\includegraphics[width=0.75\textwidth]{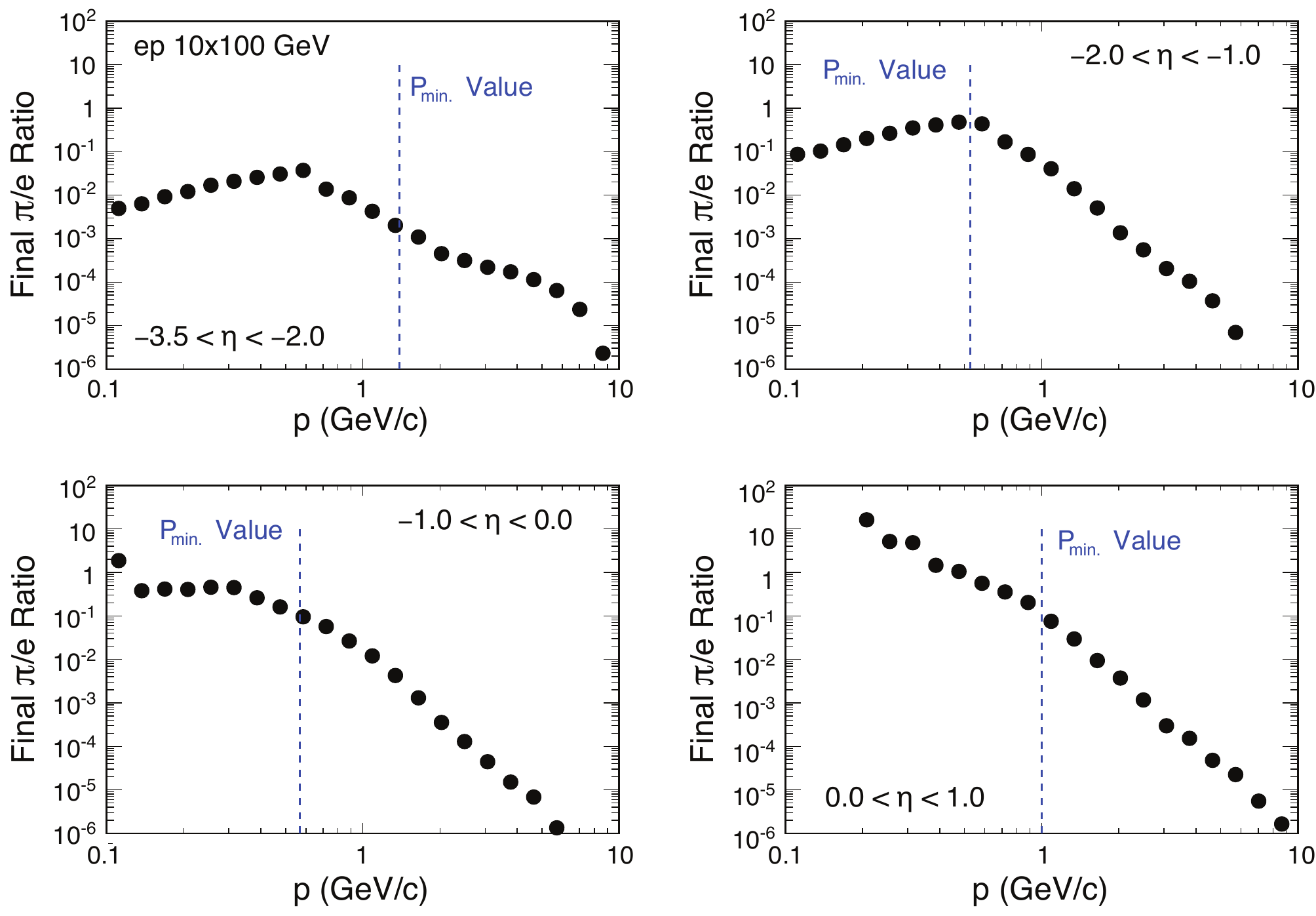}
\caption{Estimated $\pi^- / e$ ratio in NC DIS as a function of 
momentum and pseudorapidity for collisions between 10 GeV electrons and 100 GeV protons.  The pions are rejected using the basic PID performance of
the calorimeters, in combination with event-level requirements on the total $E-p_z$ and electron isolation. The vertical lines indicate the minimum momenta allowed by the requirements $Q^2 > 1 \ \mathrm{GeV}^2$ (important in the most
backward $\eta$ range) and $y < 0.95$ (important in all the other $\eta$ ranges).
\label{fig:elecpurity}
}
\end{figure}

Figure~\ref{fig:elecpurity} shows the results of the $\pi^-$ background studies as a function of 
the scattered electron momentum
for four different ranges in $\eta$ (PID subsystem information is not used). The contaminations are generally largest at
low momenta. The $1/Q^4$ factor in the cross section implies that the signal electron distribution is strongly peaked
towards the backward direction, whereas the pion spectrum is relatively
 flat in $\eta$, such that the misidentification
fractions are largest for the most central (highest $Q^2$) electrons. The estimated $\pi^-$ contamination is at or below
the 10\% level throughout the accessible kinematic range. This is because the minimum scattered electron
momenta that are allowed by a typical kinematic requirement 
(e.g., $y < 0.95$)  grow with $Q^2$.



\subsubsection{Lepton pair invariant-mass resolution}
 The resolution achieved in the reconstructed invariant mass of the lepton pairs is essential to distinguish the quarkonia excited states. For the $J/\psi$, one needs to separate the $J/\psi(1S)$ and $\psi^{\prime}$ states and for the $\Upsilon$, the three excited states $\Upsilon(1S,2S,3S)$ need to be resolved. 
 The different quarkonia and their excited states provide complementary probes of the nucleon, because of their differences in size.   

\begin{figure}[htb]
  \begin{minipage}[c]{0.65\textwidth}
    \centering
    \includegraphics[width=0.77\textwidth]{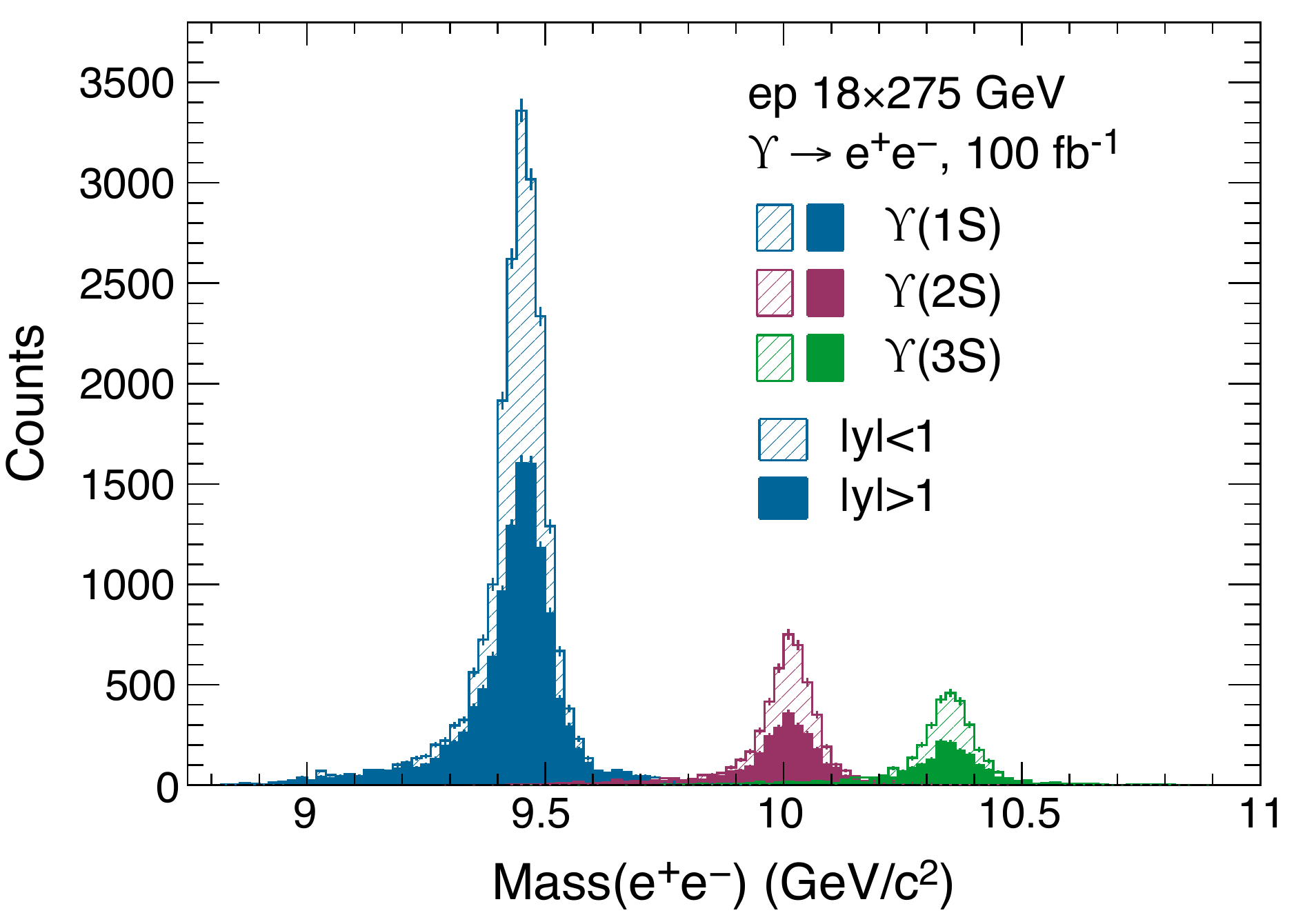}
  \end{minipage}
  \begin{minipage}[c]{0.3\textwidth}
    \caption{\label{fig:masspeak} 
    The simulated $M_{ee}$ mass spectrum for exclusive production of the three $\Upsilon$ states in collisions of 18 GeV electrons with 275 GeV protons with 100 fb$^{-1}$ of integrated luminosity, using the cross sections from eSTARlight \cite{Lomnitz:2018juf}. Spectra are shown for $\Upsilon$-production at mid-rapidity ($|y|<1$) and away from mid-rapidity ($|y|>1)$ (FullSim).
    }
  \end{minipage}
\end{figure}

Figure \ref{fig:masspeak} shows the expected ATHENA $e^+e^-$ invariant-mass spectrum, with the three $\Upsilon$ states visible.  In this simulation the relative cross sections were determined by the mass difference and their couplings to the $e^+e^-$ final state \cite{Klein:2016yzr}.  The three peaks are well-separated, and, owing to the low mass in the ATHENA tracker and beampipe, only a small low-mass shoulder from bremsstrahlung is visible.

\subsubsection{Hadron and jet reconstruction}

\begin{figure}[ht]
    \centering
  \includegraphics[width=0.8\textwidth]{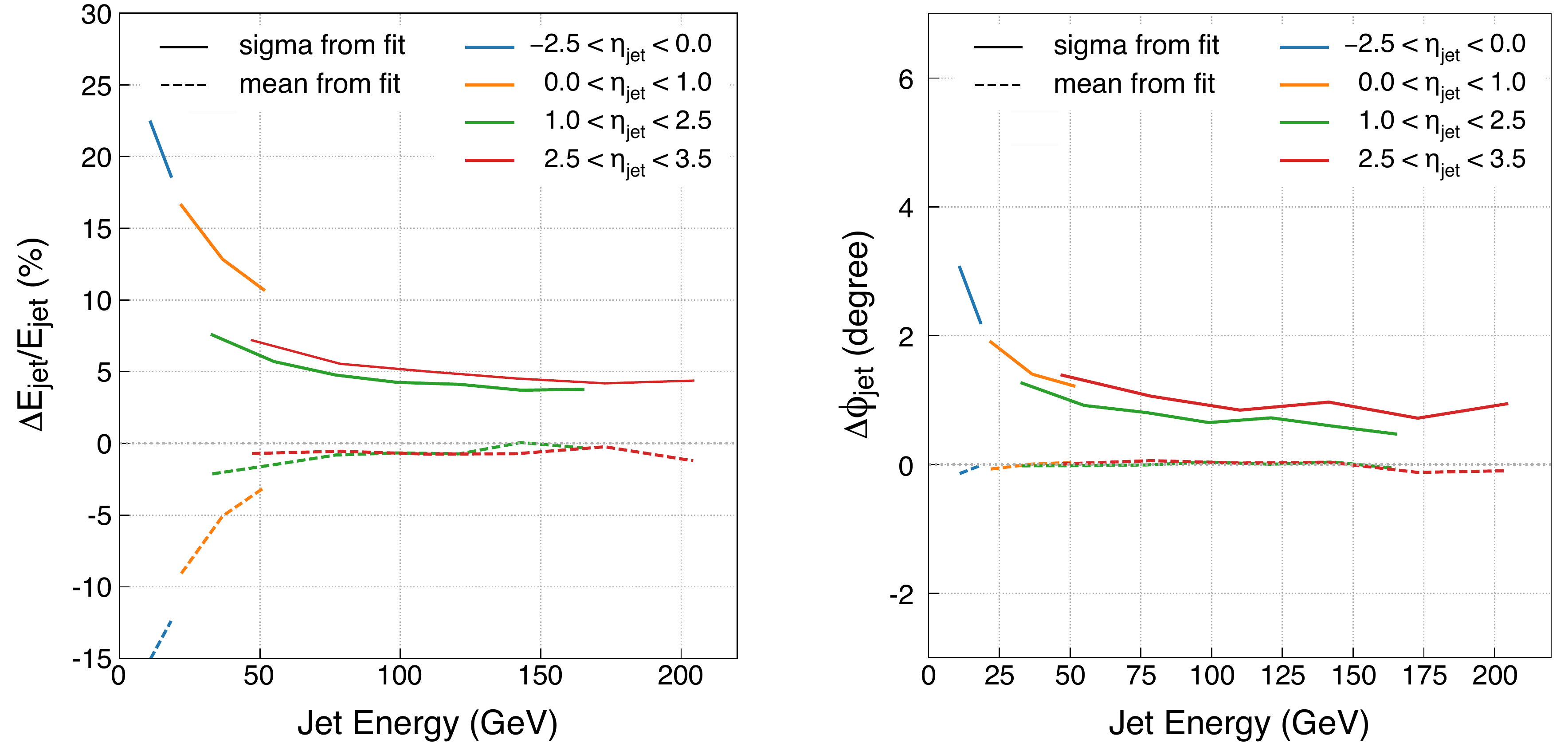} 
 \caption{Left: Relative jet energy resolution (solid lines) and jet energy scale (dashed lines) for various pseudorapidity intervals.  Right: jet azimuthal angular  resolution (solid lines) and bias (dashed lines) for various pseudorapidity intervals (FastSim).}
    \label{fig:JESJER}
\end{figure}

ATHENA's calorimetry and charged particle tracking enable precision jet measurements.
In the following figures, resolution and bias are defined as RMS and mean of the difference between reconstructed and generated values, obtained with a Gaussian fit. 
Figure~\ref{fig:JESJER} shows the resolution and bias for the jet energy (left panel) and jet azimuthal angle (right panel) as a function of $E^{\mathrm{jet}}$ in various $\eta^{\mathrm{jet}}$ intervals. Jets are reconstructed with the anti-$k_{T}$ algorithm and $R_\mathrm{jet}=1.0$, with energy-flow reconstruction. The jet energy resolution is better than 10$\%$ for $E^{\mathrm{jet}}>40$ GeV, and the jet azimuthal angular resolution is better than 1 degree for jet energies above 25~GeV.
The jet energy scale is within a few percent of unity over a wide range, except
where biases originate from thresholds and acceptance effects. 

\subsubsection{Displaced track and vertex performance}
Momentum and angular resolutions, needed in resonance reconstruction are described in Sec.~\ref{vertex-tracking}.
Track pointing and vertex resolutions enter in heavy-quark analyses, which require geometrically displaced secondary decay vertices of hadrons containing a charm or beauty quark. 
The performance depends on reconstructed displacements for individual tracks along, as well as orthogonal to, the beam directions, and the combination of single tracks into vertices.  The resolutions of the latter depend on the number of tracks and their topology.

\begin{figure}[ht!]
    \centering
     \includegraphics[width=0.9\textwidth]{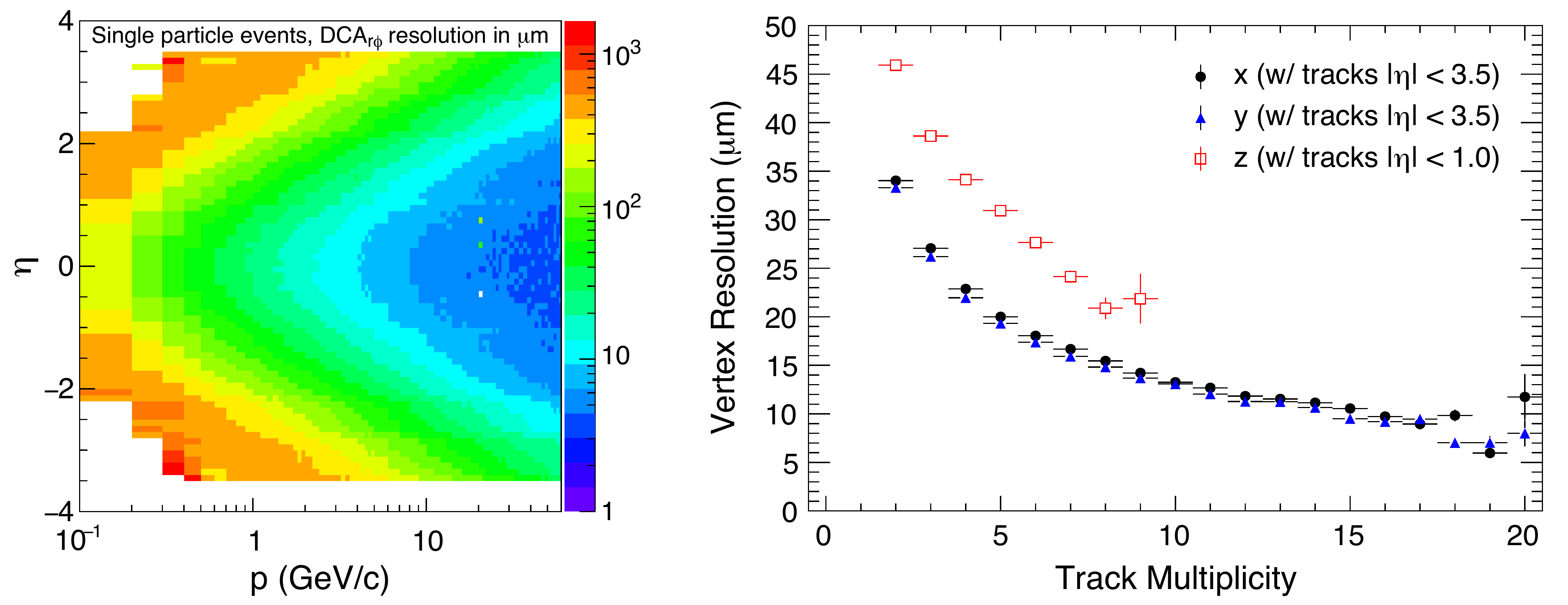} 
   \caption{Left: Single charged track distance-of-closest-approach resolution orthogonal to the beam axis (FullSim). 
    Right: Primary vertex resolution as a function of charged track multiplicity along and orthogonal to the beam axes for DIS events within the indicated detector acceptance (FastSim).}
    \label{fig:DCA}
\end{figure}

High statistics heavy-quark projections from fast simulations have been benchmarked against full GEANT simulations. The resulting finely-binned resolutions 
quantify ATHENA's measurement capabilities.
Resolutions from full simulations are shown in Fig.~\ref{fig:DCA}, left.
Charged tracks from PYTHIA have been propagated through this response and the results 
combined in weighted averages to assess the  collision vertex reconstruction performance.
Figure~\ref{fig:DCA}, right shows the primary vertex resolution as a function of the number of tracks within the indicated ATHENA acceptance.
Charged decay-prongs originating from hadrons containing charm or beauty quarks were combined in a similar way to obtain the secondary vertices.

The displaced-track resolutions allow good charm-jet tagging based on a displaced track-counting algorithm, which yields a charm efficiency that ranges from 10 to 30\% from 5 GeV/c to 30 GeV/c. The mis-tagging rate on light jets similarly varies with $p_T$ but is always below 1\%.

 \FloatBarrier

\subsection{Origin of Spin and 3-D Nucleon Imaging}
\subsubsection{Spin structure of the nucleon via polarised inclusive DIS}
\label{spin}

Understanding the spin of the proton is one of the central pillars of the EIC physics program.
Historically this question has been approached through the helicity-dependent collinear quark and gluon
distributions in the proton, following the spin sum rule:
\begin{equation}
\frac{1}{2}=
\frac{1}{2}\Delta\Sigma+\Delta G+ L_{Q}+L_{G},
\end{equation}
where $\Delta\Sigma$, $\Delta G$ and $L_{Q}$, $L_{G}$ are the contributions 
from the quark/anti-quark helicity, the gluon helicity, and
 their  angular momenta, respectively.
Data from fixed-target polarized lepton DIS experiments and polarized proton-proton experiments that provide
$\Delta\Sigma$, $\Delta G$ have jointly probed the helicity distributions in the range $0.005<x<0.6$.
EIC precision measurements, with a kinematic coverage down to $x \sim 10^{-4}$, will lead to a unprecedented knowledge of nucleon spin structure 
and a benchmark for lattice QCD calculations. 

\begin{figure}[ht]
\centering
    \includegraphics[width=\textwidth]{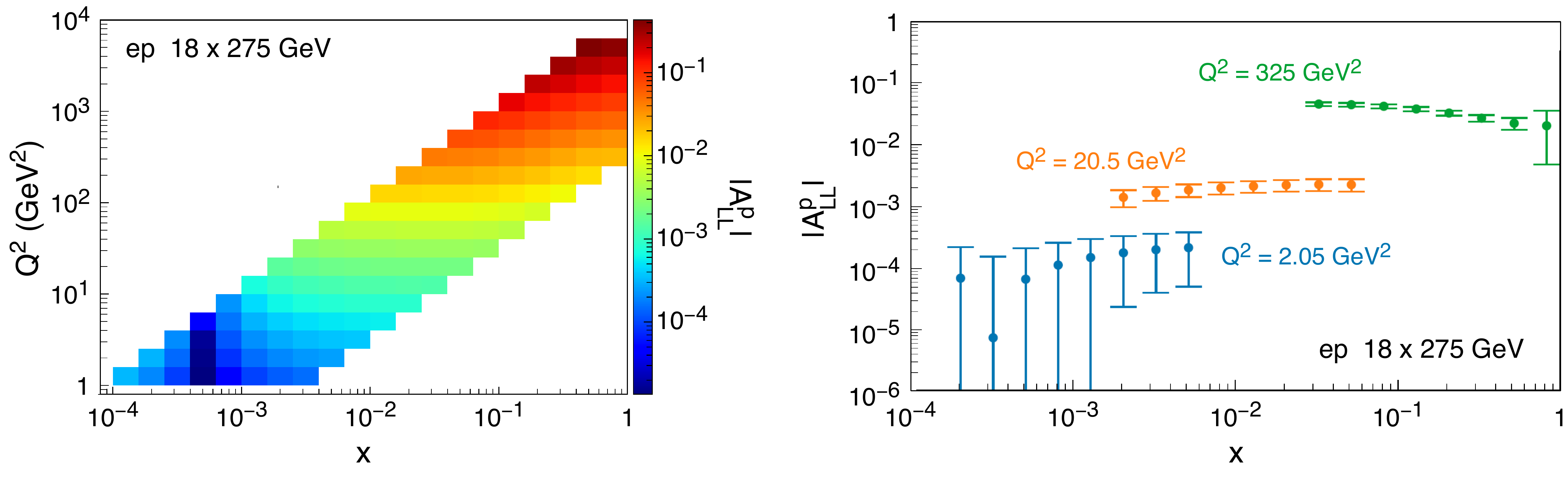}
    \caption{Left: Predicted DIS inclusive double-spin asymmetry $A_{LL}$ for 18 GeV electrons colliding with 275 GeV protons based on fits from the JAM  collaboration, in $x$ and $Q^2$ intervals corresponding to the simulated ATHENA acceptance.
    Right: Projected uncertainties (statistical and systematic) at three representative $Q^2$ values for an integrated luminosity of 10 fb$^{-1}$, where this observable is limited by the systematic uncertainties at low $Q^2$ (FastSim).
}
    \label{fig:spinerror}
\end{figure}

The basic ingredient of these studies is the double spin asymmetry $A_{LL}$, which can be measured in inclusive NC DIS data (Fig.~\ref{fig:ncdata}). 
Figure~\ref{fig:spinerror} (left) shows the absolute size of the DIS inclusive asymmetry $A_{LL}$ based on fits from the JAM collaboration~\cite{Ethier:2017zbq} evaluated over the wide $x$ and $Q^2$ ATHENA acceptance for 18 GeV electrons colliding with 275 GeV protons.
Figure~\ref{fig:spinerror} (right) compares the projected $A_{LL}$ uncertainties with the absolute size of the asymmetry at three representative $Q^2$ values for an integrated luminosity of 10 fb$^{-1}$,
where this observable becomes limited by the systematic uncertainties at low $Q^2$.

\begin{figure}[htb]
    \includegraphics[width=\textwidth]{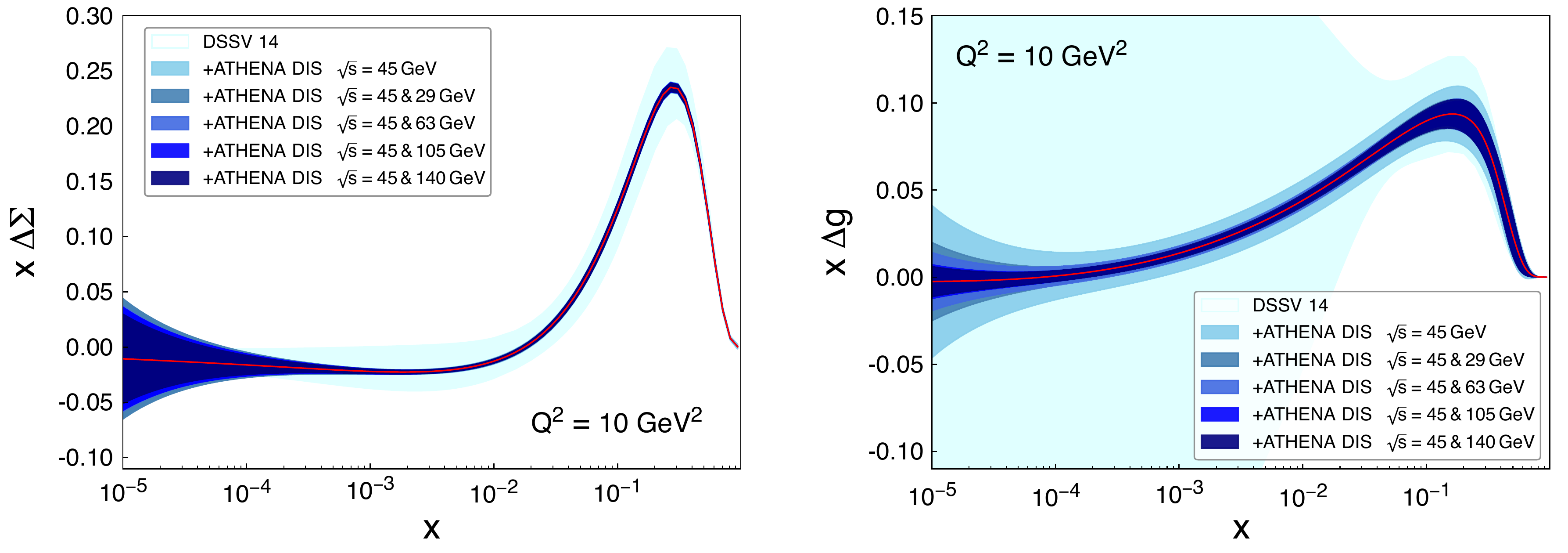}
  \caption{
  Impact of DIS inclusive $A^p_{LL}$ pseudodata from ATHENA on the  understanding of the proton spin, as expressed through helicity distributions at $Q^2 = 10 \ \mathrm{GeV^2}$ in the DSSV14 fitting framework (FastSim).
  Left: Singlet quark helicity  distribution. Right: gluon helicity distribution. 
  The outermost bands correspond to the uncertainties in DSSV14. The inner bands show the results of 
  including
  simulated ATHENA data at different center-of-mass energy combinations, as indicated.}
  \label{fig:spinplots}
\end{figure}

Figure~\ref{fig:spinplots} shows the impact of ATHENA inclusive pseudodata constraining $\Delta\Sigma$ and $\Delta G$ through a new global fit by the DSSV collaboration~\cite{Borsa:2020lsz,DeFlorian:2019xxt}. 
As indicated in the figure, the uncertainty on the gluon helicity is significantly reduced in the small-$x$ region. In Ref. \cite{PhysRevD.86.054020} the impact of systematic uncertainties on the determinations of the helicity PDFs has been investigated. A 1.5\% and a very conservative 5\% uncorrelated systematic uncertainty has been integrated in the fit. Sources of fully correlated systematic uncertainties, such as measurements of the beam polarizations, which are likely to dominate uncertainties at an EIC, only lead to a scale uncertainty in spin asymmetries but do not change the significance of the measurement. An uncorrelated systematic uncertainty of $<$ 5\% has a tolerable impact.

\subsubsection*{Spin structure of the nucleon via polarised semi-inclusive DIS}

The ATHENA particle identification subsystems enable unique capabilities to delineated the quark and anti-quark contributions to $\Delta\Sigma$ by flavor.
The sensitivity to the flavor of the struck parton in SIDIS requires the measurement of different identified hadron species in electron collision with various polarized light ion beams. 

\begin{figure}[bth]
    \centering
    \includegraphics[width=\textwidth]{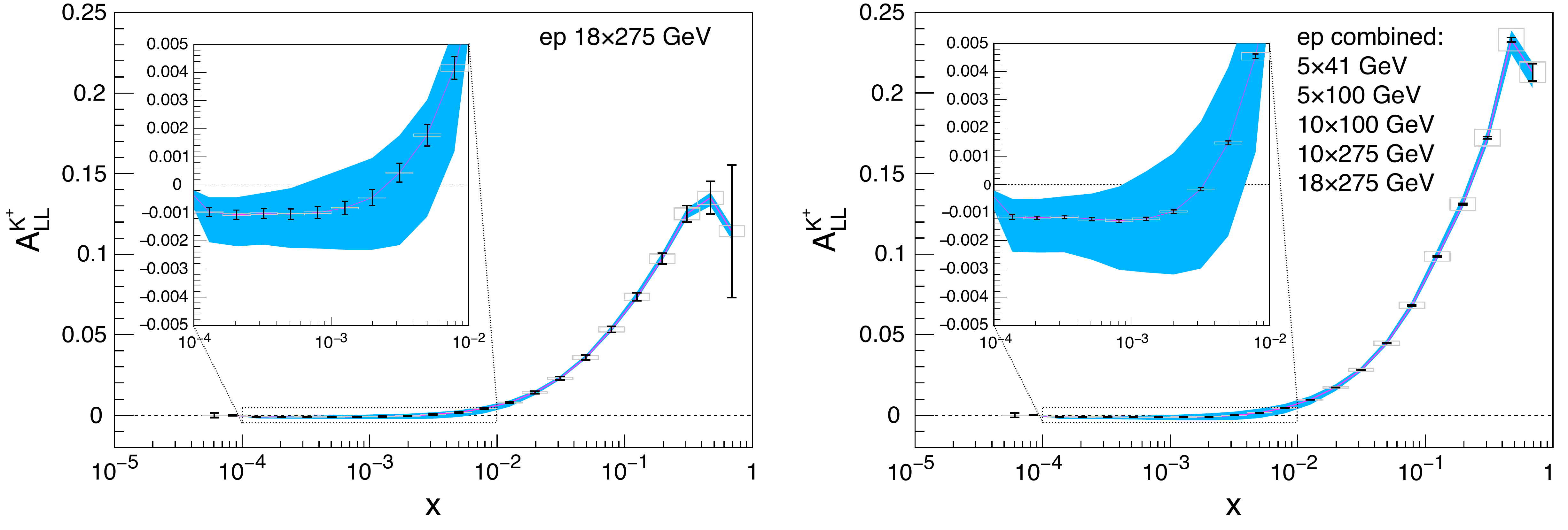}
  \caption{Projected statistical and systematic uncertainties for $A_{LL}$ for charged kaons. The curves are based on DSSV helicity PDFs~\cite{DeFlorian:2019xxt} combined with DSS FFs~\cite{deFlorian:2017lwf}. The bands reflect only the helicity PDF uncertainties. Left: The expected results for 15.5~fb$^{-1}$ of 18 GeV electrons colliding with 275 GeV protons. Right: Results combining pseudodata from 5 GeV electrons colliding with 41 and 100 GeV protons, 10 GeV electrons colliding with 100 GeV and 275 GeV protons, and 18 GeV electrons colliding with 275 GeV protons. The uncertainties of the individual datasets were scaled to the dataset with 18 GeV electrons colliding with 275 GeV protons assuming equal data taking time for each center-of-mass-energy. Pseudodata with $Q^2>1\,\mathrm{GeV}^2, z >0.2, y>0.01$ and a depolarization factor $D(y)>0.1$ were selected. Purities and efficiencies of the PID detectors have been taken into account. Systematic uncertainties were estimated to be 3\% point-to-point and 2\% scale uncertainty based on previous experience and the simulated PID performance of ATHENA. These results show a significant impact of the projected dataset, in particular, on the sea-quark helicity distributions and a constraining power over the entire $x$ range (FastSim). 
  \label{fig:ALL}}
\end{figure}

Figure~\ref{fig:ALL} shows $A_{LL}^{K^\pm}$ and its uncertainties at different $\sqrt{s}$ compared to current uncertainties from helicity PDFs~\cite{DeFlorian:2019xxt}. The pseudodata uncertainties account for the purities and efficiencies of the ATHENA PID detectors, which have been determined to be generally above 90\% and above 80\%, respectively. The possibility to measure at different $\sqrt{s}$ allows a better determination of sea-quark helicities down to $\sim 10^{-4}$ and up to $\sim 1$ in $x$ in a wide $Q^2$-range.
In particular, these measurements will clarify whether the sea-quark polarizations, especially for strange quarks, are non-vanishing in that limit. 

\begin{figure}[htb]
  \begin{minipage}[c]{0.65\textwidth}
    \includegraphics[width=0.9\textwidth]{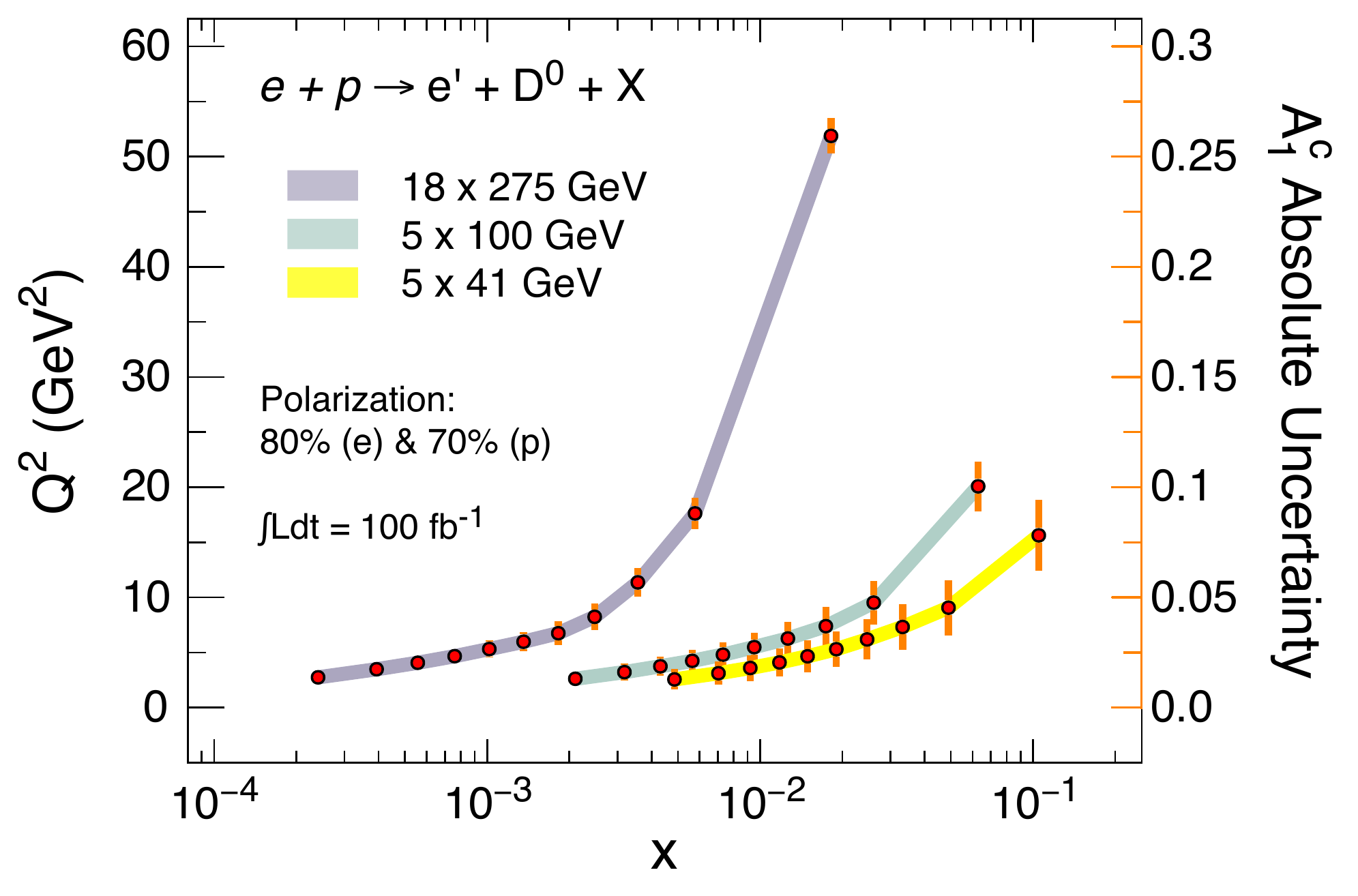}
  \end{minipage}
  \begin{minipage}[c]{0.3\textwidth}
    \caption {Projections for the spin asymmetry $A_1^c \sim g_1^c / F_1^c$ in $e + p \rightarrow e' + D^0 + X$ for an integrated luminosity of 100\,fb$^{-1}$ and longitudinal polarizations of 80\% (70\%) for the electron (proton) beams.  The points are shown at the $x$ and $Q^2$ values of the measurement, indicated by the horizontal scale and the vertical scale on the left, respectively.  The bars indicate the size of the statistical uncertainties and follow the vertical scale on the right. The bands denote the different energy combinations (FastSim).}
    \label{fig:delta-G}
  \end{minipage}
\end{figure}

Double spin asymmetries in charm production enable accessing the gluon polarization in a complementary way to the scaling violation of the inclusive structure function $g_1$.
In the EIC kinematics, 10--15\% of the inclusive DIS cross section will be from the production of charm-quark pairs. These pairs probe the shape of the gluon density 
of the nucleon at large $x$, at an effective scale determined by the charm mass. Theoretical uncertainties due to higher order corrections have been studied and HERA \ep\ data 
show good agreement with QCD expectations (see, e.g., Ref.~\cite{H1:2012xnw}).
Several impact studies of the EICs measurement have been performed to-date~\cite{Chudakov:2016ytj, Aschenauer:2017oxs, Kelsey:2021gpk,Anderle:2021hpa}.
Excellent  displaced vertex resolution is essential in achieving a large signal-to-background ratio in these measurements.

Pioneering measurements in the photo-production regime have been performed by the COMPASS collaboration at CERN~\cite{COMPASS:2012mpe}.
More recently, the formalism for \gls{NLO} perturbative QCD for such measurements in the DIS regime has become available~\cite{Hekhorn:2018ywm} and an impact study has been performed for future measurements at the EIC~\cite{Anderle:2021hpa}.
Figure~\ref{fig:delta-G} shows the projected uncertainties for the spin asymmetry $A_1^c \sim g_1^c / F_1^c$ in polarized $e + p \rightarrow e' + D^0 + X$ using the ATHENA baseline response for topological reconstruction of the $D^0$ meson through the $D^0 \rightarrow K + \pi$ decay channel.

Alternative techniques could be by invariant-mass reconstruction of three-prong $D^{*+}$ decay, single displaced $K$ tag, or by means of the significance of the signed displacement of charged particle tracks. These have not been explicitly studied for this proposal.

\subsubsection{3-D parton imaging with hadrons}
Because of confinement, partons can have momenta in the transverse plane perpendicular to their parent hadron momenta. The best studied TMD is the $f_1^q (x, k_T)$ for an unpolarized quark, carrying longitudinal, $x$, and transverse, $k_T$, momenta. At small $k_T \sim \Lambda_{QCD}$, $f_1^q$ is a non-perturbative function that can only be determined by fitting to experimental data. High precision data with a large lever arm in $Q^2$ are needed over a wide range in $x$ to determine the non-pertubative part of the evolution.

\begin{figure}[ht!]
  \includegraphics[width=\textwidth]{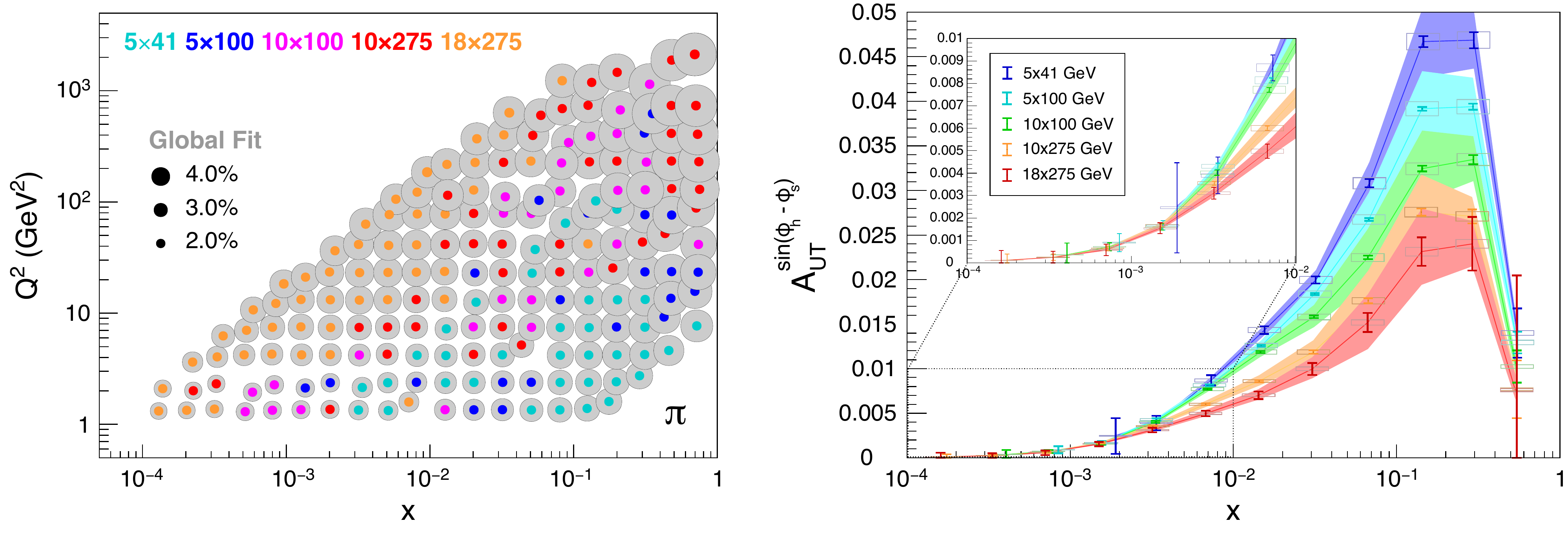}
\caption{Left: Projected uncertainties for the unpolarized cross sections measured with ATHENA compared with uncertainties of the PV17 TMD PDF extraction~\cite{Bacchetta:2017gcc} (grey dots). Over a significant part of the phase space the total uncertainty on the ATHENA pseudodata is dominated by the assumed 2\%  systematic point-to-point and 3\% scale uncertainty, whereas the theory uncertainties are dominated by the poorly known TMD 
evolution. The different colors represent the different energy combinations. The size of the markers show the uncertainties of the corresponding datasets. At each point the data and theory relative uncertainties are shown where the data has the relative highest impact.
Right: Projected Sivers asymmetries extracted from ATHENA pseudodata compared to projections from the Pavia extraction~\cite{Bacchetta:2020gko} for charged pions. Pseudodata with $Q^2>1.0\,\mathrm{GeV}^2$, $0.2 < z <0.7$, $y>0.05$, $q_T/Q<1.0$ were selected. The ATHENA data will be powerful in constraining the shape of this TMD as well as its evolution. The uncertainties of the individual datasets were scaled to 100 fb$^{-1}$ of 10 GeV electron collisions with 275 GeV protons assuming equal data taking time for each center-of-mass-energy. Based on previous experience, we assumed 2\% point-to-point uncertainty and 1.5\% scale uncertainty (FastSim).}
\label{fig:unpolTMD}
\end{figure}

At leading twist, there are eight TMDs covering all combinations of parton and hadron polarizations. The most prominent polarized TMD is the Sivers function, $f_{1T}^{\perp\, q}$, for an unpolarized parton $q$ in a transversely polarized hadron. Current extractions of the Sivers function are based on data 
with large uncertainties and covering a small part of the phase space.
Consequently, the $k_T$ and $x$ dependences of the sea-quarks are poorly constrained.
The non-perturbative contribution to the evolution of the Sivers function at low $k_T$ is driven by the same poorly known non-pertubative evolution as $f_1^q$. 
Figure~\ref{fig:unpolTMD} shows the uncertainties for the unpolarized cross sections (left) projected for ATHENA compared with uncertainties of the PV17 TMD PDF extraction~\cite{Bacchetta:2017gcc}. Over a significant part of the $Q^2$-$x$ coverage at different center-of-mass-energies the ATHENA total uncertainties are dominated by the assumed systematic 2\% point-to-point and 3\% scale uncertainty. 
The right plot shows the projected Sivers asymmetries extracted from ATHENA pseudodata compared to projections from the Pavia extraction~\cite{Bacchetta:2020gko} for charged pions. 
Data with $Q^2>1.0$, $0.2 < z <0.7$, $y>0.01$, $q_T/Q<1.0$ were selected. The ATHENA data will be powerful in constraining the shape of this TMD as well as its evolution as is clearly seen from the vanishing ATHENA uncertainties compared to the current theory uncertainties. 


\subsubsection{3-D parton imaging with heavy flavor and jets}

Jets are excellent proxies for partons, especially in the clean environment of DIS, so they are powerful probes for TMDs in a nucleon. For example, measurements of electron-jet pairs in DIS probe quark TMD PDFs without convolution with TMD fragmentation functions~\cite{Liu:2018trl,Gutierrez-Reyes:2018qez}. Similarly, dijets in DIS probe gluon TMD PDFs and offer a very promising way to constrain the magnitude of the gluon Sivers function over a wide kinematic range~\cite{Zheng:2018ssm}.
In diffractive DIS, dijet-proton correlations probe the Wigner function~\cite{Hatta:2016dxp}---the ultimate goal for nucleon imaging studies. While measurements of TMDs were highlighted in the NAS study and EIC White Paper, the potential of their measurements with jets had not been fully explored by the time the EIC White Paper was written. 

ATHENA's precision, acceptance, and particle ID performance enable high quality measurements of jets and their substructures.
The novel barrel ECal also helps hadronic final state measurements.
Combined with the excellent tracking resolution enabled by the 3T magnetic field, ATHENA will make precise energy flow and hadron-in-jet measurements.
The key requirement to probe TMDs with jets is the resolution to probe small values of lepton-jet (dijet) momentum imbalance ($q_{T}$); which is driven by jet energy and angle resolution. To access the TMD regime, the $q_{\mathrm{T}}$ value should be small relative to the total jet (dijet) $p_{T}$ and $Q$ of the event.

\begin{figure}[ht!]
    \includegraphics[width=\textwidth]{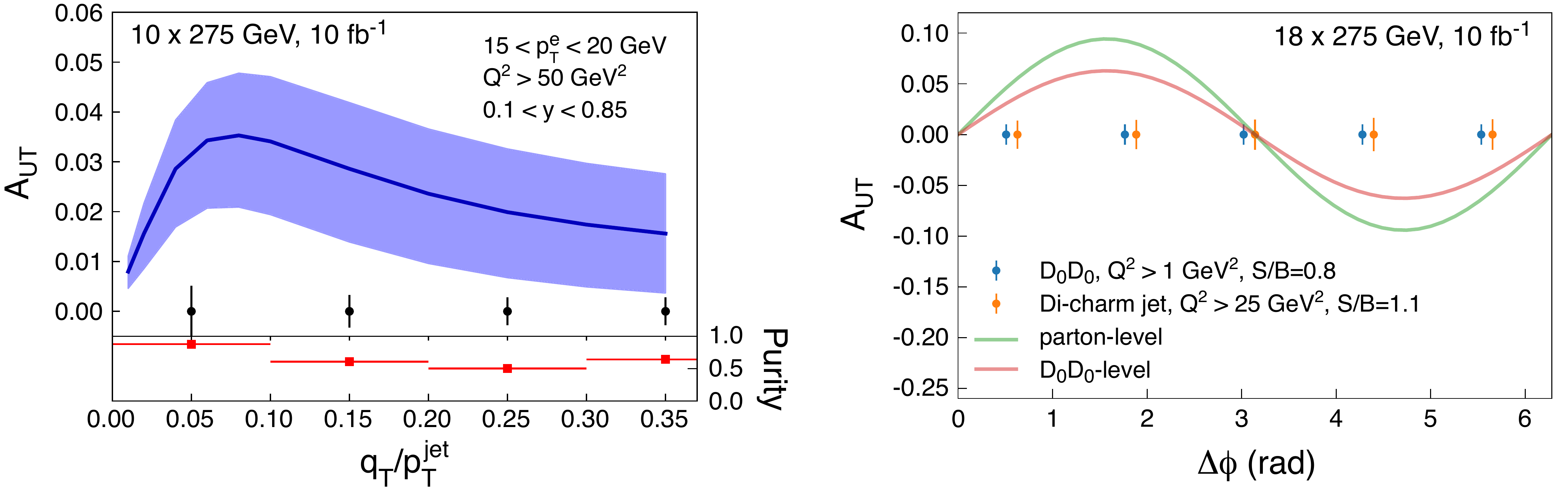}
   \caption{Left: Sensitivity for lepton-jet Sivers asymmetry (FastSim). Right: Sensitivity for di-charm Sivers asymmetry. These are representative examples of measurements probing (sea) quark TMDs and gluon TMDs, respectively (FastSim).}
   \label{fig:leptonjetSivers}
\end{figure}

Figure~\ref{fig:leptonjetSivers} shows the projected performance for the lepton-jet Sivers asymmetry (left), and di-charm Sivers asymmetry (right) measured with D$^0$ and charm jet pairs. These
illustrate ATHENA measurements with sensitivity to (anti-) quark and gluon TMD PDFs, respectively. 
The purity is defined as the ratio of the number of events where the reconstructed and generated values of $q_T/p_T^\mathrm{jet}$ are in the same bin to the number of all events reconstructed in that bin.
The purity is found to be more than 50\% for the bin widths shown, ensuring reasonable corrections for the asymmetries and unfolding for the unpolarized \ep\ baseline.
The theory predictions and uncertainties are taken from \cite{Arratia:2020nxw}.

Fragmentation function, or hadron-in-jet, measurements generalize SIDIS by providing an additional axis (the jet axis as well as the $\gamma^{*}$ axis), which can be used to probe TMD PDFs and TMD FFs in an independent and controlled way \cite{Arratia:2020nxw}.
For example, by fixing the $p_{T}$ of the jet with respect to the photon axis while varying the $p_{T}$ of the hadrons with respect to the jet axis, decouples those effects and offers great flexibility to constrain TMDs and their evolution \cite{Kang:2021ffh}.
ATHENA's precision calorimetry, tracking, and PID capabilities enable good energy flow reconstruction and are crucial for these measurements.

\begin{figure}[ht!]
  \begin{minipage}[c]{0.65\textwidth}
    \centering
\includegraphics[width=0.7\textwidth]{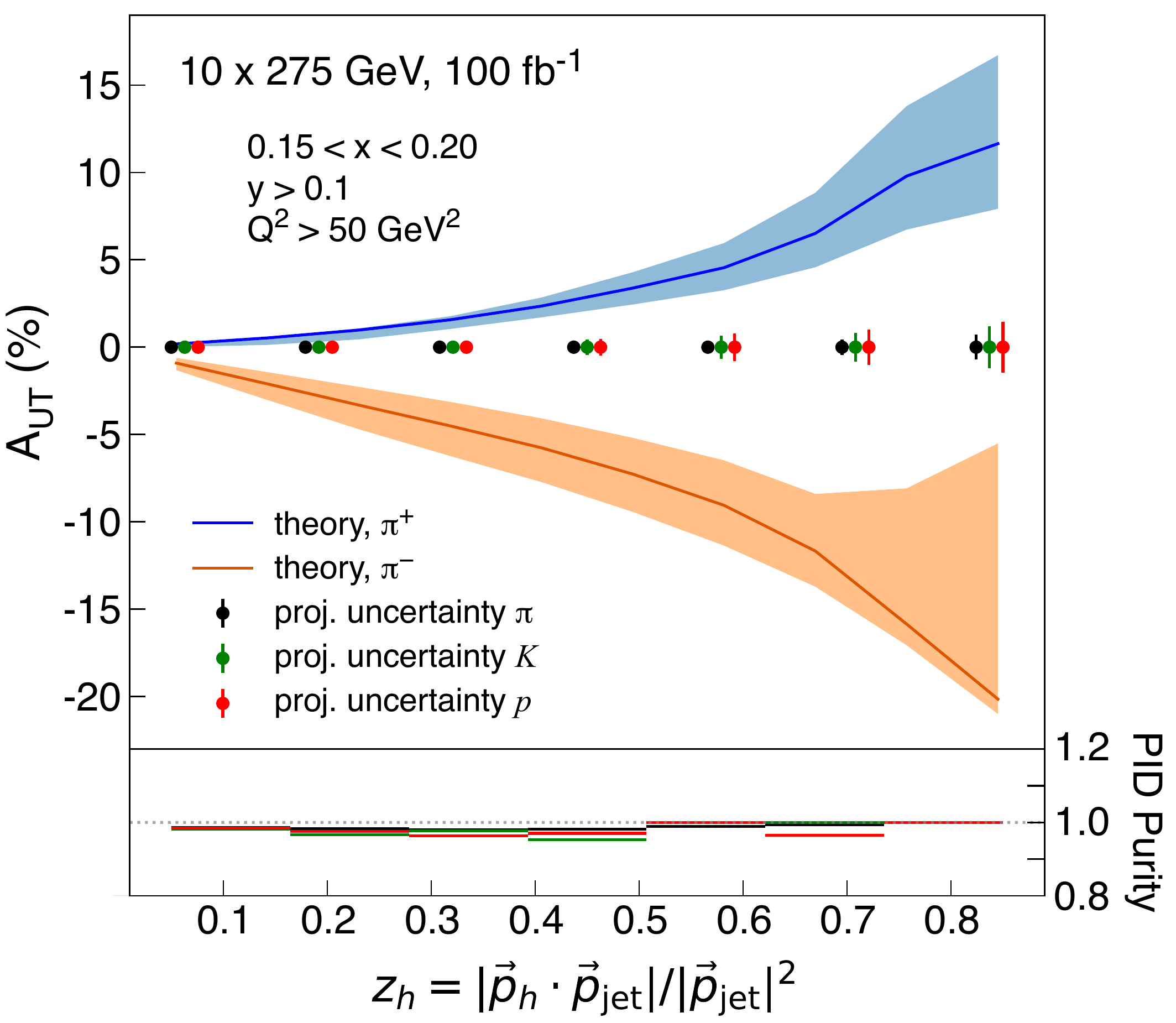}
  \end{minipage}
  \begin{minipage}[c]{0.3\textwidth}
    \caption{Projection for hadron-in-jet Collins asymmetry measurement for charged pions, kaons and protons. This is representative of the class of jet substructure measurements (FastSim).}
    \label{hadron-in-jet}
  \end{minipage}
\end{figure}

Performance requirements for fragmentation functions are similar to those for jet correlations accessing TMDs. 
However, fragmentation function measurements also require excellent PID over the entire $x$, $z$, $Q^{2}$ phase space. 
Figure~\ref{hadron-in-jet} shows the projected precision for hadron-in-jet Collins asymmetry measurements, which probe quark transversity, TMD fragmentation functions, and TMD evolution. This measurement is representative of an entire class of possible jet substructure measurements.   

\subsubsection{Transverse spatial imaging of quarks and gluons} 

Deeply virtual Compton scattering (DVCS, $\gamma^* N \rightarrow N' \gamma$) and timelike Compton scattering (TCS, $\gamma N \rightarrow N' \gamma^* \rightarrow N' \ell^+ \ell^-$) are among the most discussed exclusive reactions, allowing  the extraction of Generalized Parton Distribution (GPD) functions.
ATHENA is designed especially to reconstruct the entire final state with superior precision.
In DVCS, a virtual photon is exchanged in the scattering with its virtuality well in the perturbative regime ($Q^2 > 1$ GeV$^2$) and a high-energy real photon is emitted.
Conversely, in TCS, a real photon absorbed by a quark causes the emission of a virtual photon, which decays into a lepton pair.

Experimentally, the main distinction between the spacelike and timelike regimes is whether the scattered electron is detected, and whether it is the real photon or the lepton pair which is detected at mid-rapidity.

Exclusive reactions challenge detector designs in multiple ways, and acceptance is key.  Acceptance in $x$ can be mapped to acceptance in final state rapidity.
Resolution and PID for $\pi^0$, $\gamma$, and leptons are critical to eliminate the background from other final states and good forward acceptance for nucleons and photons is also crucial.
  
\begin{figure}[ht]
    \centering
    \begin{minipage}[t] {0.45\textwidth}
     \includegraphics[width=\textwidth]{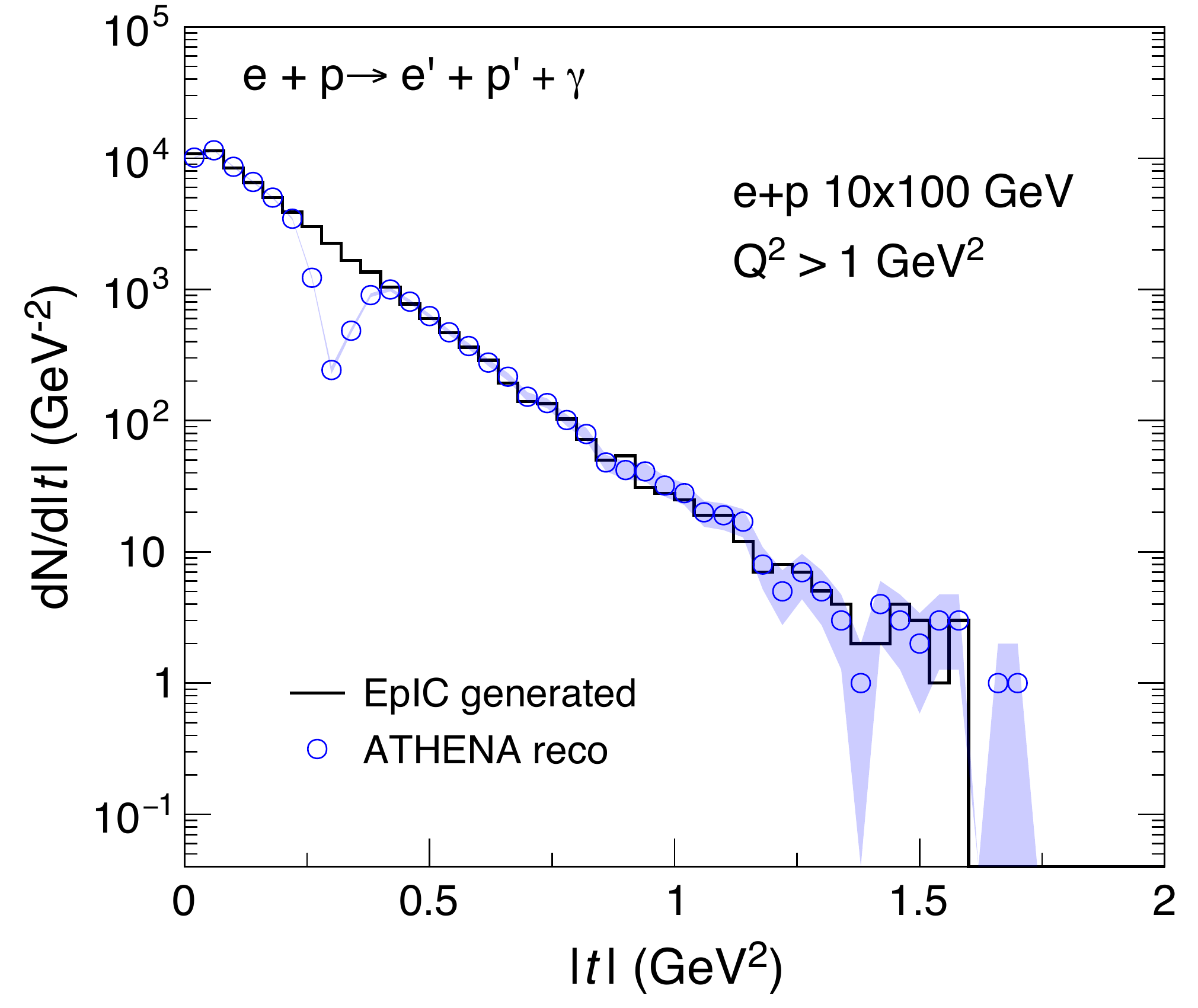}
     \caption{The $|t|$ distribution of DVCS events in \ep\ collisions from the E$p$IC Monte Carlo Generator (black line) for 10 GeV electron and 100 GeV proton beam energies, is compared with the reconstructed distribution (blue circles) (FullSim).}
     \label{fig:DVCS-t}
    \end{minipage} \hfill
    \begin{minipage}[t] {0.45\textwidth}
     \includegraphics[width=\textwidth]{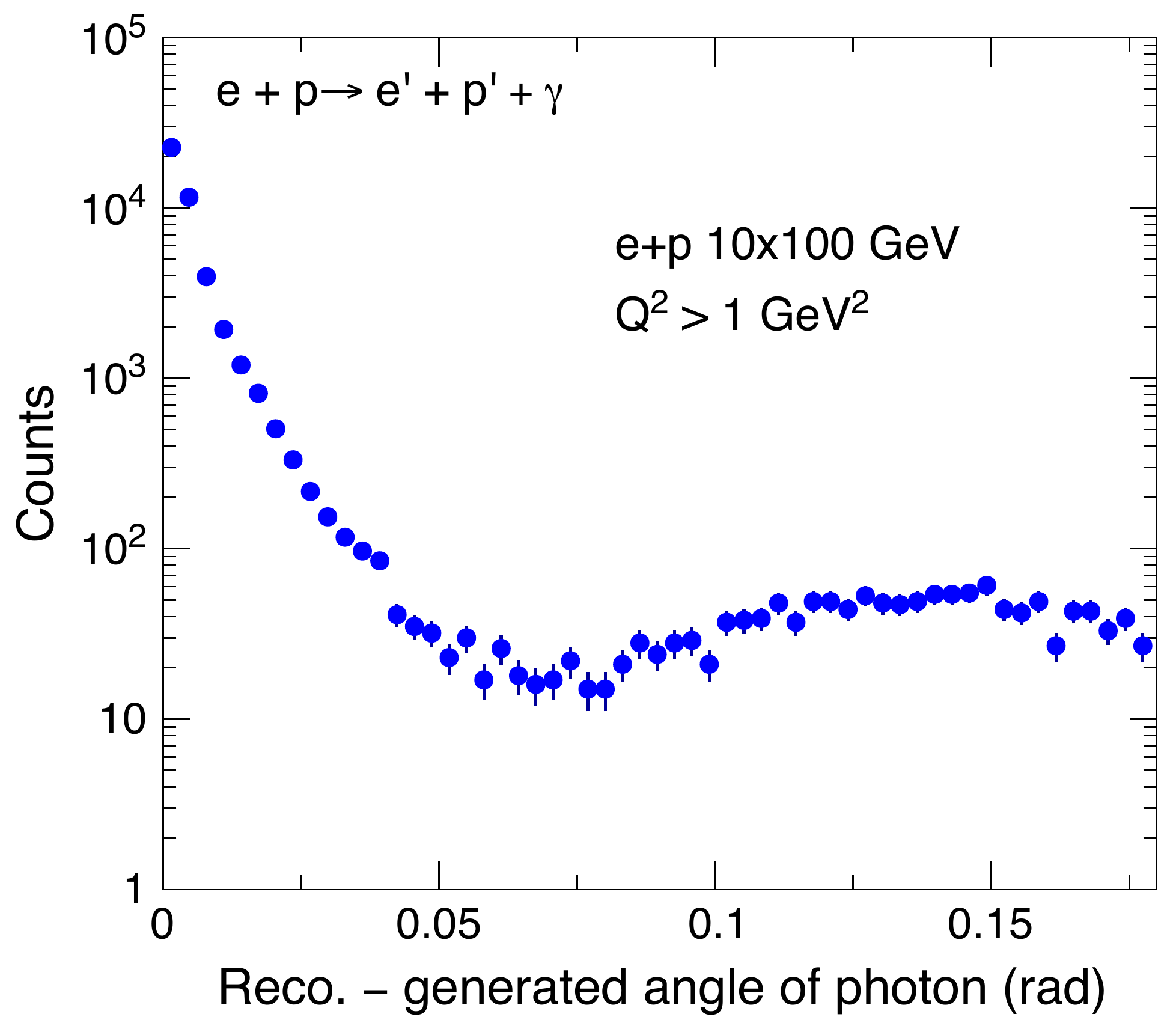}
     \caption{Difference between the generated and reconstructed angle of the produced real photon in a DVCS event simulated in \ep\ collisions for the 10 GeV electron and 100 GeV proton beam energies (FullSim).}
     \label{fig:DVCS-GammaAngReso}
    \end{minipage}
\end{figure}

DVCS and TCS give access to chiral-even GPDs, which are important for the extraction of information on nucleon tomography and the \gls{emt}.
DVCS and TCS have complementary sensitivity to the different Compton form factors, which parametrize the cross sections and asymmetries and are a stepping stone towards determining GPDs.
Therefore, a good capability in measuring both DVCS and TCS by an EIC experiment will be crucial for the partonic imaging program.   
The EIC Yellow Report Sec.~8.4.1-4~\cite{AbdulKhalek:2021gbh} identify the acceptance requirements for the forward-scattered coherent proton in a DVCS or TCS event.
An accurate $|t|$ measurement in a very wide range ($|t_{min}|\sim 0.03$~GeV$^2$ up to $|t|\sim 1.6$~GeV$^2$) is required for a good extraction of the impact parameter distributions via Fourier transform, without affecting the precision of the extracted partonic densities.
In ATHENA, such protons will be measured by the far-forward detectors (RPs and B0 tracker).
The reconstruction of $|t|$ in a DVCS event with the far-forward proton spectrometer is shown in Fig.~\ref{fig:DVCS-t}.
For intermediate ion beam energies, a small gap in $|t|$ is caused by a small acceptance mismatch between the B0 spectrometer and the RPs.
This gap is intrinsic to the IR design and does not significantly affect the quality of extracting the $|t|$-slope.
The acceptance and momentum resolution of the far-forward detectors allow for an accurate $|t|$ reconstruction 
(Fig.~\ref{fig:DVCS-t}), which translates into the capability of pinpointing different $|t|$-slopes and discriminate among the theoretical models, some of which assume an exponential dependence~\cite{Goloskokov:2005sd,Goloskokov:2007nt,Goloskokov:2009ia} and others a dipole-like dependence~\cite{Kumericki:2007sa, Kumericki:2009uq, Kumericki:2013br}.
Figure~\ref{fig:DVCS-GammaAngReso} shows the event-by-event difference between the generated and reconstructed photon angle, which is directly related to the angular resolution of the ECal.
For most DVCS events, the difference between the generated and reconstructed photon angle is well below 17 mrad, the minimum angular divergence between two decay photons of a $\pi^0$ (EIC Yellow Report~\cite{AbdulKhalek:2021gbh}, Sec. 8.4.1).
This will ensure strong suppression of the $\pi^0$ background to the DVCS process.

\begin{figure}[ht]
  \centering
  \begin{minipage}[c] {0.55\textwidth}
    \includegraphics[width=0.95\textwidth]{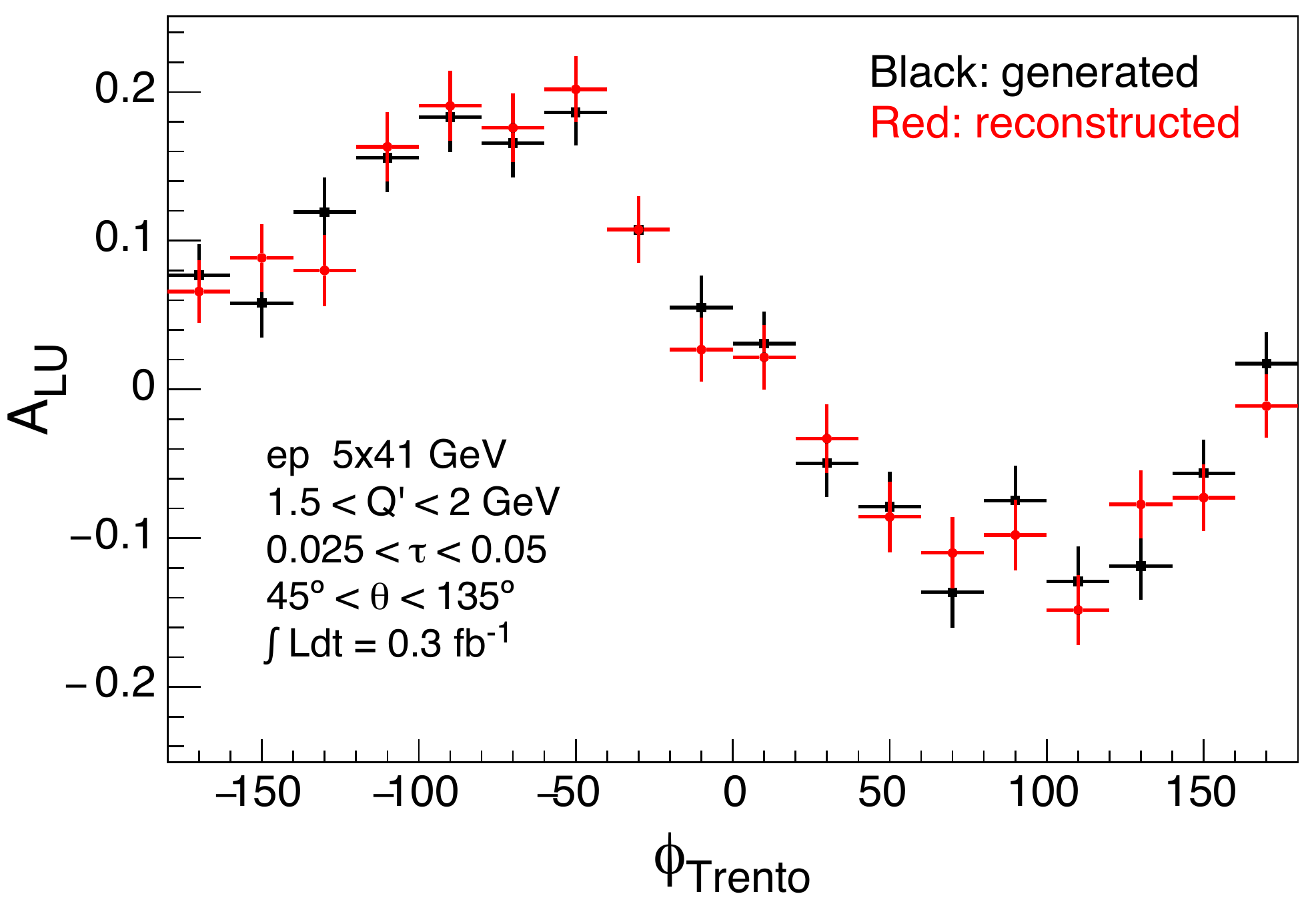}
  \end{minipage}
  \begin{minipage}[c] {0.4\textwidth}
  \caption{Beam-spin asymmetry $A_{LU}$ extracted from generated (black squares) and reconstructed (red circles) data as a function of $\phi_\mathrm{Trento}$ in polarized \ep\ collisions of 5 GeV electrons and 41 GeV protons with statistical uncertainties corresponding to $\sim$~0.3 fb$^{-1}$. The data show the kinematic bin of $1.5 < Q^{\prime} < 2$~GeV, 0.025 $< \tau <$ 0.05 and $45^o < \theta < 135^o$, where $Q^{\prime}$ is the invariant mass of the lepton pair, which provides the hard scale in the process, $\tau = Q^{\prime 2}/(s - m_p^2)$ is the equivalent of $x$ for TCS and $\theta$ is the angle between a produced lepton and the scattered proton (FullSim).}
  \label{fig:TCS-bsa}
  \end{minipage}
\end{figure}

At EIC kinematics, the TCS cross section is suppressed by two orders of magnitude compared to the experimentally indistinguishable \gls{bh} process. 
Spin asymmetries, sensitive to the interference between the BH and TCS amplitudes, recover the sensitivity to GPDs.
ATHENA detection capabilities are excellent for the extraction of \gls{bsa} in the TCS process, as can be seen from Fig.~\ref{fig:TCS-bsa}, which shows the generated and reconstructed BSA plotted as a function of Trento $\phi$. Agreement is excellent, within the statistical uncertainties. The TCS proton detection capabilities of ATHENA are identical to those for DVCS. 
The $e^+e^-$ pair from the decay of the virtual photon in TCS is produced at central rapidity, for which ATHENA has near perfect acceptance and efficiency. 
The exclusivity of TCS will be ensured in cases where the scattered electron is outside of the acceptance of the ATHENA central detector through cuts on the missing mass distribution. 
The detection of the scattered electron in the far-backward subsystems will further refine the exclusivity of the sample and suppress backgrounds in this part of the phase space.  

 \FloatBarrier
    
\subsection{Origin of Mass}
\subsubsection{Gravitational gluonic form factors through DVMP on nucleons}

The origin of the proton mass can be traced back to QCD~\cite{Shifman:1978by} through the breaking of scale invariance due to quantum effects, giving rise to a non-zero trace of the \gls{emt} known as the trace anomaly. Since the proton's constituents are either massless (gluons) or near-massless (quarks),  one needs to account for the motion of quarks and gluons. However, they still comes short to account for the proton's total mass. A remaining contribution related to the fundamental origin of mass is needed, namely a fraction of the trace anomaly, an essential ingredient to understand the origin of mass~\cite{Ji:1995sv,Ji:2021pys,Ji:2021qgo,Zahed:2021fxk}. Other interpretations of the mass decomposition have been put forward~\cite{Lorce:2018egm,Metz:2020vxd}. One important goal of the ATHENA detector is to enable measurements to probe the trace anomaly and determine the proton mass radius in the photo- and electro-production of heavy quarkonia near-threshold~\cite{Hatta:2018ina,Hatta:2018sqd,Kharzeev:2021qkd,Mamo:2019mka,Guo:2021ibg,Sun:2021gmi,Sun:2021pyw,Boussarie:2020vmu}. A number of other physics topics can also be accessed, including color correlations~\cite{Brodsky:2000zc}, $\Upsilon-p$ scattering lengths ~\cite{Gryniuk:2020mlh}, and, by studying the $Q^2$ dependence of exclusive production, and saturation~\cite{Mantysaari:2017slo}.
Short of performing elastic scattering  using a beam of $J/\psi$s and $\Upsilon$s to determine the gravitational gluonic form factors, the next best option is a multi-prong approach through \gls{dvmp} and DVCS. 
We demonstrate ATHENA's performance for 3-D gluon transverse spatial density profiles using $J/\psi$-DVMP, 
along with the $|t|$ dependence of $\Upsilon$ production near threshold.  

\begin{figure}[h]
    \includegraphics[width=\textwidth]{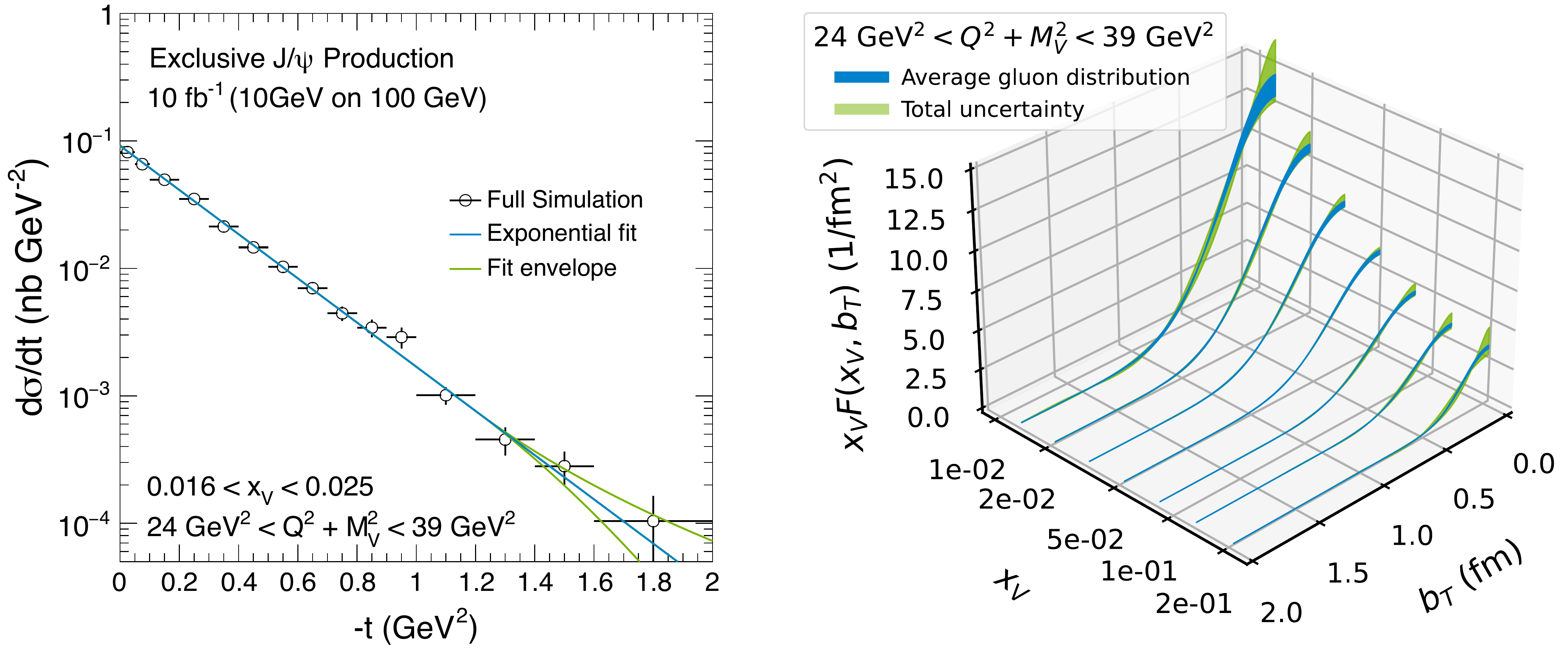}
  \caption{\label{fig:gluon-gpd-profile} Left: 
  Exclusive $J/\psi$ differential production cross section in the $e^+e^-$ decay channel for 0.016 $<x_V<$ 0.025 and 24 GeV$^2$ $< Q^2+M^2_V <$ 39 GeV$^2$.
  The blue central curve is an exponential fit to the pseudodata, while the green outer curves show the extremes of various extrapolation scenarios outside of the measured range (FullSim). 
  Right: The corresponding $b_T$ and $x_V$ dependence of the extracted gluon transverse profiles, multiplied with the gluon PDF from CT14~\cite{Dulat:2015mca}, for the same bin in $Q^2+M^2_V$. The blue band shows the statistical uncertainty, while the green band shows the total uncertainty, including the extrapolation uncertainty of the Fourier transform (FullSim). }
\end{figure}
\subsubsection{3-D gluon spatial imaging / GPDs via \texorpdfstring{J/$\psi$}{J/psi} and \texorpdfstring{$\Upsilon$}{Upsilon}}

Photo- and electro-production of quarkonia are important probes of nuclear structure, sensitive to gluon densities and 
transverse positions in nucleons and nuclei (GPDs) \cite{Accardi:2012qut,AbdulKhalek:2021gbh}. 
A key goal is to determine transverse gluon density profiles in the proton at different $x$. This is achieved by measuring, at large $\sqrt{s}$ and large $Q^2$, the elastic electro-production differential cross section $d\sigma/dt$  of $J/\psi$ and $\Upsilon$.  
A further method to extract the same gluon density profile is through DVCS measurements over a wide range of $Q^2$. 
Figure~\ref{fig:gluon-gpd-profile} (left) shows the acceptance-corrected result of such a measurement for $J/\psi$ electro-production in a single bin of $x_V= (Q^2 + M_V^2)/(2P{\cdot} q)$\footnote{This is the DVMP equivalent of Bjorken-x.} and $Q^2 + M_V^2$. This can be related to an average gluon impact parameter distribution through a Fourier transform of the $t$-dependence, assuming various scenarios to extrapolate $t$ outside of the measured region. The ATHENA detector has excellent acceptance for the vector meson decay leptons and the scattered electron over the full kinematic range of interest. For this measurement, good $t$-acceptance  in the far-forward detectors is crucial to minimize the systematic uncertainty related to this extrapolation. Figure~\ref{fig:gluon-gpd-profile} (right) shows the results for this Fourier transform for a slice of $Q^2 + M_V^2$ as a function of $x_V$ and the impact parameter $b_T$.

\subsubsection{Near-threshold \texorpdfstring{$\Upsilon$}{Upsilon} production}

\begin{figure}
\centering
\includegraphics[width=0.9\textwidth]{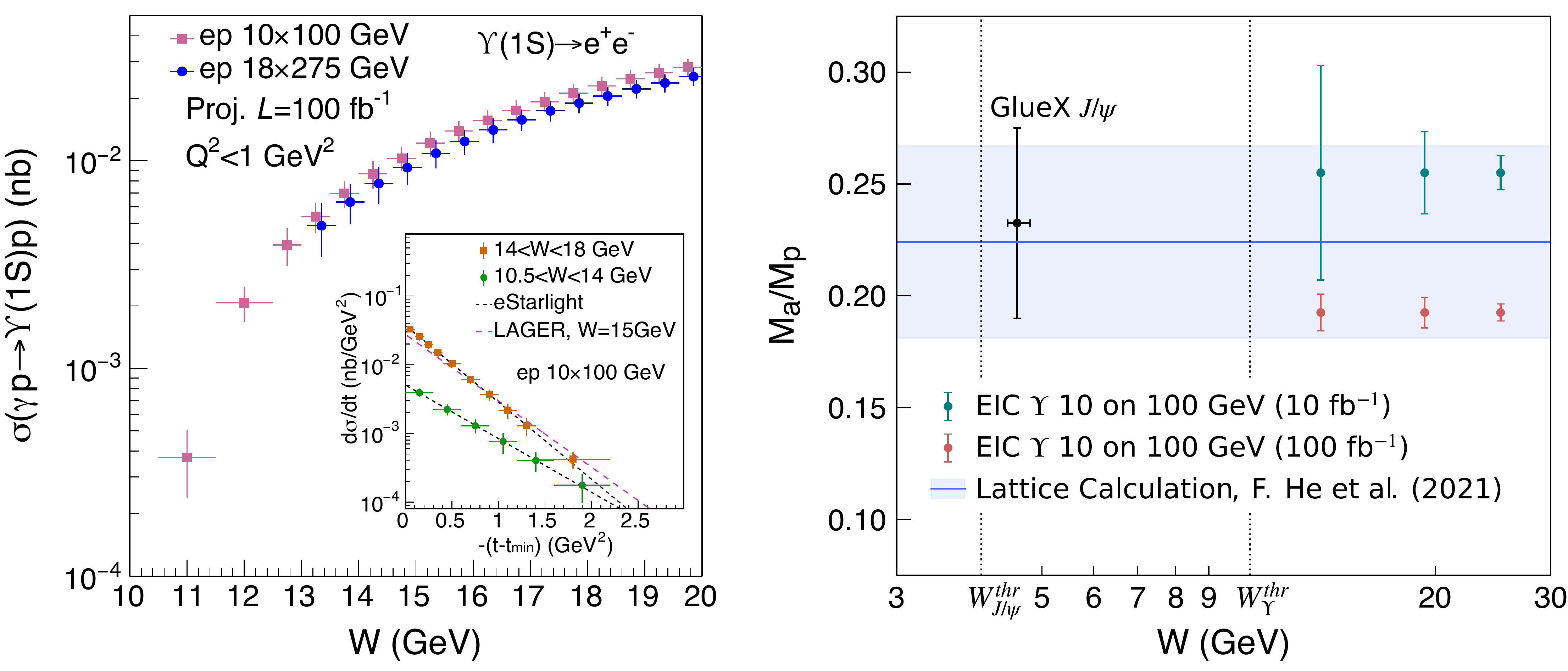}
\caption{\label{fig:sigma-anomaly} Left: The projected uncertainty of the total and differential (insertion) cross sections of $\Upsilon$(1S) near-threshold for photoproduction and electroproduction ($Q^2<1$~GeV$^2$) in \ep\ collisions via the $e^+e^-$ decay channel.  Two model predictions~\cite{Lomnitz:2018juf,Gryniuk:2020mlh} of the near threshold differential d$\sigma$/dt are also shown (FullSim). Right: The trace anomaly contribution to the proton mass in Ji's decomposition according to~\cite{Wang:2019mza,Wang:2021dis} and references therein. Green and red points correspond to 10 fb$^{-1}$ and 100 fb$^{-1}$ integrated luminosity, respectively, and are offset from each other. The band is the result of a recent lattice QCD calculation~\cite{He:2021bof} (FullSim).
}
\end{figure}
Measurement of $d\sigma/dt$ of $\Upsilon$ production near threshold may help determine
the gravitational form factors of the nucleon. 
The main requirements 
are good acceptance over the entire rapidity range, good mass resolution to separate the three $\Upsilon$ states \cite{AbdulKhalek:2021gbh}, and good $t$-resolution. The latter can be measured by either observing the outgoing proton (and measuring $t=(p_\textrm{proton}' - p_\textrm{proton})^2$), or by measuring both the photon momentum and the $\Upsilon$ momentum.  The photon momentum can be determined if the outgoing electron is observed; this requires that the photon has a non-zero $Q^2$.  These methods require precise measurements of the outgoing electron and/or proton momenta, but they are limited  by the momentum spread of the electron and proton beams. Good mass resolution ($<$ 100 MeV) is necessary to be able to separate the three $\Upsilon$ states (see Fig.\ref{fig:masspeak}).  
It is also necessary to be able to cleanly select $\Upsilon$ production events.  This requires good lepton identification, and good momentum resolution to be able to reject the continuum $\gamma\rightarrow \ell^+\ell^-$, as well as large solid angle coverage. 

Figure~\ref{fig:sigma-anomaly} (left) shows the performance achieved by ATHENA with full reconstruction in the near-threshold measurement of $\Upsilon(1S)$ differential (insert) and total cross section in a region never accessed before. While 100~fb$^{-1}$  showcases the full impact of the EIC high luminosity for this process, 10~fb$^{-1}$ is considered for the early running impact. Figure~\ref{fig:sigma-anomaly} (right) shows the constrains on the anomaly contribution to the proton mass in a model-dependent approach~\cite{Wang:2019mza,Wang:2021dis} for integrated luminosies of 10~fb$^{-1}$ and 100~fb$^{-1}$ together with prior data in $J/\psi$-production and a recent lattice evaluation~\cite{He:2021bof}.
The excellent PID performance of ATHENA for both electron and muon decay pairs of the $\Upsilon$  
was accounted for in these projections. 

 \FloatBarrier

\subsection{Gluons in Nucleons and Nuclei}
\subsubsection{Collinear parton distributions in nucleons and nuclei}
\label{nuclei}

\begin{figure}[htb]
\centering
\includegraphics[width=0.75\textwidth]{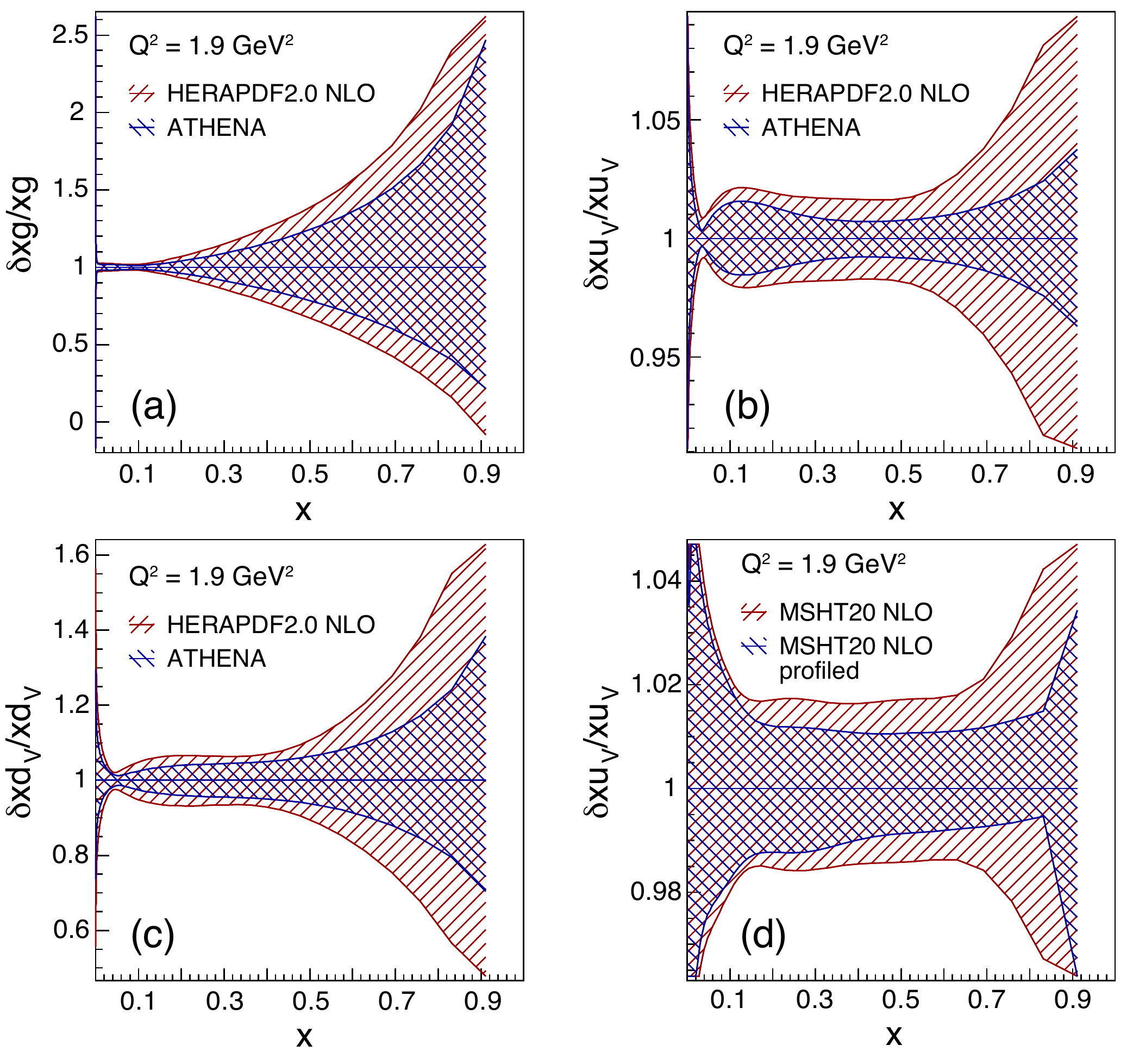}
\caption{Impact of ATHENA on the collinear parton distributions
of the proton at high-$x$. (a-c) Impact of ATHENA data when added to (DIS / HERA only) proton PDF fits~\cite{Alekhin:2014irh}, HERAPDF2.0~\cite{HERA2PDF}, showing the gluon, and the up and down valence quarks. (d) Impact of ATHENA data on the up valence density when added to an example global fit, MSHT20~\cite{MSHT20}, which includes LHC data to constrain high-$x$. 
Profiling in the xFitter framework is used.}
\label{fig:protonPDFs}
\end{figure}

The basic NC cross section data illustrated in Fig.~\ref{fig:ncdata}, together with their CC counterparts, comprise the main ingredients of DGLAP-based fits to extract the collinear parton distributions of the proton. 
Because of the integrated luminosities and the kinematic coverage at intermediate-to-large $Q^2$ and high-$x$,
data from the EIC will complement existing world data in the PDF fits in key ways. 
Charm production at the EIC will provide further constraints.
Figure~\ref{fig:protonPDFs} illustrates how inclusive DIS measurements in ATHENA will constrain the high-$x$ region, by showing the improvement relative to HERA-PDF2.0~\cite{HERA2PDF} and MHST20~\cite{MSHT20}; the ATHENA detector performance at the lowest $y$ (see Sec.~\ref{preamble}) is key. 

As the world's first \eA\ collider, the EIC will explore nuclear structure at an unprecedented level of detail up to the heaviest nuclei. ATHENA will measure nuclear PDFs (nPDFs), or nuclear modification ratios, which encode the deviations of nPDFs from simple scaling of nucleon PDFs with atomic mass $A$. 
The sensitivity  to non-linear QCD effects in inclusive DIS is closely related to the precision and kinematic acceptance of the nuclear modification ratios proposed to be measured by ATHENA.

\begin{figure}[htb]
\centering
\includegraphics[width=0.6\textwidth]{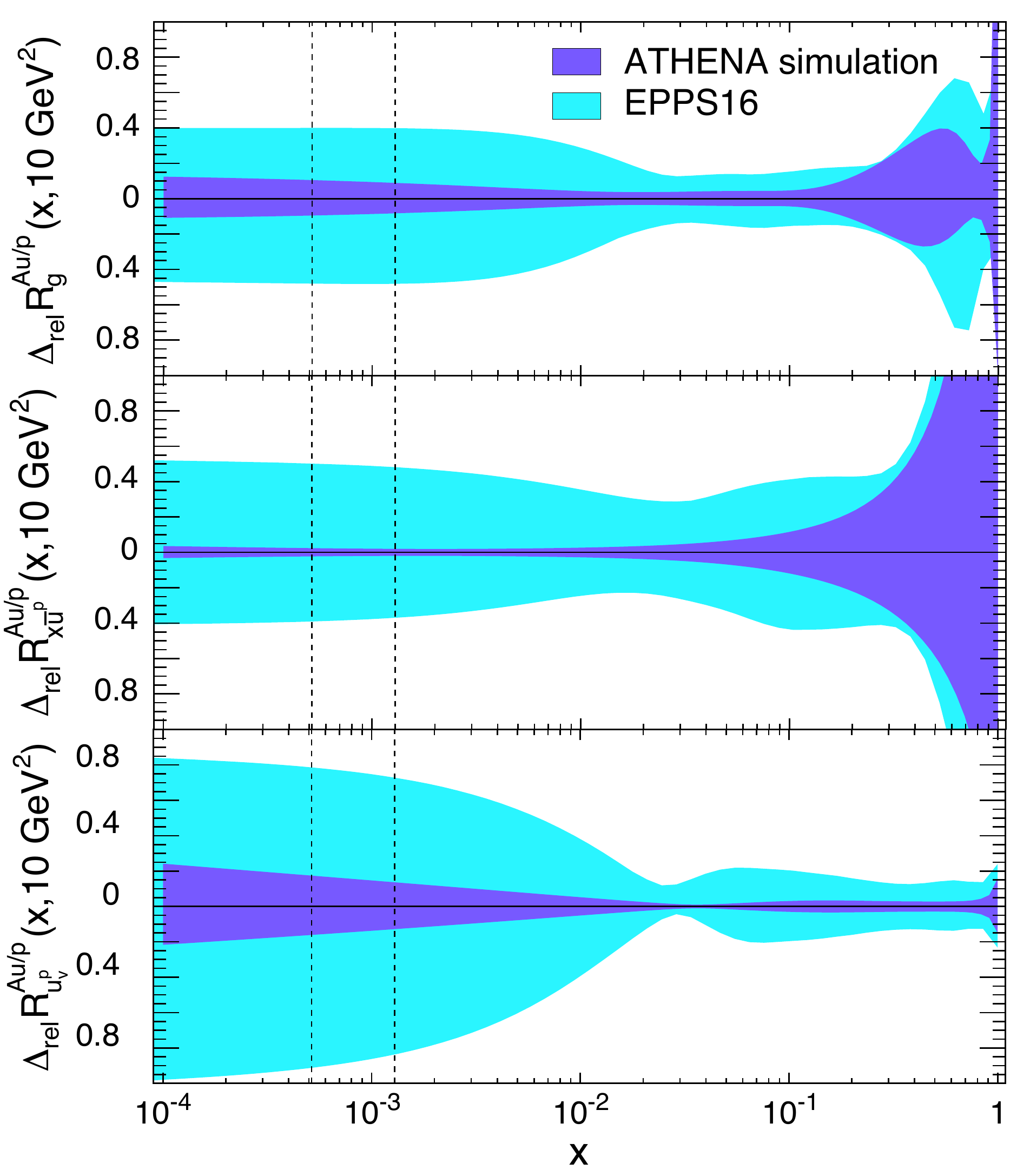}
\caption{Relative uncertainties of the nuclear modification ratios based on fits to simulated ATHENA \ep\ and \eAu\ pseudodata, compared with EPPS16~\cite{Eskola:2016oht}. Gluon density (top), sea up quark density (middle) and valence up quark density (bottom). The vertical lines indicate the minimum $x$ values of the data entering the fits for the proton (left) and nucleus (right).}
\label{fig:nPDFs}
\end{figure}

The impact of ATHENA on nuclear PDFs has been studied in the xFitter framework~\cite{Alekhin:2014irh}. 
Pseudodata from ATHENA only are used as input to fits in which the PDFs evolve according to the \gls{NLO} DGLAP equations, with minimum $Q^2 = 3.5 \ {\rm GeV^2}$, and a parameterization at the starting scale taken from the HERA2PDF studies.
Figure~\ref{fig:nPDFs} shows the results for the pivotal case of the gluon density,
as well as the sea and valence up quark densities.
The low-$x$ precision on the gluon density is $\sim$ 5\% for protons and 10\% for gold. 
The ATHENA-only projections for the nuclear modification ratio are compared
with the precision of a representative current global fit, EPPS16~\cite{Eskola:2016oht}, which
includes data from fixed target DIS and Drell-Yan experiments, hard processes
in \pA\ collisions at the LHC and $\pi^0$ data from PHENIX. The precision on the gluon
for $x \sim 0.1$ is improved by around a factor of two when using ATHENA data. 
The minimum $x$ of data points included in the EPPS16 fit is 0.008, whereas in the ATHENA fits it is approximately 0.001. ATHENA measurements will constrain the nuclear gluon density with a precision of approximately 10\%.

Inclusive DIS data can be complemented at the EIC by SIDIS production of heavy flavor pairs, that are sensitive to the gluon density and its shape through photon-gluon fusion. The large $x$-range explored at the EIC should allow for testing current understanding of gluon shadowing at $x \ll 0.1$, and exploring anti-shadowing at $x \sim 0.1$ and further suppression at $x> 0.3$ (the ``gluonic EMC effect''). 
\begin{figure}[ht]
    \centering
    \includegraphics[width=0.9\textwidth]{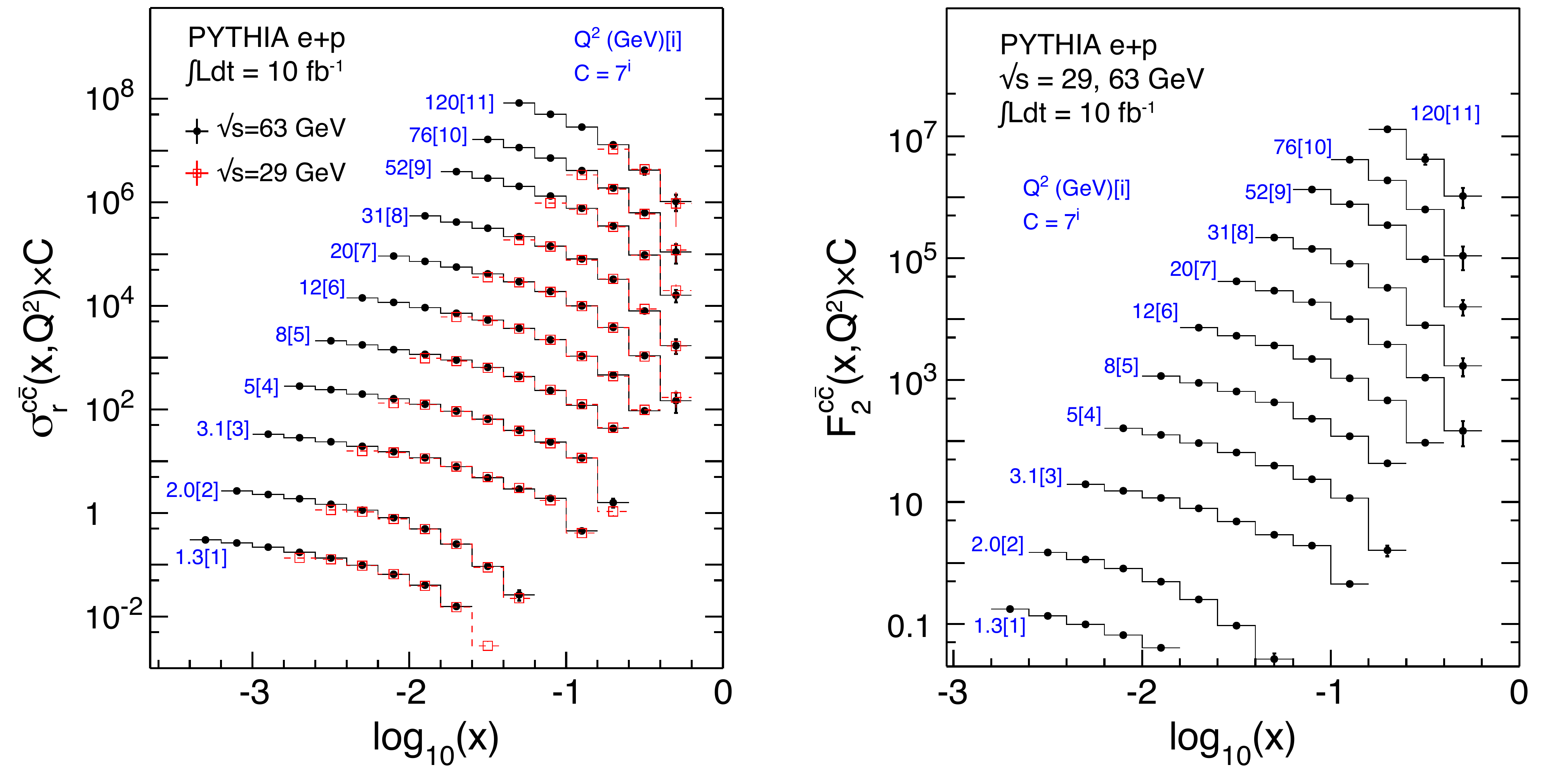} 
 \caption{The reduced charm cross section $\sigma_r^{c\bar{c}}$ (left) and corresponding structure function $F_2^{c\bar{c}}$ (right) in intervals in Bjorken-$x$ for different values of $Q^2$ from PYTHIA simulations and ATHENA response for \ep\, $\rightarrow e' + D^0 +X$ at the indicated center-of-mass energies for an integrated luminosity of 10\,fb$^{-1}$.  The vertical values for different $Q^2$ values are scaled as indicated (FastSim).}
    \label{fig:charm}
\end{figure}

Figure~\ref{fig:charm} shows the projected uncertainties for reduced charm cross section, $\sigma_r^{c\bar{c}}$, and the corresponding structure function $F_2^{c\bar{c}}$, measured via topological reconstruction of the $D^0$ meson.
The detector response includes particle-identification, momentum and single-track pointing resolutions, and primary vertex resolution in the topological reconstruction of $D^0 \rightarrow K+\pi$ decays guided by fast 
and GEANT-based simulations.
The selection criteria correspond to those in Ref.~\cite{Kelsey:2021gpk} and include the transverse displacement of the $D^0$ meson, a criterion on its direction, and the displacement between the $K$ and $\pi$ tracks. The uncertainties are up to 40\% better than those in \cite{Kelsey:2021gpk} owing to ATHENA's acceptance and performance.  Commensurate improvement and impact should be expected from \eA\ collisions via D$^0$ topological reconstruction and from other decay channels and techniques~\cite{Aschenauer:2017oxs}.

\subsubsection{Gluon saturation} 

An exciting opportunity for discovery at the EIC is in the gluon density distribution at small x. It has been predicted that at sufficiently high gluon density, the number of gluons must saturate in order to preserve unitarity of the cross section. In heavy nuclei, the gluon distributions of neighboring nucleons overlap, 
putting the saturation region within experimental reach. ATHENA's large acceptance and superb reconstruction of charged particles, neutrals, and jets provides excellent sensitivity to predicted saturation effects.

\subsubsection*{Di-hadron correlations}
\begin{figure}
  \begin{minipage}[c]{0.6\textwidth}
    \includegraphics[width=0.9\textwidth]{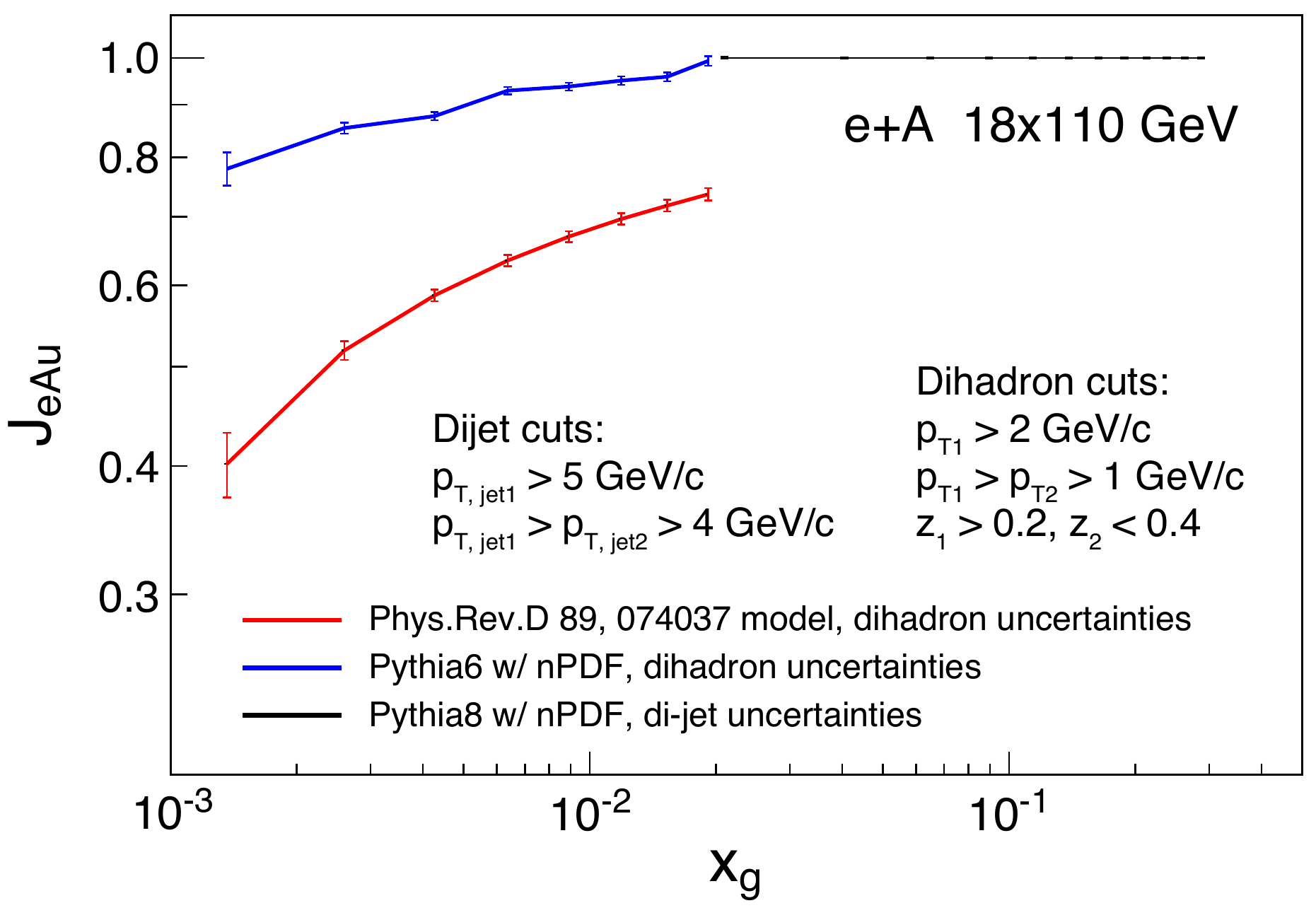}
  \end{minipage}
  \begin{minipage}[c]{0.35\textwidth}
    \caption{J$_{eAu}$ representing the suppression of the away-side peak in di-hadron and dijet production. For di-hadron production the blue curve shows the baseline without saturation effects using the CTEQ nPDFs~\cite{Kusina:2015vfa}, the red curve shows the projected effect from the saturation model in Ref.~\cite{Zheng:2014vka}. In black the projected uncertainties for dijet away-side suppression are shown. While di-hadrons will be a sensitive probe for saturation at ATHENA with reach to low $x_g$, dijets are restricted to $x_g$ above about $10^{-2}$ (FastSim).
    \label{fig:JeAu}}
  \end{minipage}
\end{figure}

In SIDIS production of two back-to-back charged hadrons from electron-nucleus collisions, the azimuthal angle difference $\Delta \phi$ allows to probe the Weizs\"acker-Williams gluon TMD~\cite{Dominguez:2010xd,Dominguez:2011wm,Zheng:2014vka,Dumitru:2018kuw}.
The away-side peak in the di-hadron $\Delta\phi$ correlation is sensitive to the back-to-back jets produced after the collision. When non-linear QCD effects set in, the away-side peak gets decorrelated. 
Figure~\ref{fig:JeAu}  depicts the predicted suppression as function of $x_g$ through J$_{eAu}$, the relative yield of correlated back-to-back hadron pairs in \eAu\ collisions compared to \ep\ collisions scaled down by A$^\frac{1}{3}$ (the number of nucleons at a fixed impact parameter). In absence of collective nuclear effects in the pair production cross section, J$_{eAu} \sim 1$, as is seen for jets and simulations with only accounting for effects due to nPDFs. While J$_{eAu} <$ 1 signifies suppression of the di-hadron correlations.
The quality of the ATHENA di-hadron measurement is illustrated by the uncertainties, shown as the vertical bars. 


\subsubsection*{Exclusive vector meson production in \eA}
Coherent diffractive vector meson production off heavy nuclei has been considered as one of the golden measurements at the EIC. It offers a clean measurement of the gluon spatial distribution in  nuclei. By using the lever arm in beam energy, $Q^{2}$, and different VM species, this measurement will be a promising experimental probe to the saturation dynamics. In hard diffractive \eA\ events, \emph{exclusive} vector meson production and DVCSs are the only processes that allow to determine $t$. Except for very light nuclei, the scattered nucleus, A$^\prime$, stays within the beam envelope and cannot be observed directly; exclusive processes allow to derive the A$^\prime$ kinematics from the rest of the event.

The measurement of the gluon spatial distribution requires a precise determination of the momentum transfer $|t|$ in coherent vector meson production. High precision $|t|$ measurements require high-resolution reconstruction of the scattered electron and the final-state vector meson decay products. This is provided by ATHENA's excellent tracking performance in its 3\,T field in conjunction with high-resolution electromagnetic calorimetry in the backward region. Furthermore, a high-purity coherent sample is needed, where the main physics background of large incoherent productions can be suppressed by vetoing the nuclear breakups using the Far-Forward (FF) region detectors \cite{Chang:2021jnu}.
Sufficient suppression of the incoherent background to a level lower than all three minima of the reconstructed coherent $|t|$ distribution is assumed in the following.
To reduce electron and ion beam effects such as angular divergence and momentum spread, which can impact the resolution of the $|t|$ reconstruction, one can remove these effects by an improved $|t|$ reconstruction method~\cite{AbdulKhalek:2021gbh}, known as ``Method~L", which is used below.

\begin{figure}[thb]
\centering
\includegraphics[width=0.9\textwidth]{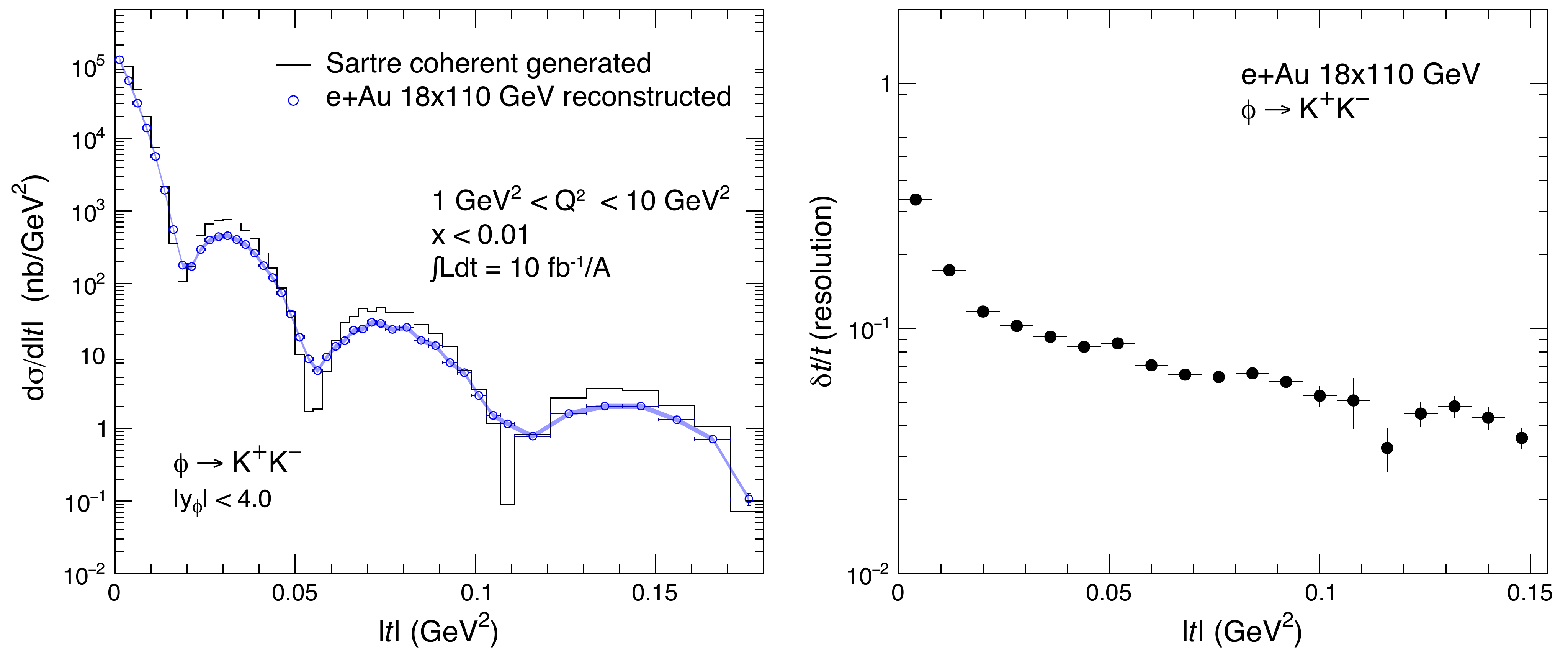}
  \caption{ \label{fig:figure_1} Left: Differential $|t|$ distribution of diffractive $\phi$-mesons in \eAu\ collisions of 18 GeV electron beams with 110~GeV Au beams.
   Distributions of the coherent differential cross section from the Sartre event generator and its reconstruction with ATHENA (``Method L") are shown. Right:
  The corresponding $|t|$ resolutions versus the generated $|t|$ (FastSim).
  }
\end{figure}

For our studies we have chosen the $e + \mathrm{Au} \rightarrow e^\prime + \mathrm{Au}^\prime +\phi$ process since it is the most challenging to measure.
Figure~\ref{fig:figure_1} left shows the  differential $\mathrm{d}\sigma/\mathrm{d}|t|$ distribution of coherent production of $\phi$ mesons in \eAu\ collisions
at 18 $\times$ 110~GeV. The plot on the right shows the corresponding $t$ resolution, $\delta t/t$, as a function of $t$.
These $|t|$ resolutions, on the order of 10--20\% at low $|t|$ and 5\% at higher $|t|$, enable ATHENA to measure the minima positions of the diffractive shape in the $|t|$ distribution. Resolving these structures is the key to measuring gluon GPDs in nuclei. 
 
\subsubsection{Properties of cold nuclear matter}

The precision and control over initial kinematics at the EIC, together with the large 
$\sqrt{s}$ range, high luminosity, and capability to collide a wide range of nuclei enable an unprecedented 
exploration of the properties of nuclear matter. 

\begin{figure}[ht!]
\begin{minipage}[c]{0.55\textwidth}
    \centering
    \includegraphics[width=0.8\textwidth]{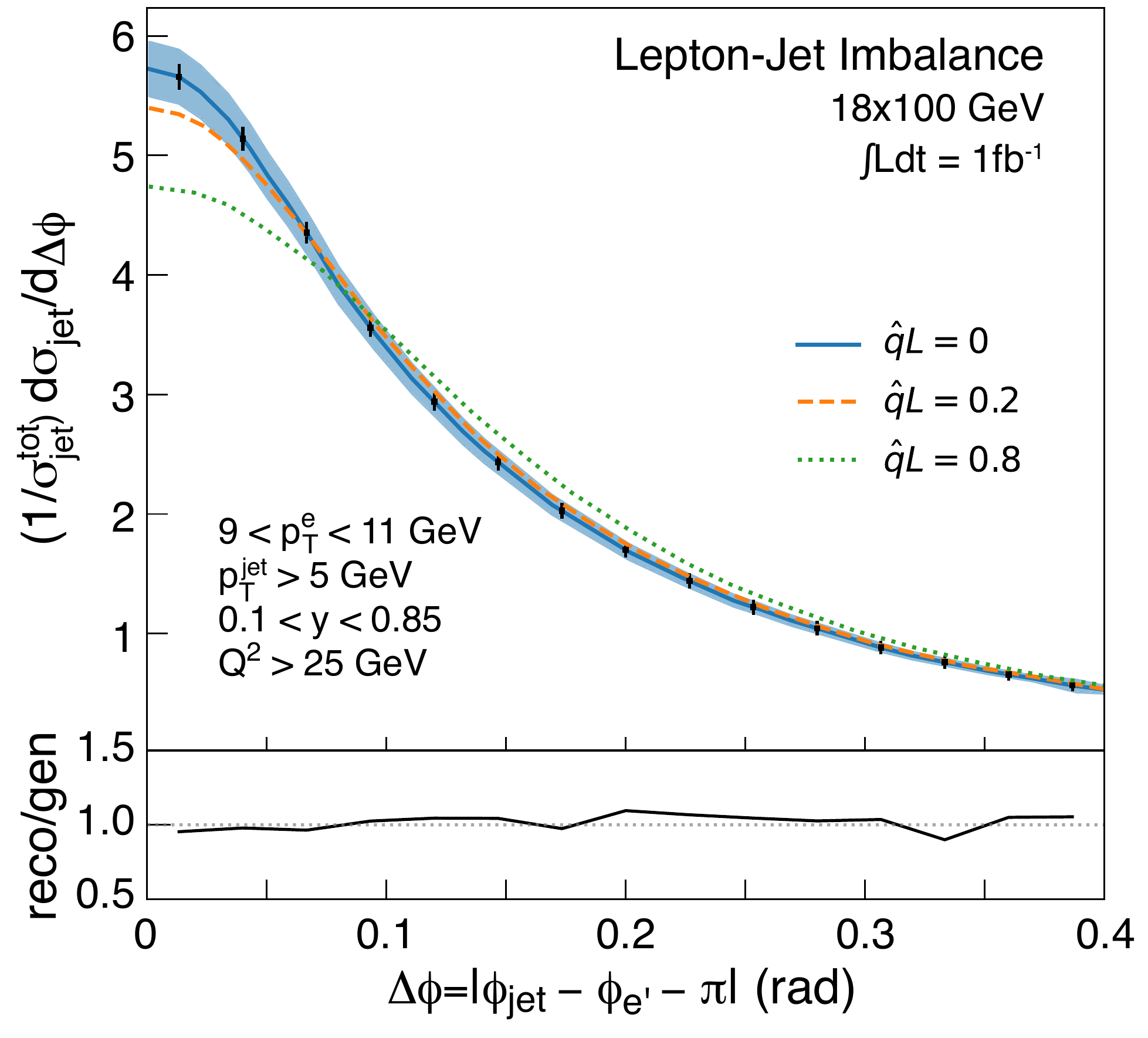}
  \end{minipage}
  \begin{minipage}[c]{0.3\textwidth}
\caption[Lepton-Jet Correlations]{Azimuthal angle between jet ($R_\mathrm{jet}=1.0$) and the scattered electron. The vertical bars show statistical uncertainties, while the band represents the estimated 4\% point-to-point systematic uncertainty. This assumes a jet-energy scale uncertainty of 1\%. Theory curves are taken from Ref.~\cite{Liu:2018trl} (FastSim).} 
\label{fig:LEPJET}  
  \end{minipage}
\end{figure}

For example, jet production in DIS can probe the properties of \gls{cnm} by comparing the momentum of the jet to that of the scattered electron, which is unaffected by the nuclear medium~\cite{Arratia:2019vju}.
Figure \ref{fig:LEPJET} shows the opening angle distribution between the jet and the scattered electron. This observable is sensitive to partonic energy loss and the jet transport coefficient, $\hat{q}$, which are both important for characterizing CNM.
ATHENA's large acceptance and excellent electron and jet energy resolutions are well-suited for these measurements. 
With their close connection to the scattered parton and well defined internal structure, jets have emerged as a key addition to the experimental `toolbox'.

\subsubsection*{Jet substructure: Shower modification and hadronization in cold nuclear matter}

\begin{figure}[ht!]
 \begin{minipage}[c]{0.65\textwidth}
    \centering
\includegraphics[width=0.95\textwidth]{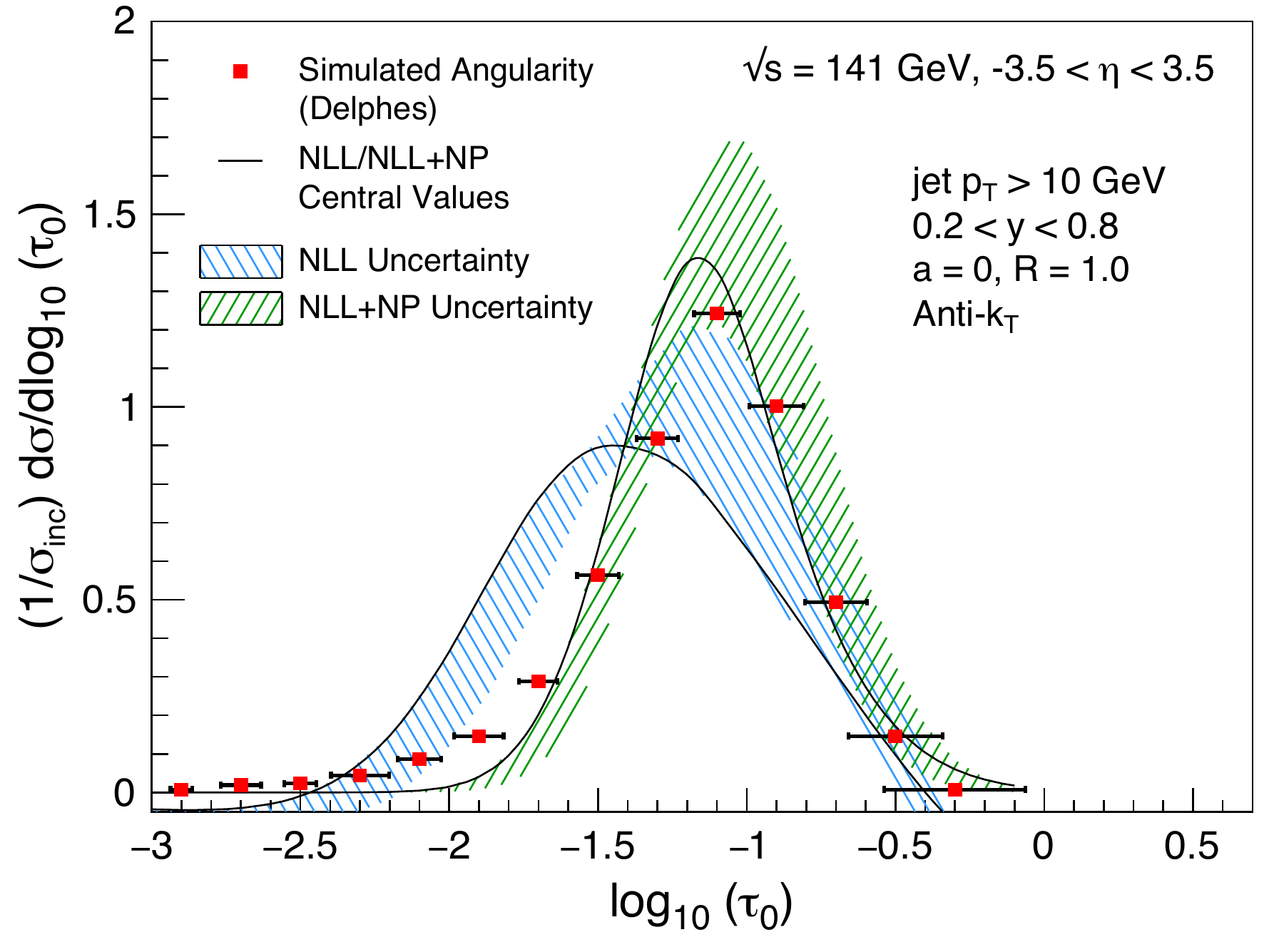}
\end{minipage}
 \begin{minipage}[c]{0.3\textwidth}
\caption[Angularity]{Jet angularity reconstructed using the ATHENA Delphes model and energy-flow algorithm compared to NLL theory predictions and NNL + non-perturbative effects~\cite{Aschenauer:2019uex}. The horizontal bars represent the angularity resolutions in each bin. Statistical uncertainties were scaled to 1~fb$^{-1}$ and are not visible on this scale (FastSim).}
\label{fig:JETANGULARITY}   
\end{minipage}
\end{figure}

Comparisons between vacuum and medium showers can shed light on the process of hadronization, in which partons shed energy and virtuality to form the final state particles we observe. Detailed studies of parton showers are possible via jet substructure observables, which are sensitive to the distribution of energy within a jet. An example substructure observable 
is jet angularity, which is defined as~\cite{Aschenauer:2019uex}:

\begin{equation}
    \tau_a=\frac{1}{p_T}\sum_{i\in j}p_{Ti}\Delta R_{ij}^{2-a} ,\,
\end{equation}

\noindent where $p_T$ is the jet transverse momentum; $p_{Ti}$ and $\Delta R_{ij}$ are the transverse momentum and distance from the jet axis of the $i^{\mathrm{th}}$ particle, respectively; $a$ is a continuous parameter. Jet angularity is sensitive to a convolution of perturbatively describable hard processes and non-perturbative hadronization physics.
The blue and green bands in Fig.~\ref{fig:JETANGULARITY} illustrate this. Precision comparisons of angularity between \ep\ and \eA\ for different $p_T$, $R$, and $a$ values will enable highly differential characterizations of CNM properties, 
shower formation and evolution, as well as vacuum vs. in-medium hadronization.

\subsubsection{Fragmentation and hadronization}

Fragmentation functions are required for the interpretation of \gls{sidis} measurements in terms of quark and gluon degrees of freedom of the initial state. Furthermore, measurements of the hadronized final state can provide insights into the mechanism(s) by which partons transfer energy and hadronize in nuclear matter.
This sensitivity to in-medium hadronization and transport properties is a highlight area in the EIC White-Paper~\cite{Accardi:2012qut}. Jet substructure techniques are being rapidly developed, and also offer a novel tool to study hadronization.

\begin{figure}[htb]
  \begin{minipage}[c]{0.65\textwidth}
    \centering
    \includegraphics[width=0.8\textwidth]{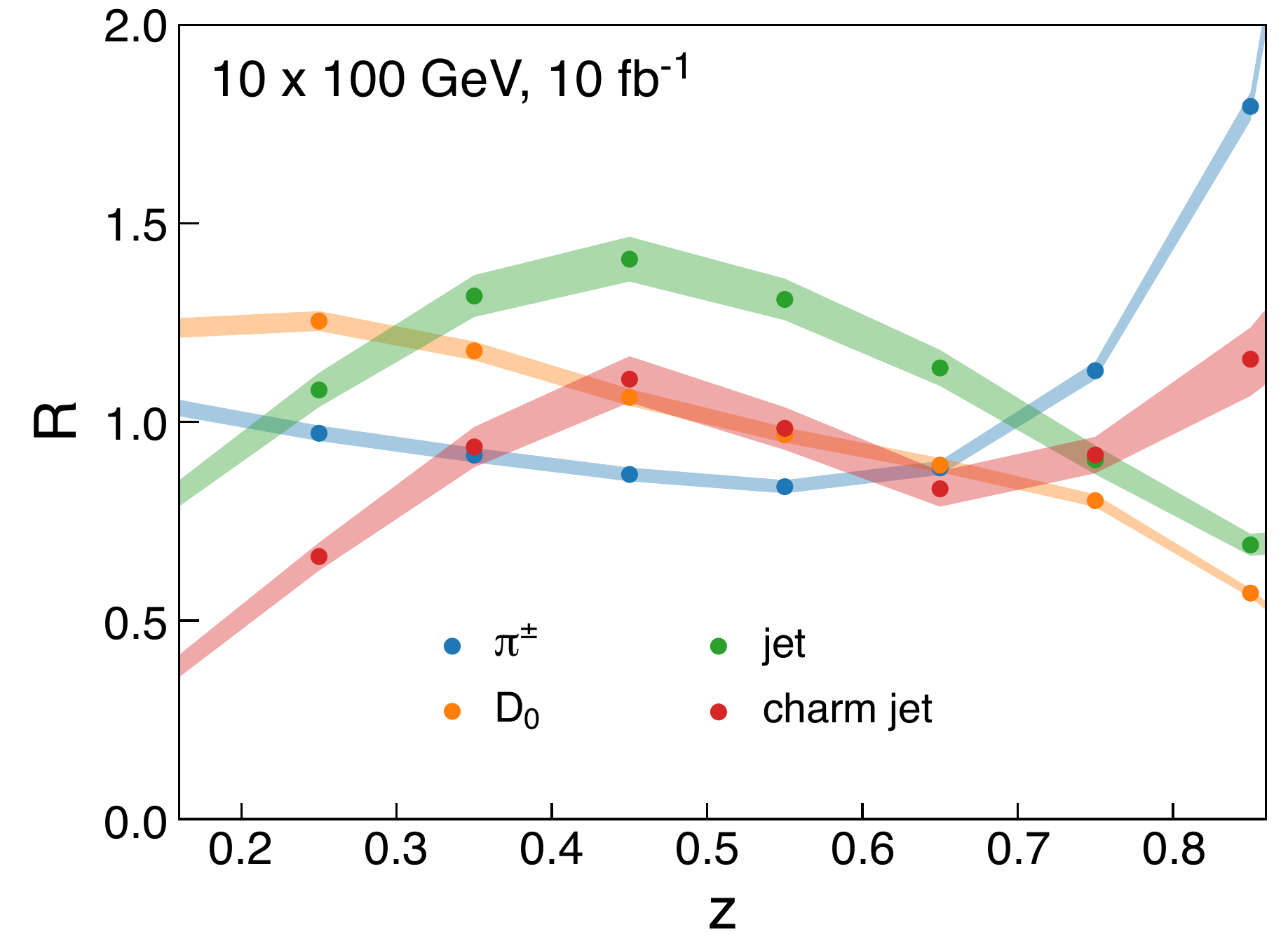}
  \end{minipage}
  \begin{minipage}[c]{0.3\textwidth}
    \caption{The double ratio $R$ of $\pi^{\pm}$, D$^0$, inclusive jet ($R_\mathrm{jet}=1.0$), and charm-tagged jet ($R_\mathrm{jet}=1.0$) production per scattered electron in \eA\ collisions to the corresponding rate in \ep\ collisions, evaluated using the BeAGLE event generator for \eA\ collisions. The band shows the dominant source of systematic uncertainty, as described in the text. Statistical uncertainties are smaller than the symbol size (FastSim).}
    \label{fig:JETREA}  
  \end{minipage}
\end{figure}

A key observable is the double ratio, $R$, of meson production to inclusive production in \eA\ DIS to that in \ep\ collisions.
Charmed mesons exhibit qualitatively different characteristics from light mesons as their hadronization differs. The BeAGLE event generator~\cite{BeAGLE} was used in Fig.~\ref{fig:JETREA} to show $D^0$ production rates per scattered electron in \eA\ compared to \ep\ collisions together with the corresponding charged pion projection versus fragmentation $z$.
The systematic uncertainty estimate corresponds that for other \gls{sidis} measurements.

ATHENA detector capabilities will also make it possible to study this observable for reconstructed jets and charm-tagged jets, which also offer the possibility to use the jet radius as an additional variable.
The jet-energy scale uncertainty partially cancels in the double ratio and the residual uncertainty is estimated to be at the level of 4\%.
The uncertainty for the charm-tagged projection was obtained by propagating the uncertainty on the signal purity.
This is anticipated to be the dominant source.

\begin{figure}[htb]
  \begin{minipage}[c]{0.65\textwidth}
    \centering
    \includegraphics[width=0.8\textwidth]{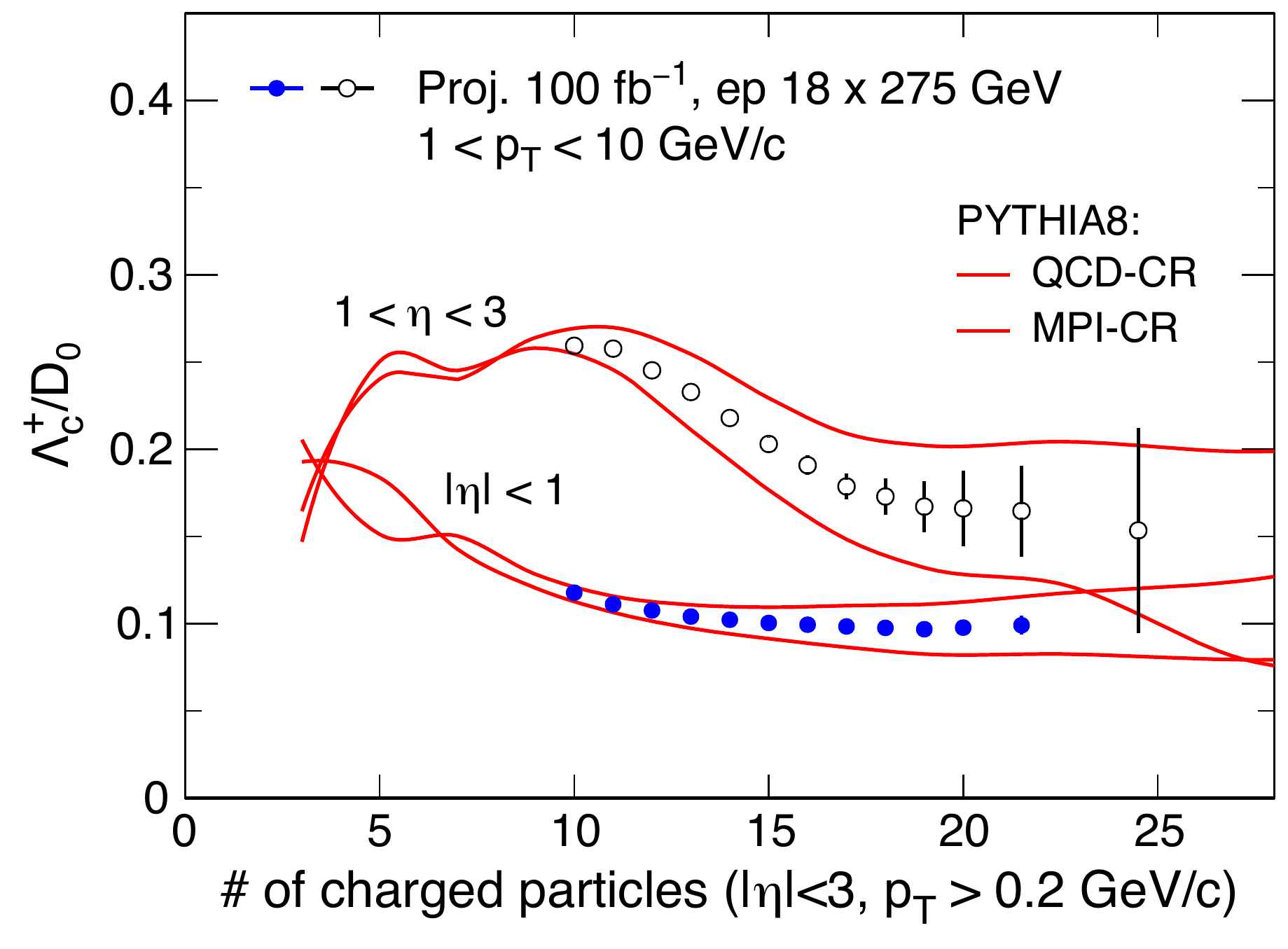}
  \end{minipage}
  \begin{minipage}[c]{0.32\textwidth}
    \caption{Projections for ATHENA measurements of the heavy-quark $\Lambda_c^+$ to $D^0$ baryon-to-meson ratio as a function of the charged track multiplicity (FastSim).}
    \label{fig:LcD0_EIC_Multi_proj}
  \end{minipage}
\end{figure}

Recent heavy-quark measurements at the LHC have revived the interest in baryon-to-meson ratios.
The cross sections are usually computed using the factorization approach as a convolution of the parton distribution functions of the initial state, the calculable QCD hard scattering cross sections at the partonic level, and fragmentation
functions into a particular meson or baryon.
The fragmentation functions are typically tuned on $e^+ + e^-$ data and are often thought to be universal.
HERA \ep\ measurements of the $\Lambda_c$ to $D^0$ baryon-to-meson ratio were indeed found consistent with those from $e^+ + e^-$ data, within their uncertainties.
However, recent measurements in \pp\ of the $\Lambda_c^+$ to $D^0$ ratio by the ALICE and CMS collaborations at $\sqrt{s} = 5\,\mathrm{TeV}$ and $\sqrt{s} = 7\,\mathrm{TeV}$ show an enhancement.
Figure~\ref{fig:LcD0_EIC_Multi_proj} shows ATHENA projections for topologically reconstructed $\Lambda_c$ and $D^0$ mesons as a function of track multiplicity compared with expectations from PYTHIA, tuned to current LHC data.

\FloatBarrier

\printnoidxglossary[type=\acronymtype, style=mcolindex]

\acknowledgments
\phantomsection \addcontentsline{toc}{section}{Acknowledgments}
We thank
Xiaocong Ai (DESY),
Corentin Allaire (CERN),
Nestor Armesto (Universidade de Santiago de Compostela),
Ignacio Borsa (Universidad de Buenos Aires), 
Paul Gessinger-Befurt (CERN),
Graham Heyes (JLab),
Charles Hetzel (BNL),
Jerome Lauret (BNL),
Renuka Rajput-Ghoshal (JLab),
Andreas Salzburger (CERN),
Rodolfo Sassot  (Universidad de Buenos Aires), 
Nobuo Sato (JLab),
Marcy Stutzman (JLab),
Katarzyna Wichmann (DESY),
Roland Wimmer (BNL),
Holger Witte (BNL), and
Yiyu Zhou (William \& Mary and JLab)
for fruitful discussions, constructive suggestions, and contributions.

This work was supported in part by the Office of Nuclear Physics within the U.S. DOE Office of Science, the U.S. National Science Foundation, the ULAB-EIC LDRD program at Argonne National Laboratory, the MRPI program of the University of California Office of the President, the LDRD program at Los Alamos National Laboratory, the  Natural Sciences and Engineering Research Council of Canada (NSERC), the National Natural Science Foundation (NSFC) and Ministry of Science and Technology of China, the Chinese Academy of Sciences, the Czech Science Foundation and Ministry of Education, Youth and Sports (MEYS), the European Union’s Horizon 2020 Research and Innovation program under Grant Agreement No.~101004761 (AIDAInnova) and Grant Agreement No.~824093 (STRONG2020), the French Centre National de la Recherche Scientifique, the French Commissariat \`{a} l’Energie Atomique, GSI Helmholtzzentrum f\"{u}r Schwerionenforschung GmbH, the Department of Atomic Energy and Department of Science and Technology of the Government of India, the Italian Ministry of Foreign Affairs and International Cooperation (MAECI) as Projects of Great Relevance within Italy/US Scientific and Technological Cooperation under Grant No MAE0065689-PGR00799, the Ministry of Education and Science of Poland, the United Kingdom Science and Technology Facilities Council (STFC), and the Ministry of Education of Taiwan.
The simulation studies have been performed making use of the computer resources of ANL-PHY, OSG, JLab, BNL, Compute Canada, ALCF, LCRC, NERSC, and INFN-CNAF.

\bibliographystyle{unsrt}
\renewcommand{\bibname}{References}
\phantomsection \addcontentsline{toc}{section}{References}
\bibliography{master.bib}

\end{document}